\documentclass[12pt]{article}  
\usepackage{jheppub}\usepackage{titlesec} 
\usepackage{amsmath}\usepackage{amsfonts}\usepackage{amssymb}\usepackage{mathtools}\usepackage{mathrsfs} \usepackage{bbm} 
\usepackage{xcolor}\usepackage{color}
\usepackage{dsfont}\usepackage{subfigure}
\usepackage{hyperref}  \usepackage{microtype}   
\usepackage{float}\usepackage{psfrag}
\usepackage{graphicx}\usepackage{graphics}
\usepackage{multirow}\usepackage{array} \usepackage{feynmp}
\usepackage{pifont} 
 \usepackage{bm}
\usepackage{slashed} \usepackage{pgf} \usepackage{epsfig}
\usepackage[margin=5pt,font=normalsize,labelfont=bf,justification=raggedright]{caption}
\newcommand{\normord}[1]{:\mathrel{\mspace{1mu}#1\mspace{1mu}}:}
\title{\Large On massive neutral lepton scattering on nucleus}
\author[1]{V.A.~Bednyakov}
\emailAdd{bedny@jinr.ru}
\affiliation[1]{Dzhelepov Laboratory of Nuclear Problems, JINR, 141980, Dubna, Russia}
\abstract{The paper presents a theoretical approach to the description of the relativistic scattering of a massive (neutral) lepton on a nucleus, in which the latter retains its integrity. The measurable cross section of this process includes the elastic (or coherent) contribution, when the nucleus remains in its original quantum state and the inelastic (incoherent) contribution, when the nucleus goes into another (excited) quantum state. 
Transition from the elastic scattering regime to the inelastic scattering regime is regulated {\em automatically}\/  by the dependence of the nucleon-nucleus form factors on the momentum transferred to the nucleus. 
At small momentum transfers elastic scattering dominates. 
AS the transferred momentum increases, the contribution of the inelastic scattering increases, 
and the latter becomes dominant at sufficiently large transferred momenta.
The interaction of a pointlike lepton with structureless nucleons of the target nucleus 
is parameterized  with four effective coupling constants, 
reflecting the (axial)vector nature of the weak interaction. \par
 The scattering of massive (anti)neutrinos interacting with nucleons through the $V\mp A$ currents 
 of the Standard Model is considered in detail. 
 Because of the nonzero masses, an additional channel arises for elastic and inelastic scattering of 
 these (anti)neutrinos on nuclei due to the possibility of changing the helicity of these (anti)neutrinos.
For example, despite the smallness of the masses at (kinetic) energies of (anti)neutrinos much lower than the neutrino masses (for example, relic ones), the cross section of their interaction with the nucleus turns out to be many times enhanced, at least due to the  "nucleus coherence effect"\/.  \par
The expressions obtained  for the cross sections are applicable to any precision data analysis involving neutrinos and antineutrinos, especially when non-zero neutrino masses can be taken into account.
These expressions can also be used in the analysis of experiments on direct detection of (neutral) massive weakly interacting  {\em relativistic}\/ dark matter particles since, unlike the generally accepted case, 
they simultaneously take into account both elastic and inelastic interactions of the particles. 
The presence of an "inelastic signal"\/ with its characteristic signature may be the only registrable 
evidence of interaction of the dark matter particle with the nucleus.}
\begin{document}
\maketitle

\section{\large  Introduction}\label{1chiA-IntroDuction} 
In \cite{Bednyakov:2018mjd,Bednyakov:2019dbl,Bednyakov:2021ppn}, 
an approach was formulated to describe the interaction of {\em massless}\/ neutrinos and antineutrinos with a nucleus resulting in that the nucleus either remains in its original state or goes into an excited state while maintaining the integrity  $\nu(\bar\nu ) +A\to \nu(\bar\nu) +A^{(*)}$.
In \cite{Bednyakov:2022dmc}, this approach was generalized to the case of nonrelativistic scattering of a massive neutral weakly interacting lepton on nuclei.
On this basis, the balance of coherence and incoherence was studied in the problem of direct dark matter particle search with allowance for specific kinematic conditions and excitation energy levels of the nucleus.
It was noted that one usually underestimates the role of inelastic processes in this problem \cite{Bednyakov:2023dmc}.
\par
Recall that the approach \cite{Bednyakov:2018mjd,Bednyakov:2019dbl,Bednyakov:2021ppn}
is based on the description of the nucleus as a bound state of its structureless nucleons, whose interaction with the lepton is given by the effective 4-fermionic Lagrangian and is described by means of the scalar products of the lepton and nucleon currents, which allows one to control the spin states of nucleons in the initial and
 final states of the nucleus.
Due to the constructive use of the completeness of the wave functions of nuclear states
(conservation of probability), it is possible to represent the observable cross section as a sum of two terms.
One term corresponds to the elastic interaction, in which the initial quantum state of the nucleus is preserved.
The other term is the total contribution to the observable cross section of all other (potentially possible) inelastic processes, which involve a change in the quantum state of the nucleus.
\par
It was found that as the 3-momentum $\bm{q}$ transferred to the nucleus increases, 
the nuclear form factors of the proton/neutron $F_{p/n}(\bm{q})$ {\em  automatically regulate}\/ the transition from 
the elastic regime to the inelastic regime of the lepton-nucleus interaction.
This phenomenon is the {\em fundamental difference}\/ of the approach
\cite{Bednyakov:2018mjd,Bednyakov:2019dbl,Bednyakov:2021ppn} from Friedman's concept of coherence
\cite{Freedman:1973yd,Freedman:1977xn}, when, based on the result of comparing of the radius of the target nucleus and the characteristic momentum transferred to it, one decided which formula --- coherent or incoherent --- should be used to calculate the cross section for (anti)neutrino scattering by the nucleus.
  \par
It was also noted that elastic and inelastic neutrino $\nu(\bar\nu) A$ processes turn out to be experimentally indistinguishable 
when the recoil energy of the nucleus is the only observable quantity.
Therefore, in experiments aimed at studying the coherent scattering of (anti)neutrinos
(at sufficiently high energies) by detecting only the recoil energy of the nucleus \cite{Akimov:2017ade}, an incoherent background can occur that is indistinguishable from the signal 
when $\gamma$-quanta de-exciting the nucleus cannot be registered.
\par
A similar situation takes place in the direct detection of dark matter particles.
The incoherent (inelastic) contribution to the expected event rate dominates in a number of kinematic regions 
at quite admissible values of the New Physics parameters
\cite{Bednyakov:2022dmc,Bednyakov:2023dmc}.
This incoherent contribution is of independent interest because it is an additional source 
of important information about the nature (unknown in the case of dark matter) of the interaction.
It can be measured directly by detection of photons emitted by the target nuclei excited 
as a result of inelastic processes 
\cite{Bednyakov:2021bty}.
The number of such photons is proportional to the ratio of the cross section of the inelastic interaction channel
to the cross section of the elastic channel, and these photons should have an energy spectrum characteristic of the nucleus $A$, which, as a rule, is much larger than the recoil energy of the nucleus, which could greatly simplifiy their detection.
\par
In this context, the goal of this paper is to further  generalize  both the {\em massless neutrino}\/
\cite{Bednyakov:2018mjd,Bednyakov:2019dbl,Bednyakov:2021ppn}
and the {\em massive lepton}\/ \cite{Bednyakov:2022dmc,Bednyakov:2023dmc} approaches
to the {\em relativistic}\/ case of interaction of {\em massive}\/ neutral weakly interacting particles
($\chi$ leptons) with nuclei $\chi A\to \chi A^{(*)}$,
and elucidate new regularities that arise within the generaliztion.

\section{\large Kinematics and cross section of ${\chi A}$ scattering}\label{2chiA-Kinematics}
\begin{figure}[h]
\includegraphics[scale=1.3]{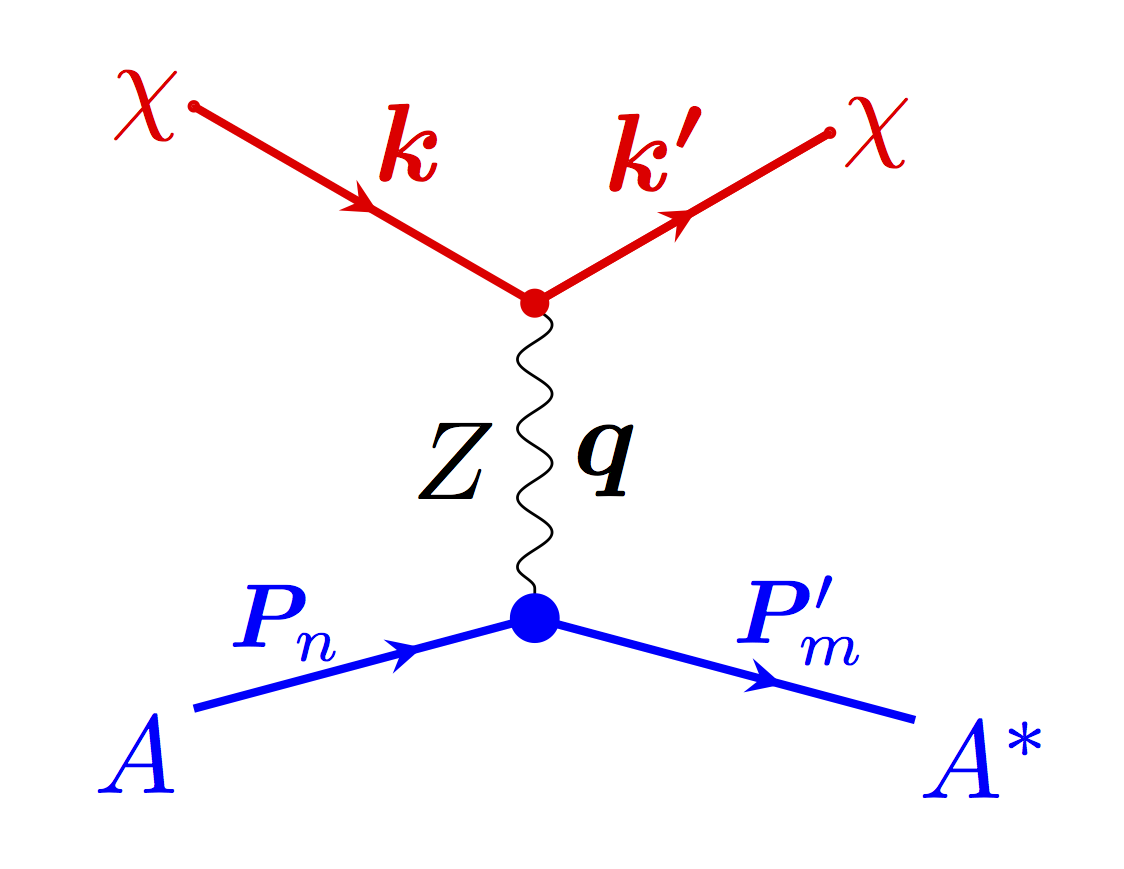}\includegraphics[scale=0.3]{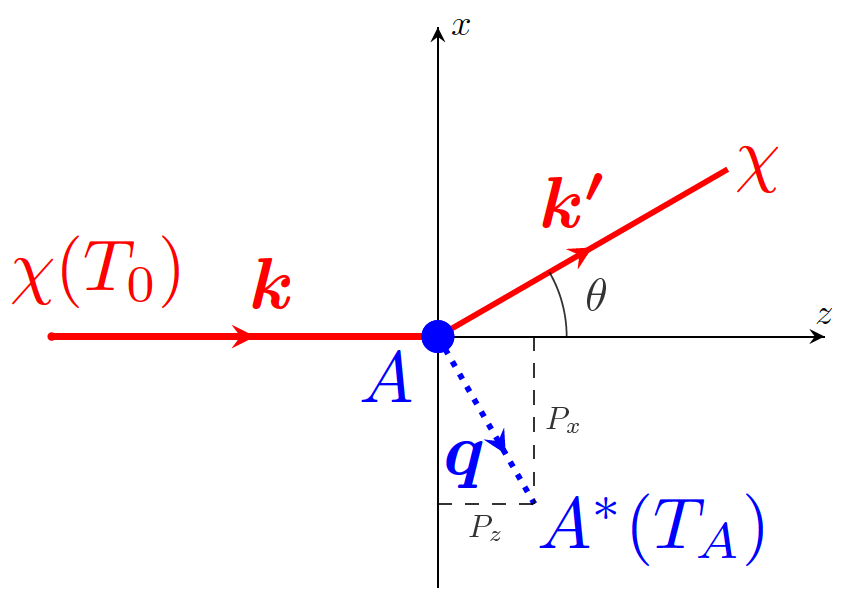}
\caption{\small An example of $\chi A$-interaction due to the exchange of the neutral $Z$-boson (left). 
Kinematics of this process in the laboratory frame where the nucleus $A$ is at rest (right). } \label{fig:DiagramCENNS}
\end{figure}

When two particles interact with the formation of two particles, $\chi+A\rightarrow \chi+A^{(*)}$,
the 4-momenta of the incoming and outgoing $\chi$ particles are denoted as $k=(k_0=E_\chi,\bm{k})$ and $k'=(k'_0=E_\chi',\bm{k}')$, and the 4-momenta of the initial and the final state of the nucleus, respectively, as $P_n=(P^0_n,\bm{P}_n)$ and $P'_m=(P^0_m, \bm {P}_m)$ (Fig.~\ref{fig:DiagramCENNS}, left).
The total energy of the nuclear state $|P_n\rangle$ is $P_n^0 = E_{\bm{P}}+\varepsilon_n$,
where $\varepsilon_n$ is the internal energy of the $n$th quantum state of the nucleus.
If the $\chi$ particle with mass $m_\chi$  and momentum $\bm{k}$ hits the nucleus $A$ at rest along the $z$ axis 
and flies away at angle $\theta$ to the $x$-axis with momentum $\bm{k}'$,
then all 4-momenta can be written as 
\begin{eqnarray*}
k&=&\big(k_0=\sqrt{m^2_\chi+|\bm{k}|^2}, 0, 0, k_z=|\bm{k}|\big), \qquad
P_n = \big(P_n^0=m_A +\varepsilon_n, 0, 0, 0\big), \\
k'&=&\big(k'_0=\sqrt{m^2_\chi + |\bm{k}'|^2}, k'_x= |\bm{k}'|\sin\theta, 0, k'_z = |\bm{k}'|\cos\theta\big), \\
P'_m &=& \big(P_m^0=\varepsilon_m + \sqrt{m_A^2 + \bm{q}^2}, - |\bm{k}'|\sin\theta, 0, |\bm{k}|- |\bm{k}'|\cos\theta\big),
\end{eqnarray*}
where $m_A$ is the mass of the $A$ nucleus, and $\varepsilon_m$ is the excitation energy of the $m$th level of this nucleus
\cite{Bednyakov:2022dmc}.
The 4-momentum $q =(q_0, \bm{q})$ transferred to the nucleus is related to these quantities in the following way:
\begin{eqnarray} \nonumber\label{eq:Kinematics-momentum-transfer}
q^2&\equiv &(k-k')^2 =2 \big(m^2_\chi -\sqrt{(m^2_\chi + |\bm{k}'|^2)(m^2_\chi + |\bm{k}|^2)} + |\bm{k}| |\bm{k}'|\cos\theta\big),
\\ q_0 &=& k_0 -k'_0 = P^0_m - P^0_n = \Delta\varepsilon_{mn}+ T_A,
\\ \bm{q}^2 &=& (\bm{k}-\bm{k}')^2 = (-|\bm{k}'|\sin\theta)^2+(|\bm{k}|- |\bm{k}'|\cos\theta)^2=
  |\bm{k}|^2 + |\bm{k}'|^2 - 2 |\bm{k}| |\bm{k}'|\cos\theta.
\nonumber\end{eqnarray}
The notation for the energy difference between the nuclear $|m\rangle$ and $|n\rangle$ states is introduced:
\begin{equation} \label{eq:Kinematics-delta_Epsilon}
\Delta\varepsilon_{mn} \equiv \varepsilon_m-\varepsilon_n.
\end{equation}
The kinetic energy of the motion of the recoil nucleus is defined as
\begin{equation} \label{eq:Kinematics-Recoil-kinetic-energy}
T_A(|\bm{k}'|, \cos\theta)=\sqrt{m^2_A+|\bm{k}'|^2+|\bm{k}|^2 - 2 |\bm{k}| |\bm{k}'|\cos\theta}-m_A
.\end{equation}
The maximum value of the recoil energy, $T_A^{\max}$, corresponds to the maximum momentum transferred, which is obtained at $\cos\theta =-1$, i.e. $\bm{q}^2_{\max}= (|\bm{k}|+|\bm{k}'|)^2 $.
As a rule, $T_A^{\max}\le 200$ keV, and for the most interesting target nuclei $\Delta\varepsilon_{mn}$ is noticeably less than 1 MeV.
\par
Since in the lab frame the target nucleus is at rest in some
$|n\rangle$-state, the differential cross section of the process $\chi A_n\to \chi A^{(*)}_m$
(see, for example, 
\cite{Tanabashi:2018oca,Peskin:1995ev,Bilenky:1995zq,Bednyakov:2018mjd,Bednyakov:2019dbl,Bednyakov:2021ppn,Bednyakov:2022dmc})
takes the form 
  \begin{equation}  \label{eq:Kinematics-CrossSection-DM-n-2-DM-m}
\frac{d^2\sigma_{mn}}{d|\bm{k}'| d\cos\theta}= \frac{-|{i\cal M}_{mn}|^2 |\bm{k}'|^2 }{2^5\pi\sqrt{m^2_\chi + |\bm{k }'|^2}}
\frac{\delta\Big(k_0-\sqrt{m^2_\chi + |\bm{k}'|^2}-\Delta\varepsilon_{mn}-T_A(|\bm{k}'|, \cos\theta)\Big)}
{(m_A+\varepsilon_m+T_A(|\bm{k}'|,\cos\theta)) \sqrt{k^2_0(m_A+\varepsilon_n)^2 -m_\chi^2 m_A^2}}.
\end{equation}
This expression and the recoil energy of the nucleus (\ref{eq:Kinematics-Recoil-kinetic-energy}) both depend on  
momentum $|\bm{k}'|$ and $\cos\theta$.
Integration over $\cos\theta$ is performed using the $\delta$-function of the energy conservation law 
from (\ref{eq:Kinematics-CrossSection-DM-n-2-DM-m}) represented as a function of $\cos\theta$
\begin{equation}
\label{eq:Kinematics-Delta-function-4-cosT-rel}
\delta(f(\cos\theta))=\sum^{}_{i}\frac{\delta(\cos\theta-\cos\theta_i)}{|d f(\cos\theta_i)/d \ cos\theta|}.
\end{equation}
Here $\cos\theta_i$ are the solutions (may be not only one) of the equation
\begin{equation}
\label{eq:Kinematics-Under-Delta-function-cosT-0}
f(\cos\theta_i)\equiv k_0-\sqrt{m^2_\chi + |\bm{k}'|^2}-\Delta\varepsilon_{mn}-T_A(|\bm{k}'| ,\cos\theta_i)=0
.\end{equation}
The derivative of this function included in (\ref{eq:Kinematics-Delta-function-4-cosT-rel}) is
$\dfrac{df(\cos\theta)}{d \cos\theta} = \dfrac{|\bm{k}'| |\bm{k}|}{T_A(|\bm{k}'|,\cos\theta)+m_A}.$
After integrating  expression (\ref{eq:Kinematics-CrossSection-DM-n-2-DM-m}) over $\cos\theta$ one has
\begin{eqnarray} \label{eq:Kinematics-dSigma-po-dk-prime}
\frac{d \sigma_{mn}}{d|\bm{k}'|}&=& \frac{-1}{2^5\pi \sqrt{k^2_0(m_A+\varepsilon_n)^2 - m_\chi^2 m_A^2}}
\frac{|{i\cal M}_{mn}|^2 |\bm{k}'| }{ |\bm{k}|\sqrt{m^2_\chi + |\bm{k}'|^2}}
\frac{m_A+T_A(|\bm{k}'|,\cos\theta_i)}{(m_A+T_A(|\bm{k}'|,\cos\theta_i)+\varepsilon_m)}.
\qquad \end{eqnarray}
Using the Jacobian of the transition from the variable $|\bm{k}'|$ to the observable variable $T_A$
\begin{eqnarray} \label{eq:Kinematics-dT_A-po-dk-prime}
\dfrac{d T_A (|\bm{k}'|, \cos\theta) }{d |\bm{k}'|}= -\dfrac{|\bm{k}'|}{\sqrt{ m^2_\chi + |\bm{k}'|^2} },
\end{eqnarray}
from formula (\ref{eq:Kinematics-dSigma-po-dk-prime}) one gets the 
expression for the differential cross section of $\chi A$ scattering
\begin{eqnarray}
\label{eq:Kinematics-dSigma-po-dT-A-kin}
\dfrac{d \sigma_{mn}}{d T_A} = \dfrac{d \sigma_{mn}}{d|\bm{k}'|} \frac{d |\bm{k}'|}{ dT_A}
  =\frac{|{i\cal M}_{mn}|^2}{2^5\pi \sqrt{w}}\frac{1 }{ |\bm{k}|}\frac{T_A+ m_A}{T_A+ m_A+\varepsilon_m}
, \end{eqnarray} where the notation for the initial particle flux is 
\begin{equation}  \label{eq:Kinematics-Flux-definition}
\sqrt{w} \equiv \sqrt{(m^2_\chi + |\bm{k}|^2)(m_A+\varepsilon_n)^2 -m_\chi^2 m_A^2}
\simeq |\bm{k}| m_A \sqrt{1+ \dfrac{2\varepsilon_n }{m_A}\Big(1+ \dfrac{ m^2_\chi }{|\bm{k}|^2 }\Big) }
.\end{equation}
Finally, the relativistic cross section of the process $\chi A_n\to \chi A_m$ has the form
\begin{eqnarray}
\label{eq:Kinematics-dSigma-po-dT-A-kin-rel}
\dfrac{d \sigma_{mn}}{d T_A} \big(\chi A_n\to \chi A_m\big) = \frac{|{i\cal M}_{mn}|^2}{2^ 5\pi |\bm{k}|^2 m_A} C_{mn}(T_A), \qquad \text{where}
\end{eqnarray}
$$ C_{mn}(T_A)=\frac{1}{ \sqrt{1+ \dfrac{2\varepsilon_n }{m_A}\Big(1+ \dfrac{ m^2_\chi }{|\bm{ k}|^2 }\Big) }}
\frac{T_A+m_A}{T_A+ m_A+\varepsilon_m}
\simeq O(1), $$ since $m_A\gg T_A +\varepsilon_{m}$ in the considered approximation.

\section{\large The amplitude of ${\chi}$ particle scattering on nucleus}\label{3chiA-ScatteringAmplitude}
The probability amplitude of relativistic $\chi$ lepton scattering on a nucleus, as in the nonrelativistic case
\cite{Bednyakov:2022dmc}, is based on the assumption that the interaction is between the lepton and the structureless nucleon.
This allows one to use the effective 4-fermion Lagrangian \cite{Bednyakov:2018mjd,Bednyakov:2021ppn}
written as the product of the lepton $L_\mu(x)=h_\chi \overline{\psi}_\chi(x) O_\mu \psi_\chi(x)$
and nucleon $H^\mu(x)=\sum_{f=n,p} h_f \overline{\psi}_f(x) O^\mu \psi_f(x)$
currents, where the $O^{\mu}$ are combinations of $\gamma$-matrices.
\par
\begin{figure}[h]
\centering\includegraphics[scale=0.3]{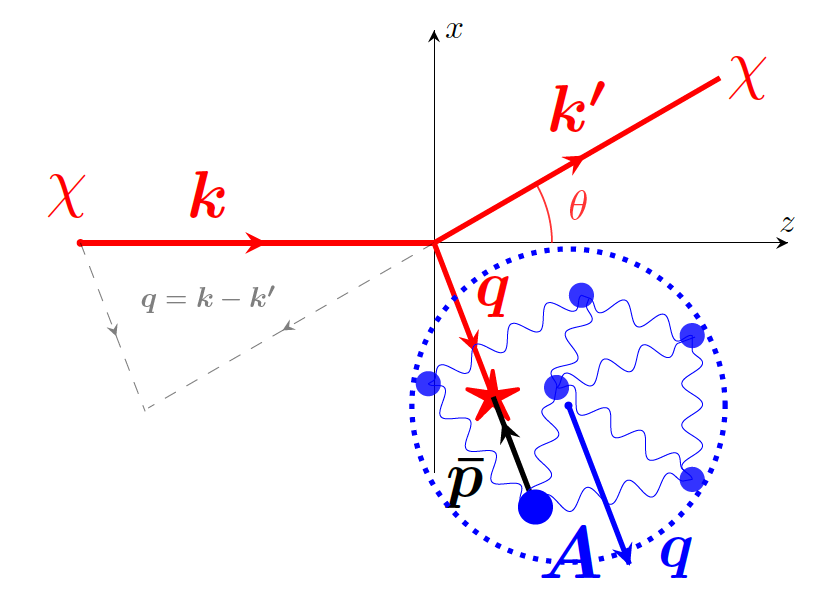}
\caption{\small The requirement of simultaneous fulfillment of the energy conservation law and
integrity of the nucleus "chooses"\/  the momentum of the active nucleon $\bar{\bm{p}}= ( p_L(\bm{q}), p_T)$,
inter\-action with which occurs at the point marked with a red star. } \label{fig:Identity-Conservation}
\end{figure}

The element of the $\mathbb{S}$-matrix, $\langle P'_m,k'|\mathbb{S}|P_n,k\rangle$, 
which determines  the probability for the nucleus and the $\chi$ particle to transform from  
the initial state $|P_n,k\rangle$ to the final state $\langle P'_m,k'|$ due to their interaction
is written in the standard way 
\begin{equation}
\label{eq:ScatteringAmplitude-MatrixElevent}
\langle P'_m,k'|\mathbb{S}|P_n,k\rangle= (2\pi)^4\delta^4(q+P_n-P'_m)i\mathcal{M}_{mn }
=\dfrac{i G_{\rm F}}{\sqrt{2}} \int d^4 x \, H^\mu_{mn}(x) \, L_\mu^\chi(x)
, \end{equation}
where $H^\mu_{mn}(x)\equiv \langle P'_m| H^\mu(x)|P_n\rangle$ is the matrix element of the transition of the nucleus from the state $|P_n\rangle$ to the state $\langle P'_m|$ due to the hadronic current $H^\mu(x) $.
Here, the nuclear state wave function  is determined by the expression
  \cite{Bednyakov:2018mjd,Bednyakov:2019dbl,Bednyakov:2021ppn,Bednyakov:2022dmc}
 \begin{equation}
   \label{eq:ScatteringAmplitude-P_n}
  |P_n\rangle=\int\Big(\prod^{A}_{i}d\widetilde{\bm{p}}^\star_i\Big)\frac{\widetilde{\psi}_n(\{p^\star\})}{\sqrt{{A!}}} \Phi_n(p)|\{p^\star\}\rangle, \quad \text{where} \quad\bm{\widetilde{p}}^\star_i = \frac{d\bm{p}^\star_i}{(2\pi)^3 \sqrt{2E_{\bm{p}^\star_i}}}
. \end{equation}
The function $\Phi_n(p)=(2\pi)^3\sqrt{2P_n^0}\delta^3(\bm{p}-\bm{P})$
corresponds to the nucleus with a certain 3-momentum $\bm{P}$ and energy $P^0_n=E_{\bm{p}}+\varepsilon_n$.
In formula (\ref{eq:ScatteringAmplitude-P_n}) one has $\{p^\star\}\equiv (p^\star_1,\dots ,p^\star_n)$, and 
$p^\star_i$ is thr 4-momentum of the $i$th nucleon in the center-of-mass system of the nucleus (at rest).
After transforming the right side of (\ref{eq:ScatteringAmplitude-MatrixElevent}), the
matrix element defined in  (\ref{eq:Kinematics-dSigma-po-dT-A-kin-rel}),
is written as follows:
\begin{eqnarray} \label{eq:ScatteringAmplitude-M-Element-via-H-and-L}
i \mathcal{M}_{mn}=\dfrac{i G_{\rm F}}{\sqrt{2}} \sqrt{4P^{0'}_m P^0_n}\, \, l_\mu (k',k,s',s)\, h^\mu_{mn} (\bm{q})
,\end{eqnarray}
where the lepton current is denoted as
\begin{equation}\label{eq:ScatteringAmplitude-LeptonicCurrent}
  l_\mu(k',k,s',s) \equiv \overline{u}_\chi(\bm{k}',s') O_{\mu} u_\chi (\bm{k}, s)
.\end{equation}
The hadronic current $h^\mu_{mn}(\bm{q})=\langle m|H^\mu( 0)|n\rangle$ 
defined in terms of nuclear $|n\rangle$ state in the rest frame of the nucleus has the form
\begin{eqnarray}\nonumber
\label{eq:ScatteringAmplitude-h-mu-mn-with-delta-p_k}
h^\mu_{mn}(\bm{q})&=& \sum^A_{k} \frac{\overline{u}(\bm{\bar{p}}^\star_k+\bm{q} ,r'_{k})\, O^\mu_k\, u(\bm{\bar{p}^{\star}}_k,r_{k})}
{\sqrt{4E_{\bm{\bar{p}}^\star_k}E_{\bm{\bar{p}}^\star_k+\bm{q}}}}
\times\\&&\times \int \prod^{A}_{i=1}\frac{d\bm{p}^\star_i \delta\big(f(\bm{p^{\star}_k })\big)}{(2\pi)^3}
\widetilde{\psi}^{*}_m(\{p^{(k)}_\star\},\bm{p^{\star}}_k\!+\!\bm{q} )\widetilde{\psi}_n(\{p^\star\})
(2\pi)^{3} \delta^3(\sum^A_{i=1}\bm{p}^\star_i)
, \end{eqnarray}
where $\bm{\bar{p}^{\star}_k}(\bm{q})$ is the solution of the equation
\footnote{It is assumed here that the mass of the proton 
is equal to the mass of the neutron, i.e. $m_p=m_n\equiv m$.}
\begin{equation}\label{eq:ScatteringAmplitude-Energy-and-Identity-Conservation}
  f(\bar{\bm{p}}) \equiv \sqrt{m^2+ \bar{\bm{p}}^2} - \sqrt{m^2+ (\bar{\bm{p} }+\bm{q})^2_{}} - T_A - \Delta\varepsilon_{mn}= 0
. \end{equation}
It is the condition for the simultaneous fulfillment of the energy conservation law  and the nuclear integrity 
(Fig.~\ref{fig:Identity-Conservation}).
It "selects"\/ the momentum of the active nucleon $\bar{\bm{p}}= ( p_L(\bm{q}), p_T)$
  {\em depending on $\bm{q}$}\/ in the form \cite{Bednyakov:2018mjd,Bednyakov:2021ppn}
\begin{eqnarray}\label{eq:ScatteringAmplitude-Energy-and-Identity-Conservation-Solution}
p_L(\bm{q}) = -\frac{|\bm{q}|}{2}\Bigg[1-\sqrt{\beta}\sqrt{1+\frac{4m_{T}^2}{\bm{q}^ 2(1-\beta)}}\Bigg],
\text{~where~}\beta=\frac{(T_A+\Delta\varepsilon_{mn})^2}{\bm{q}^2}, \quad m_{T}^2 = m^2 +p_T^2
. \qquad \end{eqnarray}
In formula (\ref{eq:ScatteringAmplitude-h-mu-mn-with-delta-p_k}), expression 
$\{p^{(k)}_\star\}$ coincides with  $\{p_\star\}$ except for the $k$th element,
which is equal to $(\bm{p}_{k}^\star+\bm{q},s_k)$.
\par
As before in \cite{Bednyakov:2018mjd,Bednyakov:2019dbl,Bednyakov:2021ppn,Bednyakov:2022dmc},  one assumes that
the wave function $\widetilde{\psi}_n$ has the form of the product of the momentum $\widetilde{\psi}_n$ and spin $\chi_n$ components independent of each other
\begin{equation} \label{eq:ScatteringAmplitude-factorize_spin}
\widetilde{\psi}_n(\{p^\star\}) =\widetilde{\psi}_n(\{\bm{p}_\star\})\chi_n(\{r\}).
\end{equation}
Here the momentum $\{\bm{p}_\star\}=(\bm{p}_1^\star,\dots{}, \bm{p}_A^\star)$
and spin $\{r\}=(r_1,\dots , r_A)$ variables are introduced.
Spin functions are normalized by the conditions
\begin{equation} \label{eq:ScatteringAmplitude-spin_functions_norm}
  \chi_m^{*}(\{r\})\chi_n(\{r\}) = \delta_{nm} \quad\text{and}\quad \chi_{n}^{*}(\{r '\})\chi_n(\{r\}) = \delta_{\{r'\} \{r\}}
, \end{equation}
which mean that if (after the interaction)
all the spins of the nucleons remained unchanged,  the spin state of the nucleus also did not change,
and vice versa, if the spin state of the nucleus has not changed, all the spins of the nucleons must
also remain unchanged.
\par
Since after the interaction the spin $r'_k$ of the active nucleon in the $|m\rangle$ state 
may turn out to be different from the spin $r_k$ of this nucleon in the $|n\rangle$ state,  
we define the product of spin functions
\begin{equation}
\label{eq:ScatteringAmplitude-lambda_def}
  \lambda^{mn}(r',r) \equiv \lambda^{mn}( \{r'\}, \{r\})\equiv \chi^*_m(\{r^{(k) }\})\chi_n(\{r\})
  = \delta_{mn}\delta_{r'r}+(1-\delta_{mn})\lambda^{mn}_{r'r}
  , \end{equation}
  where $\{r^{(k)}\}\equiv \{r'\}$ is the same as $\{r\}$ except for the $k$th element, which is equal to $r'_k$, and
  the notation $\lambda^{mn}_{r'r}\equiv \lambda^{mn}(r',r)$ for $m\ne n$ is introduced.
  Given (\ref{eq:ScatteringAmplitude-lambda_def}) and $\bar{\bm{p}}(\bm{q})$ from
(\ref{eq:ScatteringAmplitude-Energy-and-Identity-Conservation}), 
expression (\ref{eq:ScatteringAmplitude-h-mu-mn-with-delta-p_k}) is rewritten as follows:
\begin{equation}
\label{eq:ScatteringAmplitude-h-mu-mn-with-delta-p_k-spin-factorization}
h^\mu_{mn}(\bm{q}) = \sum_{k=1}^{A} 
\frac{\bar{u}(\bar{\bm{p}}+\bm{q} ,r'_k) O^\mu_k u(\bar{\bm{p}},r_k)}{\sqrt{4E_{\bar{\bm{p}}}E_{\bar{\bm{p}}+\bm{q}}}} 
\lambda^{mn}(r',r) \ \langle m|e^{i\bm{q}\hat{\bm{X}}_{k}} |n\rangle
,\end{equation}
where the multidimensional integral from (\ref{eq:ScatteringAmplitude-h-mu-mn-with-delta-p_k}) 
can be written as a nuclear matrix element of the operator  $\hat{\bm{X}}_k$ that implements the three-dimensional shift of the $k$th nucleon
\begin{eqnarray}
\label{eq:ScatteringAmplitude-f^k_mn-definition}
f^k_{mn}(\bm{q})\equiv\langle m|e^{i\bm{q}\hat{\bm{X}}_{k}}|n\rangle\!=\!\!\!
\int\!\!\prod^{A}_{i=1}\frac{d\bm{p}^\star_i }{(2\pi)^3}\delta\big (f(\bm{p^{\star}_k})\big)\widetilde{\psi}_m^*(\{\bm{p}^{(k)}_\star\},\bm{ p^{\star}}_k\!+\!\bm{q} ) \widetilde{\psi}_n(\{\bm{p}_\star\})(2\pi)^3 \delta^ 3\big(\sum_{l=1}^A \bm{p}^\star_l\big)
. \qquad \end{eqnarray}
Here $\delta\big(f(\bm{p^{\star}_k})\big)$ ensures
 the integrity of the nucleus after the $\hat{\bm{X}}_k$-operator shift  of the momentum of the active $ k$th nucleon caused by an external action.
\par
Finally, for the matrix element (\ref{eq:ScatteringAmplitude-M-Element-via-H-and-L}) that defines
the probability amplitude of the process $\chi_s A_n\to \chi_{s'} A^{(*)}_m$, one gets the {\em final}\/ expression
\begin{equation}
\label{eq:ScatteringAmplitude-Matrix_element-via-ScalarProducts}
i\mathcal{M}^{s' s,r'r }_{mn}(\bm{q}) = i\frac{G_F}{\sqrt{2}} \frac{m_A}{m}C_ {1,mn}^{1/2} \sum_{k=1}^{A}
  f^k_{mn}(\bm{q}) \lambda^{mn}(r',r) (l_{s's}\, h^k_{r'r}), \quad\text{where } 
\end{equation}
\begin{equation}\label{eq:ScatteringAmplitude-ScalarProduct-lh}
(l_{s's}\, h^k_{r'r})\equiv l_\mu(k',k,s',s) \  \bar{u}(\bar{\bm{p}}+\bm{q},r'_k) O^\mu_k u(\bar{\bm{p}},r_k)
\end{equation}
is the scalar product of the lepton current and the $k$th nucleon current,
containing all the specifics of the interaction between them.
In formula (\ref{eq:ScatteringAmplitude-Matrix_element-via-ScalarProducts}),  the auxiliary factor
is  introduced, 
\begin{equation} \label{eq:ScatteringAmplitude-C1_def}
C_{1,mn}\equiv\frac{P^0_n\, P^{'0}_m }{E_{\bar{\bm{p}}} \, E_{\bar{\bm{p}} +\bm{q} } } \frac{m^2}{m_A^2}
\simeq O(1)
, \end{equation} 
whose value is close to 1 with good accuracy, and the dependence on $n,m$ and $T_A$ 
manifests itself in it only at a level $O(10^{-3})$
\footnote{Formulas (\ref{eq:ScatteringAmplitude-Matrix_element-via-ScalarProducts})
and (\ref{eq:ScatteringAmplitude-C1_def}) are discussed in Appendix \ref{ARchiA-Appendix}. }. 

\section{\large Cross sections of ${\chi}$ particle-nucleus scattering}\label{4chiA-CrossSection}  
\subsection{\normalsize\em Coherent and incoherent contributions
to the  ${\chi A\to \chi A^{(*)}}$ scattering} \label{41chiA-CrossSection-Coh-vs-InCoh} 
The {\em observable}\/ differential cross section of the process $\chi_s A\to \chi_{s'} A^{(*)}$ can be 
obtained by averaging over all possible initial $|n\rangle$ states and summation over all final 
$|m\rangle$ states of the nuclear cross section (\ref{eq:Kinematics-dSigma-po-dT-A-kin-rel}).
After summation over the spin $r,r'$ indices of the (active) nucleon {\em at the level of}\/
the matrix element (\ref{eq:ScatteringAmplitude-Matrix_element-via-ScalarProducts}),
the observable  cross section can be written as
\cite{Bednyakov:2018mjd,Bednyakov:2019dbl,Bednyakov:2021ppn,Bednyakov:2022dmc}
\begin{eqnarray} \label{eq:41chiA-CrossSection-Coh-vs-InCoh-CrossSection-with-s-sprime-and-Tmns}
\frac{d\sigma_{s's}}{d T_A}(\chi A \to \chi A^{(*)}) 
 &=&  \dfrac{G^2_F m_A}{2^6\pi |\bm{k}^\chi_l|^2 m^2 }  \Big[T^{s's}_{m=n} + T^{s's}_{m\ne n}\Big], 
\quad\text{where}
\end{eqnarray} 
\begin{eqnarray}  \label{eq:41chiA-CrossSection-Coh-vs-InCoh-CrossSection-Term-nn-definition-with-s-sprime}
T^{s's}_{m=n}&=&g^{}_\text{c}\sum^A_{k,j}\sum_{n}\omega_n  \Big[ f^k_{nn}f^{j*}_{nn}\sum_r  (l_{s's},h^k_{rr})\sum_{x} (l_{s's}\,h^j_{xx})^{*}\Big], 
\\ 
\label{eq:41chiA-CrossSection-Coh-vs-InCoh-CrossSection-Term-mn-definition-with-s-sprime}
T^{s's}_{m\ne n}&=& g^{}_\text{i} \sum^A_{k,j}\sum_{n}\omega_n   \Big[ \sum_{m\ne n} 
f^k_{mn} f^{j*}_{mn}  \sum_{r'r}\lambda^{mn}_{r'r}(l_{s's}, h^k_{r'r})\Big(\sum_{x'x}\lambda^{mn}_{x'x}(l_{s's}\, h^j_{x'x}) \Big)^{\dag} \Big].
\end{eqnarray}
Here $\sum_n \omega_n=1 $ is the probability sum of all possible initial states of nucleus $A$.
For the sake of completeness, the correction kinematic coefficients 
$g^{}_\text{c}= C^{}_{1,nn}C^{}_{nn}$ and $g^{}_\text{i}= C^{}_{ 1,mn}C^{}_{mn}$ are kept,
whose values are close to 1 at the $O(10^{-3})$ level.
\par 
Summation over index the $n$ in formula
(\ref{eq:41chiA-CrossSection-Coh-vs-InCoh-CrossSection-Term-nn-definition-with-s-sprime})
determines the nucleon form factors averaged over all possible initial states of the nucleus
\cite{Bednyakov:2018mjd,Bednyakov:2019dbl,Bednyakov:2021ppn,Bednyakov:2022dmc}
\begin{equation}
\label{eq:41chiA-CrossSection-Coh-vs-InCoh-form-factors-with-s-sprime}
\sum_n \omega_n f_{nn}^k f_{nn}^{j*}= \left\{\begin{matrix}
|F_{p/n}(\bm{q})|^2,  &~~~~~~~~~~~~\text{ when }(k,j)=(p,p) \text{~~or~} (n,n);\\
F_{p}(\bm{q})F_{n}^*(\bm{q}),   &\text{ when }(k,j) = (p,n);\\
F_{n}(\bm{q})F_{p}^*(\bm{q}),   &\text{ when }(k,j)= (n,p).\\
\end{matrix}\right.
\end{equation}
With this definition in mind, expression
(\ref{eq:41chiA-CrossSection-Coh-vs-InCoh-CrossSection-Term-nn-definition-with-s-sprime})
for the contribution to the cross section corresponding to the preservation of the initial state of the nucleus,
when the projection of the spin of the active nucleon does not change,
can be represented as the square of the modulus of the sum of the contributions of protons and neutrons
\begin{eqnarray} 
\label{eq:41chiA-CrossSection-Coh-vs-InCoh-cross_section_coherent_term-1-with-s-sprime}
T^{s's}_{m=n}(\bm{q}) &=& g^{}_\text{c} \Big|\sum_{f=p,n}\sum^{A_f}_{k=1}\sum_{r} (l_{s's},h^f_{rr}(\bm{q}))F_f(\bm{q})\Big|^2.
\end{eqnarray}
Here $A_f$ denotes the total number of $f$-type nucleons in the nucleus.
\par
To sum over the nuclear indices $m,n$ in formula
(\ref{eq:41chiA-CrossSection-Coh-vs-InCoh-CrossSection-Term-mn-definition-with-s-sprime}), 
note that from the spin function normalization (\ref{eq:ScatteringAmplitude-spin_functions_norm}), 
for {\em the same}\/ $k$th nucleon there follows the relation
\begin{equation}\label{eq:41chiA-CrossSection-Coh-vs-InCoh-NuclearSpinAmplitudeProduct}
\lambda^{mn}_{r'_kr^{}_k} [\lambda^{mn}_{x'_kx^{}_k}]^* \equiv \delta_{r'_kx'_k} \delta_{r^{}_kx^{}_k}|\lambda^{mn}_{r'_kr^{}_k}|^2, \quad \text{where }\quad |\lambda^{mn}_{r'_kr^{}_k}|^2 =1
.\end{equation}
Therefore, one can assume that $\lambda^{mn}_{r'r}$ do not depend on indices $m$ and $n$;
however, $\lambda^{mn}_{r'r} \simeq \lambda^{p/n}_{r'r}$ may differ for protons and neutrons.
In other words, for any initial spin orientation of the active nucleon (index $r$) 
any orientation of the spin of this nucleon (index $r'$) is possible after the nuclear $|n\rangle\to |m\rangle$ transition
\footnote{Discussion of this approximation is given in Appendix \ref{ARchiA-Appendix}.}. 
As a result, $\lambda^{mn}_{r'r}$ can be taken out of the summation over $m,n$ in 
(\ref{eq:41chiA-CrossSection-Coh-vs-InCoh-CrossSection-Term-mn-definition-with-s-sprime}).
Then, if the indices $k$ and $j$ in formula
(\ref{eq:41chiA-CrossSection-Coh-vs-InCoh-CrossSection-Term-mn-definition-with-s-sprime})
 “indicate”\/  the same nucleon, e.g., the proton, the summation gives
\begin{eqnarray}
\label{eq:41chiA-CrossSection-Coh-vs-InCoh-cross_section_incoherent_term1}
\sum_n \omega_n\sum_{m\ne n} f_{mn}^k f_{mn}^{k*} =
\sum_n \omega_n \Big[ \langle n|e^{i\bm{q}\bm{X}_k}\sum_{m}|m\rangle\langle m|e^{-i\bm{q}\bm{X}_k}|n\rangle\Big] -|F_{p}(\bm{q})|^2 =1-|F_{p}(\bm{q})|^2. \qquad 
 \end{eqnarray}
If $k\ne j$, but they still indicate protons (index $p$), it can be written that
 \begin{equation}		
 \label{eq:41chiA-CrossSection-Coh-vs-InCoh-Averaged-Covariance}
 \sum_n \omega_n\sum_{m\ne n} f_{mn}^k f_{mn}^{j*}  = 
 \sum_n \omega_n \text{cov}_{nn}(e^{i\bm{q}\hat{\bm{X}}_k},e^{-i\bm{q}\hat{\bm{X}}_j}) \equiv
 \langle\text{cov}(e^{i\bm{q}\hat{\bm{X}}_k}, e^{-i\bm{q}\hat{\bm{X}}_j})\rangle_p, 
 \end{equation}
where the covariation operator of the shift operators  $e^{-i\bm{q}\hat{\bm{X}}_j}$ 
and $e^{i\bm{q}\hat{\bm{X}}_k}$  with respect to the state $|n\rangle$  is introduced in the form
\begin{equation} \label{eq:41chiA-CrossSection-Coh-vs-InCoh-nn-Covariance}
\text{cov}_{nn}(e^{i\bm{q}\hat{\bm{X}}_k},e^{-i\bm{q}\hat{\bm{X}}_j}) \equiv \langle n|e^{i\bm{q}\hat{\bm{X}}_k}\, e^{-i\bm{q}\hat{\bm{X}_j}} |n\rangle - \langle n|e^{i\bm{q}\hat{\bm{X}}_k}|n\rangle \langle n|e^{-i\bm{q}\hat{\bm{X}}_j}|n\rangle
.\end{equation}
Expression (\ref{eq:41chiA-CrossSection-Coh-vs-InCoh-Averaged-Covariance})
vanishes at small and at large transfer momenta.
A similar consideration applies to neutrons and can be generalized to the common case of protons and neutrons.
 It is believed that when correlations between nucleons in the nucleus are sufficiently weak, 
 one can neglect the covariance functions like (\ref{eq:41chiA-CrossSection-Coh-vs-InCoh-Averaged-Covariance}).
For example, in nuclear shell models where multiparticle wave functions of nuclei
are constructed in the form of a product of single-particle
wave functions \cite{Blokhintsev:1963,Bohr:1974}, covariation (\ref{eq:41chiA-CrossSection-Coh-vs-InCoh-nn-Covariance}) 
  identically vanishes. 
 Therefore, in what follows as in \cite{Bednyakov:2018mjd,Bednyakov:2019dbl,Bednyakov:2021ppn,Bednyakov:2022dmc},
covariance contributions  like (\ref{eq:41chiA-CrossSection-Coh-vs-InCoh-Averaged-Covariance})
are {\em completely neglected}.
In view of the above the term 
(\ref{eq:41chiA-CrossSection-Coh-vs-InCoh-CrossSection-Term-mn-definition-with-s-sprime})
can be finally presented  as a sum over all nucleons
\begin{eqnarray}
\label{eq:41chiA-CrossSection-Coh-vs-InCoh-CrossSection-Term-mn-with-s-sprime-final}
T^{s's}_{m\ne n} &=& g^{}_\text{i}\sum_{f=p,n}\big[1-|F_f(\bm{q})|^2\big]
\sum^{A_f}_{k=1} \sum_{r'r}  \big| (l_{s's}\, h^f_{r'r}(\bm{q})) \big|^2
.\end{eqnarray}
Thus, the observable differential cross section 
(\ref{eq:41chiA-CrossSection-Coh-vs-InCoh-CrossSection-with-s-sprime-and-Tmns})
for the process $\chi A\to \chi A^{(*)}$ can be written in the form
of two fundamentally different terms
\begin{eqnarray}
\label{eq:41chiA-CrossSection-Coh-vs-InCoh-CrossSection-TwoTerms-with-s-sprime}
\frac{d\sigma_{s's}}{d T_A}&=&
\frac{d\sigma^{s's}_\text{coh}}{d T_A}(\chi A\to \chi A^{})+\frac{d\sigma^{s's}_\text{inc}}{d T_A}(\chi A\to \chi A^{*}),  \quad \text{where }
\\ \nonumber
\frac{d\sigma^{s's}_\text{coh}}{d T_A} &=& c_A 
g_\text{c}\Big|\sum_{f=n,p}\sum_{k=1}^{A_f} \sum_{r} (l_{s's}\, h^f_{rr}(\bm{q}))F_f(\bm{q})\Big|^2
\quad \text{and} 
\\ \nonumber    
\frac{d\sigma^{s's}_\text{inc}}{d T_A} &=&c_A  
g_\text{i}\!\sum_{f=n,p}\sum_{k=1}^{A_f} \sum_{r'r} |(l_{s's}\,h^f_{r'r}(\bm{q}))|^2[1-|F_f(\bm{q})|^2] 
,\end{eqnarray}
respectively correspond to the elastic and inelastic interactions of the $\chi$ particle with nucleus $A$.
For convenience,  in formulas (\ref{eq:41chiA-CrossSection-Coh-vs-InCoh-CrossSection-TwoTerms-with-s-sprime})
universal factors are introduced, 
\begin{equation} \label{eq:41chiA-CrossSection-Coh-vs-InCoh-GeneralFactor-with-G2_F}
c_A \equiv \dfrac{G^2_F m_A}{2^6\pi |\bm{k}^\chi_l|^2 m^2 } \quad \text{and}\quad 
\hat c_A \equiv \dfrac{G^2_F m_A}{4\pi }\dfrac{1}{|\bm{k}^\chi_l|^2 m^2 }  = 4^2 c_A
, \end{equation}
where $\bm{k}^\chi_l$ is the momentum of the incident $\chi$ particle in the rest frame of the target nucleus.
The helicities of the $\chi$ particle (in the general case, the projection of the spin onto some direction) in the initial 
$s$ and final $s'$ states are assumed to be fixed in 
(\ref{eq:41chiA-CrossSection-Coh-vs-InCoh-CrossSection-TwoTerms-with-s-sprime}).
In the future, they can be averaged (summated).
\par
Since the scalar products $(l_{s's}\,h^f_{r'r})$
in formulas (\ref{eq:41chiA-CrossSection-Coh-vs-InCoh-CrossSection-TwoTerms-with-s-sprime})
depend only on the type of the active nucleon and do not depend on its number
(that is, from the summation index $k$), then the summation over $k$
in formulas (\ref{eq:41chiA-CrossSection-Coh-vs-InCoh-CrossSection-TwoTerms-with-s-sprime})
gives the total number of protons and neutrons in the nucleus.
As a result, the differential cross sections for
 $\chi$ particle scattering on the nucleus $\chi_s A\to \chi_{s'} A^{(*)}$ take the form
\begin{eqnarray}  \nonumber
\label{eq:41chiA-CrossSection-Coh-vs-InCoh-CS-with-s-sprime-for-proton-neutron}
\frac{d\sigma^{s's}_\text{inc}}{d T_A} &=&c_A  g_\text{i} 
\sum_{f=p,n}[1-|F_f(\bm{q})|^2]
\Big[A^f_+  \sum^{}_{r'=\pm}|(l_{s's}\, h^{\eta,f}_{r'+})|^2+A^f_- \sum^{}_{r'=\pm}|(l_{s's}\,h^{\eta,f}_{r'-})|^2 \Big], \qquad 
\\  \frac{d\sigma^{s's}_\text{coh}}{d T_A} &=&c_A  g_\text{c} \Big| \sum_{f=p,n}F_f(\bm{q})  [A^f_+(l_{s's}\,h^{\eta,f}_{++})+ A^f_-(l_{s's}\,h^{\eta,f}_{--})] \Big|^2
.\end{eqnarray}
Here the sums over the initial projections of the nucleon spin on the direction of arrival of the $\chi$ particle, 
which is marked by the index $\eta$ for hadronic currents, are explicitly distinguished.
In  (\ref{eq:41chiA-CrossSection-Coh-vs-InCoh-CS-with-s-sprime-for-proton-neutron})
  $A^f_\pm$ is a number of $f$-type nucleons with the spin projection $\pm 1$ onto this direction.
\par 
It is convenient to {\em finally}\/
rewrite formulas (\ref{eq:41chiA-CrossSection-Coh-vs-InCoh-CS-with-s-sprime-for-proton-neutron})
 in terms of the total number of $f$-type nucleons,
$A_f = A^f_+ + A^f_- $, and the difference in the number of nucleons, $\Delta A_f = A^f_+ - A^f_-$, having positive and negative spin projections on the selected direction
\begin{eqnarray}\label{eq:43RchiA-CrossSection-via-ScalarProducts-both-CS-main} 
\frac{d\sigma^{s's}_\text{coh}(\bm{q})}{g_\text{c}  d T_A}&=& c_A 
 \Big| \sum^{}_{f=p,n} F_{f}(\bm{q}) \dfrac{A_f}{2}  Q_f^{s's}  \Big|^2 
 ,\\  \frac{d\sigma^{s's}_\text{inc}(\bm{q})}{g_\text{i}  d T_A} &=& c_A 
\sum_{f=p,n} \widehat F^2_f(\bm{q})  \dfrac{ A_f  }{2}   \Big[ \, Q^{s's}_+  + \dfrac{\Delta A_f }{A_f } \, Q^{s's}_-  \Big]
, \quad \text{where}\quad \widehat F^2_f(\bm{q}) \equiv  \big[1-F^2_f(\bm{q})\big] 
\nonumber. \end{eqnarray}
Thus, coherent and incoherent cross sections are defined, respectively, by the following combinations of scalar products:
\begin{equation}\label{eq:43RchiA-CrossSection-via-ScalarProducts-CohCS-factors}
Q_f^{s's}\equiv \hat Q^{s's}_{+} + \dfrac{\Delta A_f}{A_f} \hat Q^{s's}_{-} , 
\quad \text{where} \quad  \hat Q^{s's}_{\pm}\equiv (l_{s's}\,h^{f}_{++}) \pm (l_{s's}\,h^{f}_{--}) 
, \end{equation}
\begin{equation}\label{eq:R43chiA-CrossSection-via-ScalarProducts-InCohCS-factors}
\text{and}\quad  Q^{s's}_\pm\equiv 
\sum^{}_{r'=\pm}|(l_{s's}\, h^{f}_{r'+})|^2 \pm  \sum^{}_{r'=\pm}|(l_{s's}\,h^{f}_{r'-})|^2 
.\end{equation}
As a rule, if a nucleus has spin zero, then $ \Delta A_f= 0$.
Almost always ${\Delta A_f }\ll {A_f }$, except for such nuclei as
helium \cite{Lyon:2022sza}, which, perhaps, deserves special attention.
To complete the derivation of the differential cross section of 
the process $\chi A\to \chi A^{(*)}$, 
 it is necessary to have explicit expressions for scalar products $(l_{s's}\,h^{p/n}_{r'r})$.
These values are given in section \ref{42RchiA-CrossSection-ScalarProducts-ChiEta-All}.

\subsection{\normalsize\em Scalar products for ${\chi A\to \chi A^{(*)}}$ scattering} 
\label{42RchiA-CrossSection-ScalarProducts-ChiEta-All} 
Scalar products of $\chi$-lepton and nucleon currents obtained in \cite{Bednyakov:2021pgs}
are necessary for self-consistent calculations of coherent and incoherent $\chi A\to \chi A^{(*)}$ 
scattering cross sections.
They are defined by the expression
$$(l^{i}_{s's} \, h^{k}_{r'r}) \equiv \sum^{4}_{\mu,\nu} J^{i,\mu}_{ s's}(\bm{k}')\, g_{\mu\nu}\, J^{k,\nu}_{r'r}(\bm{p}'),
$$
where indices $s$ and $r$ denote the initial spin states of the lepton and the nucleon,
and the indices $s'$ and $r'$ denote their final states.
The indices $i$ and $k$ denote the vector ($J^{v,\mu}\equiv V^\mu$), axial vector ($J^{a,\mu}\equiv A^\mu$)
lepton (argument $\bm{k}'$) and nucleon (argument $\bm{p}'$) currents.
The scalar products are expressed in term of  the masses of the nucleon and the $\chi$ particle
$$m, \quad m_\chi, \quad \text{and parameters} \quad \lambda_\pm=\sqrt{E_p\pm m}, \qquad \xi_\pm=\sqrt{E_\chi\pm m_\chi} ,$$ as well as the scattering angle $\theta$  of the $\chi$ particle in the lepton and nucleon center-of-mass system (c.m.s.),
where $\bm{k}+\bm{p}=\bm{k}'+\bm{p}'=0$. 
Here $p$ and $k$ are the 4-momenta of the nucleon and $\chi$ particles, in addition, 
\begin{eqnarray}\nonumber
\label{eq:ScalarProducts-ChiEta-All-Energies-etc} 
E_\chi &\equiv& \sqrt{m^2_\chi+|\bm{k}|^2} = \sqrt{m^2_\chi+|\bm{k}'|^2} = \frac{s+m^2_\chi-m^2}{2\sqrt{s}}
,\\E_p &\equiv& \sqrt{m^2+|\bm{p}|^2} =\sqrt{m^2+|\bm{p}'|^2}  = \frac{s+m^2-m^2_\chi}{2\sqrt{s}} 
,\\|\bm{p}| &=& \sqrt{E^2_p-m^2} = \lambda_+ \lambda_-  =|\bm{k}| = \sqrt{E^2_\chi- m^2_\chi} = \xi_+ \xi_-  
= \dfrac{\lambda(s, m^2, m^2_\chi)}{2\sqrt{s}}, \text{~~where}  \nonumber
\\&&\lambda^2(s, m^2, m^2_\chi)\equiv [s-(m+m_\chi)^2][s-(m-m_\chi)^2]
  \nonumber.\end{eqnarray}
The invariant square of the total energy in the c.m.s. has the form
\begin{eqnarray*}\nonumber 
s_{\text{c.s.m.}}^{}&=&(p+k)^2 = (p_0+k_0)^2 - (\bm{p}+\bm{k})^2 = (E_\chi+ E_p)^2
. \end{eqnarray*} 
In this case, the momenta of the particles do not change: $|\bm{k}'| = |\bm{k}| = |\bm{p}'|= |\bm{p}|$.
Therefore,  the only free variable of $\chi$-lepton elastic scattering on the nucleon in c.m.s. takes place. 
It is the angle $\theta$ between the direction of the initial lepton momentum $\bm{k}$ and the direction of its final
momentum $\bm{k}'$. It is present in the 3-momentum transferred to the nucleon
\begin{equation}\label{eq:ScalarProducts-CMS-q-definition}
\bm{q}^2 = 2 |\bm{k}|^2(1-\cos\theta) \equiv \bm{q}^2_{\max} \sin^2\dfrac{\theta}{2}.
\end{equation}
In the lab frame, where the nucleon is at rest, $p_l =(m, \bm{p}_l=\bm{0})$, one has
\begin{equation}\label{eq:ScalarProducts-Lab-s-definition}
 s= (k_l+p_l)^2 =m^2_\chi +m^2 + 2 m\sqrt{m^2_\chi + [\bm{k}^l_\chi]^2}
=m^2_\chi +m^2 + 2 m m_\chi \Big[1 + \dfrac{2 T_0}{m_\chi }\Big]^{1/2}, 
\end{equation}
where $T_0 = \dfrac{|\bm{k}^l_\chi|^2}{2 m_\chi}$ is the kinetic energy of the lepton incident on the nucleon at rest.
The following relation is also useful:
\begin{equation}\label{eq:ScalarProducts-Lab-k^2-to-k^2_lab}
\frac{|\bm{k}|^2}{|\bm{k}^l_\chi|^2 } = \frac{m^2}{s}.
\end{equation}
Considering  the definition of the angle $\theta$ in terms of the recoil energy of the nucleus $T_A$
  \footnote{Since $\bm{q}^2 \ll m_A^2 $, then $\bm{q}^2 \simeq 2 m_A T_A$, and from (\ref{eq:Kinematics-Recoil-kinetic-energy}) follows  $T_A^{\max} \simeq m_A\Big(1+\frac{4|\bm{k}^l_\chi|^2}{2m^2_A}\Big) -m_A  
 = \dfrac{4 m_\chi T_0}{m_A}$.} 
\begin{equation}
\label{eq:ScalarProducts-Lab-sinTheta-definition}
\sin^2\dfrac{\theta}{2} = \dfrac{\bm{q}^2_{} }{\bm{q}^2_{\max} }\simeq \dfrac{T_A^{}}{T_A^{\max}}, 
\quad \text{where}\quad T_A^{\max} \simeq \dfrac{4 m_\chi T_0}{m_A},
\end{equation}   
the parameters $T_0$ (the initial kinetic energy of the $\chi$ particle) through relations
(\ref{eq:ScalarProducts-ChiEta-All-Energies-etc}), and (\ref{eq:ScalarProducts-Lab-s-definition})
and $T_A$ (the kinetic energy of the recoil of the nucleus) determine
all quantities necessary for calculating the cross sections.
\par
Recall in this regard that according to the condition of preserving the integrity of the nucleus 
(\ref{eq:ScatteringAmplitude-Energy-and-Identity-Conservation}), {\em in the rest frame of the nucleus}\/ 
the active nucleon "meets"\/ the $\chi$ particle incident on the nucleus  
with a nonzero momentum from
(\ref{eq:ScatteringAmplitude-Energy-and-Identity-Conservation-Solution}),
and therefore the $s$ invariant in the rest frame of the nucleus (lab frame) should be recalculated
\begin{eqnarray} \label{eq:42DM-CrossSection-ScalarProducts-ChiEta-All-s-with-T_A-recoil}  
s&=& (k_l+p_l)^2 = m^2_\chi +m^2 + 2 m m_\chi\Bigg\{ \sqrt{1 + \dfrac{ |\bm{k}^l_\chi |^2}{m_\chi^2 }} 
\sqrt{1+ \dfrac{p_L^2}{m^2} } - \dfrac{|\bm{k}^l_\chi |}{m_\chi} \dfrac{p_L}{m}\Bigg\}.
\end{eqnarray}
In other words, this value depends both on the kinetic energy of the incident $\chi$ particle $T_0$ (or $|\bm{k}^l_\chi |$) and on the kinetic energy of the recoil of the nucleus $T_A$ .
However, as a rule, one has ${p_L} \le 0.1 \, {m}$, and this correction is small enough.
\par
The scalar products of the lepton and nucleon currents with 
the spin projection of the nucleon ($r',r=\pm1$) and the $\chi$-particle ($s',s=\pm1$), 
given below are defined as follows
\cite{Bednyakov:2021pgs}:
\begin{eqnarray}
\label{eq:42RchiA-CrossSection-ScalarProducts-ChiEta-Weak-definition} 
(l^w_{s's}\, h^{w,f}_{r'r})   &=& \alpha_f (l^v_{s's}\, h^v_{r'r}) + \beta_f (l^v_{s's}\, h^a_{r'r}) 
+ \gamma_f (l^a_{s's}\, h^v_{r'r})  +  \delta_f  (l^a_{s's}\, h^a_{r'r})
.\end{eqnarray} 
Recall that the index $f$ denotes a neutron or a proton.
\par
When the $\chi$-lepton helicity and the nucleon spin projection are preserved, 
these scalar products have the form
\begin{eqnarray*}
(l^w_{\pm\pm}\, h^{w,f}_{\pm\pm})^{}&=& 4 \cos\frac{\theta}{2}\Big[ (\alpha_f - \delta_f) \big[\xi_+ \xi_-  \lambda_+ \lambda_- \cos^2\frac{\theta}{2}  +  (m_\chi + \xi^2_-)(m + \lambda^2_- \cos^2\frac{\theta}{2})\big]  
\\  &&\pm \, \lambda_+ \lambda_- (\gamma_f - \beta_f)  \big[  (m_\chi+\xi^2_- )   \cos^2\frac{\theta}{2} 
+ m +\lambda^2_- \cos^2\frac{\theta}{2}\big] \Big]
, \\   (l^w_{\pm\pm}\, h^{w,f}_{\mp\mp})^{} &=& 4 \cos\frac{\theta}{2}\Big[(\alpha_f + \delta_f)  (m_\chi + \xi^2_-)  m + \xi_+ \xi_-  \lambda_+ \lambda_-  \big(\alpha_f + \alpha_f \sin^2\frac{\theta}{2}+ \delta_f\cos^2\frac{\theta}{2}  \big)
 + \\&&  +( m_\chi+ \xi^2_- ) \lambda^2_- \big(\delta_f + \delta_f\sin^2\frac{\theta}{2}+\alpha_f \cos^2\frac{\theta}{2} \big)  \pm \lambda_+ \lambda_- \big\{ (\beta_f+\gamma_f )  m 
\\&& +  \lambda^2_-  \big(\beta_f+\beta_f \sin^2\frac{\theta}{2} + \gamma_f \cos^2\frac{\theta}{2}\big) 
  + ( m_\chi + \xi^2_- ) \big(\gamma_f + \gamma_f \sin^2\frac{\theta}{2}+\beta_f \cos^2\frac{\theta}{2} \big)\big\} \Big]
.\end{eqnarray*} 
When the helicity of the $\chi$ lepton is preserved, but the projection of the nucleon spin changes:
\begin{eqnarray*} 
(l^w_{\pm\pm}\, h^{w,f}_{\pm\mp})^{} &=& \mp 4 \sin\frac{\theta}{2} e^{\mp i \phi}  
\Big[  2m\delta_f  (m_\chi + \xi^2_-) +\xi_+ \xi_- \lambda_+ \lambda_- (\alpha_f + \alpha_f \sin^2\frac{\theta}{2} + \delta_f \cos^2\frac{\theta}{2})  
\\&& +  (\xi^2_- + m_\chi) \lambda^2_- (\delta_f +\delta_f \sin^2\frac{\theta}{2}+\alpha_f \cos^2\frac{\theta}{2})
\pm \lambda_+ \lambda_- \big\{2m \beta_f +
\\&& +(\xi^2_- + m_\chi )(\gamma_f+ \gamma_f\sin^2\frac{\theta}{2}+\beta_f \cos^2\frac{\theta}{2}) 
+  \lambda^2_- (\beta_f+\beta_f  \sin^2\frac{\theta}{2} +\gamma_f \cos^2\frac{\theta}{2} ) \big\} \Big]
, \\ 
(l^w_{\pm\pm}\, h^{w,f}_{\mp\pm})^{}   &=& 
4 \sin\frac{\theta}{2} \cos^2\frac{\theta}{2} e^{\pm i\phi} \Big[
(\gamma_f-\beta_f) \lambda_+ \lambda_-  [m_\chi   +\lambda^2_- + \xi^2_-]
\\&&\pm  (\alpha_f-\delta_f) [\xi_+ \xi_- \lambda_+ \lambda_-  +  (\xi^2_- + m_\chi)\lambda^2_- ]    \Big]
. \end{eqnarray*} 
When the helicity of the $\chi$ lepton changes, but the projection of the nucleon spin is preserved:
\begin{eqnarray*} 
(l^w_{\pm\mp}\, h^{w,f}_{\mp\mp})^{}&=&4 m_\chi \sin\frac{\theta}{2}  e^{\mp i\phi} \Big[(\beta_f-\gamma_f) \lambda_+ \lambda_-  \cos^2\frac{\theta}{2} \pm (\alpha_f-\delta_f)  (m + \lambda^2_- \cos^2\frac{\theta}{2})  \Big]
, \\ (l^w_{\pm\mp}\, h^{w,f}_{\pm\pm})^{}   &=&4 m_\chi  \sin\frac{\theta}{2} e^{\mp i\phi} \Big[
(\gamma_f - \beta_f ) \lambda_+ \lambda_- \cos^2\frac{\theta}{2}  
\pm \big\{ (\alpha_f+ \delta_f)  m   + (\alpha_f- \delta_f ) \lambda^2_- \cos^2\frac{\theta}{2}\big\}  \Big]
. \end{eqnarray*}  
Finally, when both the helicity of the $\chi$ lepton and the projection of the nucleon spin change:
\begin{eqnarray*} 
(l^w_{\pm\mp}\, h^{w,f}_{\mp\pm})^{}   &=& 
4 m_\chi \cos\frac{\theta}{2} \Big[(\alpha_f- \delta_f) \lambda^2_- \sin^2\frac{\theta}{2} - 2 \delta_f m 
\pm (\gamma_f - \beta_f)\lambda_+ \lambda_-  \sin^2\frac{\theta}{2}\Big]
,  \\  (l^w_{\pm\mp}\, h^{w,f}_{\pm\mp})^{}&=&
 - 4 m_\chi \lambda_-  \cos\frac{\theta}{2} \sin^2\frac{\theta}{2}  e^{\mp 2 i \phi}  \big[  
 (\alpha_f- \delta_f)   \lambda_- \mp (\gamma_f - \beta_f) \lambda_+ \big] 
. \end{eqnarray*}
It is convenient to have scalar products in the form of an expansion in the coupling constants of weak currents 
$\alpha, \beta,\gamma, \delta$ (the index $f$ is temporarily omitted) in the following compact form
\begin{eqnarray}\nonumber
\label{eq:42RchiA-CrossSection-via-ScalarProducts-via-alpha-beta-gamma-delta}
\frac{(l^w_{\pm\pm}\, h^{w,f}_{\pm\pm})}{4 \cos\frac{\theta}{2}}&=&
\alpha {f_1(\theta)} - \delta {f_1(\theta)} \pm \gamma {f_2(\theta)} \mp \beta {f_2(\theta)} 
, \\  \nonumber
\frac{(l^w_{\pm\pm}\, h^{w,f}_{\mp\mp})}{4 \cos\frac{\theta}{2}} &=&
\alpha {f_3(\theta)}+ \delta{f_4(\theta)} \pm  \beta{f_5(\theta)} \pm \gamma{f_6(\theta)}  
,\\  \nonumber
\frac{(l^w_{\pm\pm}\, h^{w,f}_{\pm\mp})}{\mp 4 \sin\frac{\theta}{2} e^{\mp i \phi}} &=&     
\alpha {f_7(\theta)}+\delta {f_8(\theta)}\pm \beta{f_9(\theta)} \pm \gamma {f_{10}(\theta)} 
, \\ \nonumber 
\frac{(l^w_{\pm\pm}\, h^{w,f}_{\mp\pm})}{4 \sin\frac{\theta}{2} e^{\pm i\phi}} &=& 
\pm \alpha{f_{11}(\theta)}  \mp \delta{f_{11}(\theta)}+\gamma {f_{12}(\theta)} -\beta {f_{12}(\theta)} 
; \\ \frac{(l^w_{\pm\mp}\, h^{w,f}_{\mp\mp})}{4 \sin\frac{\theta}{2}  e^{\mp i\phi}}&=& 
\pm \alpha {f_{13}(\theta)} \mp \delta{f_{13}(\theta)}+\beta{f_{14}(\theta)} -\gamma {f_{14}(\theta)}
, \\  \nonumber
\frac{(l^w_{\pm\mp}\, h^{w,f}_{\pm\pm})}{4 \sin\frac{\theta}{2} e^{\mp i\phi} }   &=&
\pm \alpha{f_{13}(\theta)} \pm  \delta{f_{15}(\theta)} +\gamma  {f_{14}(\theta)} - \beta  {f_{14}(\theta)}
,\\ \nonumber
\frac{(l^w_{\pm\mp}\, h^{w,f}_{\mp\pm})}{4 \cos\frac{\theta}{2} } &=& 
+\alpha{f_{16}(\theta)}- \delta{f_{17}(\theta)}\pm \gamma_f {f_{14}(\theta)} \mp \beta_f {f_{14}(\theta)}
,  \\ \nonumber 
\frac{ (l^w_{\pm\mp}\, h^{w,f}_{\pm\mp})}{4 \cos\frac{\theta}{2}  e^{\mp 2 i \phi} } &=&
- \alpha{f_{16}(\theta)}  + \delta {f_{16}(\theta)}   \pm\gamma{f_{14}(\theta)}  \mp \beta {f_{14}(\theta)}
. \nonumber \end{eqnarray}
The following "form-factor-type"\/ notations are introduced here:
\begin{eqnarray}\nonumber
\label{eq:42RchiA-CrossSection-via-ScalarProducts-f_1--f_17}
f_1(\theta) &\equiv& E_\chi m + (E_\chi+\lambda_+^2) \lambda^2_- \cos^2\frac{\theta}{2}
,\qquad  f_2(\theta)\equiv \lambda_+ \lambda_-  \big[m  + (E_\chi +\lambda^2_- )\cos^2\frac{\theta}{2}\big]
,\\ \nonumber
f_3(\theta)&\equiv& E_\chi(m+ \lambda^2_-  \cos^2\frac{\theta}{2}) +  \lambda^2_+ \lambda^2_- (1+ \sin^2\frac{\theta}{2})
,\\ \nonumber
f_4(\theta)&\equiv&  E_\chi m + E_\chi \lambda^2_- (1+ \sin^2\frac{\theta}{2})  + \lambda^2_+ \lambda^2_-  \cos^2\frac{\theta}{2}
 ,\\ \nonumber
 f_5(\theta) &\equiv&  \lambda_+ \lambda_- [m+  \lambda^2_- (1+\sin^2\frac{\theta}{2})   +  E_\chi \cos^2\frac{\theta}{2}]
,\\ \nonumber
f_6(\theta)&\equiv& \lambda_+ \lambda_- [m +  \lambda^2_- \cos^2\frac{\theta}{2}   + E_\chi (1+ \sin^2\frac{\theta}{2})]
, \\ 
f_7(\theta)&\equiv& \lambda^2_+ \lambda^2_- (1+\sin^2\frac{\theta}{2})+E_\chi \lambda^2_- \cos^2\frac{\theta}{2}
,\\ \nonumber
f_8(\theta)&\equiv& 2 E_\chi m +  \lambda^2_+ \lambda^2_- \cos^2\frac{\theta}{2}  +  E_\chi \lambda^2_- (1+\sin^2\frac{\theta}{2})
,\\ \nonumber
f_9(\theta)&\equiv& \lambda_+ \lambda_-  [2m  + E_\chi  \cos^2\frac{\theta}{2} +  \lambda^2_- (1+ \sin^2\frac{\theta}{2})]
\\ \nonumber
f_{10}(\theta)&\equiv&  \lambda_+ \lambda_- [E_\chi (1+\sin^2\frac{\theta}{2})+\lambda^2_- \cos^2\frac{\theta}{2} ]
,\qquad    f_{11}(\theta)\equiv (E_\chi+\lambda^2_+ )\lambda^2_-  \cos^2\frac{\theta}{2}
,\\ \nonumber
f_{12}(\theta)&\equiv&  \lambda_+ \lambda_-  (E_\chi   +\lambda^2_- ) \cos^2\frac{\theta}{2}
,\qquad\qquad\qquad f_{13}(\theta)\equiv m_\chi  (m + \lambda^2_- \cos^2\frac{\theta}{2})
,\\ \nonumber
f_{14}(\theta)&\equiv&m_\chi \lambda_+ \lambda_-  \cos^2\frac{\theta}{2}
,\quad\qquad  f_{15}(\theta)\equiv m_\chi  ( m   -  \lambda^2_- \cos^2\frac{\theta}{2}) 
= f_{13}(\theta) - 2 m_\chi \lambda^2_- \cos^2\frac{\theta}{2}
,\\ f_{16}(\theta)&\equiv& m_\chi  \lambda^2_- \sin^2\frac{\theta}{2}
,\qquad \qquad f_{17}(\theta)\equiv m_\chi(2m+ \lambda^2_- \sin^2\frac{\theta}{2}) =2 m_\chi m + f_{16}(\theta)
\nonumber
.\end{eqnarray}
The squares of these scalar products are written as follows:
\begin{eqnarray*}
\frac{|(l^w_{\pm\pm}\, h^{w,f}_{\pm\pm})|^2}{16\cos^2\frac{\theta}{2}}&=&
(\alpha^2 +  \delta^2) f_1^2 - 2 \alpha\delta f_1^2   + (\beta^2 + \gamma^2) f_2^2 - 2\beta \gamma f_2^2
\mp 2 (\alpha\beta + \gamma \delta ) f_1 f_2  \pm 2 (\alpha\gamma+ \beta  \delta) f_1 f_2 
, \\   \frac{|(l^w_{\pm\pm}\, h^{w,f}_{\mp\mp})|^2}{16 \cos^2\frac{\theta}{2}} &=&
 \alpha^2 f_3(\theta)^2 + \beta^2 f_5(\theta)^2 + \gamma^2 f_6(\theta)^2 +  \delta^2 f_4(\theta)^2
 +2 \alpha\delta f_3(\theta) f_4(\theta) +  2 \beta\gamma f_5(\theta) f_6(\theta)  
\\&&\pm 2 \alpha\beta f_3(\theta) f_5(\theta) \pm 2\alpha\gamma f_3(\theta) f_6(\theta) 
\pm 2\beta\delta  f_4(\theta)f_5(\theta)    \pm 2\gamma\delta f_4(\theta)f_6(\theta) 
,\\  \frac{|(l^w_{\pm\pm}\, h^{w,f}_{\pm\mp})|^2}{16 \sin^2\frac{\theta}{2} } &=&     
 \alpha^2   f_7(\theta)^2+ \delta^2  f_8(\theta)^2+ 2 \alpha \delta f_7(\theta) f_8(\theta)  
+ \beta^2 f_9(\theta)^2+ \gamma^2 f_{10}(\theta)^2 +  2 \beta \gamma f_9(\theta)f_{10}(\theta)  
\\&& \pm 2 \alpha\beta f_7(\theta)f_9(\theta)  \pm 2 \alpha\gamma f_7(\theta)f_{10}(\theta)
 \pm 2  \beta\delta f_9(\theta)f_8(\theta)  \pm 2 \gamma \delta f_8(\theta)f_{10}(\theta)
, \\  \frac{|(l^w_{\pm\pm}\, h^{w,f}_{\mp\pm})|^2}{16 \sin^2\frac{\theta}{2} } &=& 
(\alpha^2 + \delta^2- 2 \alpha\delta) f_{11}(\theta)^2  + (\beta^2 + \gamma^2- 2 \beta\gamma) f_{12}(\theta)^2 \\&& \mp 2 (\alpha\beta+ \gamma\delta)  f_{11}(\theta)f_{12}(\theta)  \pm  2 (\alpha\gamma +\beta\delta ) f_{11}(\theta)f_{12}(\theta) 
; \end{eqnarray*} 
\label{PAGE:42RchiA-CrossSection-via-ScalarProducts-via-alpha-beta-gamma-delta-Squared}
\begin{eqnarray*} 
\frac{|(l^w_{\pm\mp}\, h^{w,f}_{\mp\mp})|^2}{16 \sin^2\frac{\theta}{2}}&=& 
(\alpha^2 + \delta^2- 2 \alpha \delta) f_{13}(\theta)^2  + (\beta^2 + \gamma^2 - 2 \beta \gamma) f_{14}(\theta)^2  
\\&&\pm 2(\alpha\beta +\gamma \delta)  f_{13}(\theta)f_{14}(\theta)
\mp 2 (\alpha\gamma  +\beta\delta) f_{13}(\theta)f_{14}(\theta)   
, \\  \frac{|(l^w_{\pm\mp}\, h^{w,f}_{\pm\pm})|^2}{16 \sin^2\frac{\theta}{2} }   &=&
\alpha^2   f_{13}(\theta)^2 + \delta^2 f_{15}(\theta)^2 + 2 \alpha\delta f_{13}(\theta) f_{15}(\theta)  
+  (\beta^2  + \gamma^2- 2 \beta \gamma) f_{14}(\theta)^2 
\\&&\mp 2 (\alpha\beta -\alpha\gamma) f_{13}(\theta)f_{14}(\theta)
\mp 2 (\beta\delta - \gamma\delta) f_{14}(\theta)f_{15}(\theta)
,\\  \frac{|(l^w_{\pm\mp}\, h^{w,f}_{\mp\pm})|^2}{16 \cos^2\frac{\theta}{2} } &=& 
 \alpha^2 f_{16}(\theta)^2+ \delta^2 f_{17}(\theta)^2- 2 \alpha\delta f_{16}(\theta)f_{17}(\theta)  
+ (\beta^2 + \gamma^2 - 2 \beta\gamma) f_{14}(\theta)^2
\\&&\pm 2 \alpha(\gamma-\beta)f_{14}(\theta)f_{16}(\theta)\pm 2 \delta(\beta-\gamma) f_{14}(\theta)f_{17}(\theta) 
, \\  \frac{|(l^w_{\pm\mp}\, h^{w,f}_{\pm\mp})|^2}{16 \cos^2\frac{\theta}{2} } &=&
(\alpha^2  +  \delta^2 - 2 \alpha \delta) f_{16}(\theta)^2+(\beta^2 + \gamma^2  - 2 \beta \gamma) f_{14}(\theta)^2 
\\&& \pm 2 (\alpha \beta +\gamma \delta)f_{14}(\theta)f_{16}(\theta) 
\mp 2 (\alpha\gamma+\beta\delta)f_{14}(\theta)f_{16}(\theta) 
. \end{eqnarray*}
These scalar products of the nucleon and lepton currents are the basis for calculation of coherent (elastic) and incoherent (inelastic) cross sections for the $\chi$ particle-nucleus interaction in the relativistic approximation.

\subsection{\em Relativistic cross sections of the process ${\chi A\to \chi A^{(*)}}$}\label{43chiA-CrossSection-via-ScalarProducts}
\paragraph{\em Coherent cross sections.}
According to definition (\ref{eq:43RchiA-CrossSection-via-ScalarProducts-both-CS-main}), the 
coherent $\chi A$ cross sections are given by $\hat Q^{s's}_{\pm}$ combinations of scalar products
(\ref{eq:43RchiA-CrossSection-via-ScalarProducts-CohCS-factors}), which look like
\begin{eqnarray*}
\frac{\hat Q^{\mp\mp}_{+}}{8 \cos\frac{\theta}{2}}&\equiv& \alpha f_{\alpha+}(\theta) + \delta f_{\delta-} (\theta) 
\mp \beta f_{\beta-} (\theta) \mp \gamma f_{\gamma+} (\theta)
,\\ \frac{\hat Q^{\mp\mp}_{-}}{8 \cos\frac{\theta}{2}}&\equiv&\pm [\alpha f_{\alpha-} (\theta)+\delta f_{\delta+} (\theta)  \mp \beta f_{\beta+} (\theta) \mp \gamma f_{\gamma-} (\theta)]
;\\  \frac{\hat Q^{\mp\pm}_{+}}{8 \sin\frac{\theta}{2}  e^{\pm i\phi}}&=&  \mp \alpha \hat f_{\alpha}(\theta) \pm \delta \hat f_{\delta-}(\theta) 
,\qquad  \frac{\hat Q^{\mp\pm}_{-}}{8\sin\frac{\theta}{2}  e^{\pm i\phi}} 
= \mp (\gamma-\beta)  \hat f_{\beta\gamma}(\theta) + \delta \hat f_{\delta+}(\theta) 
.\end{eqnarray*} 
Then the $\displaystyle Q_{f}^{s's} $ values from (\ref{eq:43RchiA-CrossSection-via-ScalarProducts-CohCS-factors}) are
\begin{eqnarray}\nonumber
\label{eq:43RchiA-CrossSection-via-ScalarProducts-WeakCohQ_f^s's}
\frac{ Q_{f}^{\mp\mp}}{8 \cos\frac{\theta}{2}}&=&
\alpha \big[ f_{\alpha+}(\theta) \pm \dfrac{\Delta A_f}{A_f}  f_{\alpha-} (\theta)\big]
+ \delta \big[ f_{\delta-} (\theta) \pm \dfrac{\Delta A_f}{A_f}  f_{\delta+} (\theta)\big]
\\&& \qquad  \mp \beta\big[ f_{\beta-} (\theta) \pm \dfrac{\Delta A_f}{A_f}  f_{\beta+} (\theta)\big] 
\mp \gamma\big[ f_{\gamma+} (\theta)\pm \dfrac{\Delta A_f}{A_f}  f_{\gamma-} (\theta)\big]
,  \\ \nonumber
\frac{Q_{f}^{\mp\pm}}{8\sin\frac{\theta}{2}  } &=&
\mp e^{\pm i\phi}  \big[ \alpha \hat f_{\alpha}(\theta)- \delta \hat f_{\delta-}(\theta)
+(\gamma-\beta) \dfrac{\Delta A_f}{A_f}  \hat f_{\beta\gamma}(\theta) \mp\delta \dfrac{\Delta A_f}{A_f} \hat f_{\delta+}(\theta) \big]
.\end{eqnarray}
Here the following notations for form-factor functions are introduced:
\begin{eqnarray}\nonumber
\label{eq:43RchiA-CrossSection-via-ScalarProducts-WeakCohCS-fs}
f_{\alpha+} (\theta)&\equiv&\frac12 [f_1(\theta)+f_3(\theta)] = E_\chi (m + \lambda^2_- \cos^2\frac{\theta}{2})+ \lambda^2_+ \lambda^2_- =E_\chi (m\sin^2\frac{\theta}{2} +E_p\cos^2\frac{\theta}{2})+ |\bm{k}|^2 
, \\\nonumber   f_{\delta+} (\theta)&\equiv&\frac12 [f_4(\theta)+f_1(\theta)]=E_\chi (m+ \lambda^2_- ) +\lambda^2_+ \lambda^2_-  \cos^2\frac{\theta}{2}= E_\chi E_p +|\bm{k}|^2  \cos^2\frac{\theta}{2}
,\\\nonumber f_{\beta+} (\theta)&\equiv& \frac12 [f_5(\theta)+f_2(\theta)] = \lambda_+ \lambda_- (m + E_\chi  \cos^2\frac{\theta}{2} + \lambda^2_-) =|\bm{k}|(E_p + E_\chi  \cos^2\frac{\theta}{2})
,\\\nonumber  f_{\gamma+} (\theta)&\equiv&\frac12 [f_6(\theta)+f_2(\theta)]=\lambda_+ \lambda_-  ( m  +  E_\chi+ \lambda^2_-  \cos^2\frac{\theta}{2} ) =|\bm{k}|(E_\chi+  m\sin^2\frac{\theta}{2}+E_p\cos^2\frac{\theta}{2})
,\\\nonumber f_{\alpha-} (\theta)&\equiv &\frac12 [f_3(\theta)-f_1(\theta) ]=\lambda_+^2 \lambda^2_- \sin^2\frac{\theta}{2} =|\bm{k}|^2 \sin^2\frac{\theta}{2}
,\\\nonumber f_{\delta-} (\theta)&\equiv&\frac12 [f_4(\theta) -f_1(\theta)]= E_\chi \lambda^2_- \sin^2\frac{\theta}{2}
= E_\chi (E_p-m) \sin^2\frac{\theta}{2}
,\\\nonumber f_{\beta-} (\theta)&\equiv & \frac12 [f_5(\theta)-f_2(\theta)] = \lambda_+ \lambda_- \lambda^2_-  \sin^2\frac{\theta}{2} = |\bm{k}| (E_p-m)  \sin^2\frac{\theta}{2}
,\\\nonumber f_{\gamma-} (\theta)&\equiv& \frac12 [f_6(\theta)-f_2(\theta)]
=\lambda_+ \lambda_- E_\chi \sin^2\frac{\theta}{2}=|\bm{k}| E_\chi \sin^2\frac{\theta}{2}
;\\ \hat f_{\alpha}(\theta)&\equiv&  f_{13}(\theta)=  m_\chi  m + m_\chi \lambda^2_- \cos^2\frac{\theta}{2}    
=  m_\chi  (m +(E_p-m) \cos^2\frac{\theta}{2})    
,\\\nonumber \hat f_{\beta\gamma}(\theta)&\equiv& f_{14}(\theta)= m_\chi \lambda_+ \lambda_-  \cos^2\frac{\theta}{2} 
= m_\chi |\bm{k}| \cos^2\frac{\theta}{2} 
,\\\nonumber \hat f_{\delta-}(\theta)&\equiv&\frac12 [ f_{13}(\theta)-f_{15}(\theta)] =  m_\chi \lambda^2_- \cos^2\frac{\theta}{2} = m_\chi  (E_p-m) \cos^2\frac{\theta}{2}
,\\\nonumber  \hat f_{\delta+}(\theta)&\equiv&\frac12 [ f_{15}(\theta)+f_{13}(\theta)] =  m_\chi m. 
\end{eqnarray}
Substitution of $Q_{f}^{s's} $ parameters from (\ref{eq:43RchiA-CrossSection-via-ScalarProducts-WeakCohQ_f^s's}) into formula
(\ref{eq:43RchiA-CrossSection-via-ScalarProducts-both-CS-main}) gives the following 
expressions for {\em relativistic}\/ cross sections of {\em coherent} $\chi A$ scattering of the massive $\chi$ particle on the nucleus (process $\chi A\to \chi A$)
due to weak interaction (\ref{eq:42RchiA-CrossSection-ScalarProducts-ChiEta-Weak-definition}):
\begin{eqnarray}\nonumber
\label{eq:43RchiA-CrossSection-via-ScalarProducts-WeakCohCS}
\frac{d\sigma^{\mp\mp}_\text{coh}(\bm{q})}{g_\text{c} \hat c_A  d T_A}&=&
\cos^2\frac{\theta}{2} \Big| \sum^{}_{f=p,n} G^f_{\alpha}(\bm{q}) f_{\alpha+}(\theta) \pm \Delta G^f_{\alpha}(\bm{q})  f_{\alpha-} (\theta)  +  G^f_{\delta}(\bm{q}) f_{\delta-} (\theta) \pm \Delta G^f_{\delta}(\bm{q})   f_{\delta+} (\theta) 
\\&& \qquad \nonumber \mp  G^f_{\beta}(\bm{q}) f_{\beta-} (\theta) - \Delta G^f_{\beta}(\bm{q}) f_{\beta+} (\theta) \mp G^f_{\gamma}(\bm{q})  f_{\gamma+} (\theta)  - \Delta G^f_{\gamma}(\bm{q})   f_{\gamma-} (\theta) \Big|^2 
 , \\  \frac{d\sigma^{\mp\pm}_\text{coh}(\bm{q})}{g_\text{c} \hat c_A  d T_A}&=& 
\sin^2\frac{\theta}{2}\Big| \sum^{}_{f=p,n}  G^f_{\alpha}(\bm{q})  \hat f_{\alpha}(\theta)- G^f_{\delta}(\bm{q}) \hat f_{\delta-}(\theta) \mp  \Delta G^f_{\delta}(\bm{q})  \hat f_{\delta+}(\theta) 
\\&& \nonumber \qquad \qquad + [\Delta G^f_{\gamma}(\bm{q}) - \Delta G^f_{\beta}(\bm{q})]\hat f_{\beta\gamma}(\theta)  \Big|^2
\nonumber .\end{eqnarray}
In formulas (\ref{eq:43RchiA-CrossSection-via-ScalarProducts-WeakCohCS}), 
coherent structural factors that accumulate dependence on both the nucleus and the weak interaction constants
are introduced:
\begin{eqnarray}\nonumber
\label{eq:43RchiA-CrossSections-via-ScalarProducts-Coh-G-and-DeltaG-s}
G^f_{\alpha}(\bm{q}) &\equiv& \alpha_f F_{f}(\bm{q}) {A_f}
,\qquad  \Delta G^f_{\alpha}(\bm{q}) \equiv  \alpha_f  F_{f}(\bm{q}) \Delta A_f 
,\\ \nonumber G^f_{\beta}(\bm{q}) &\equiv& \beta_f F_{f}(\bm{q}) {A_f} 
,\qquad    \Delta G^f_{\beta}(\bm{q})\equiv \beta_f  F_{f}(\bm{q}) \Delta A_f 
,\\ G^f_{\gamma}(\bm{q}) &\equiv &\gamma_f F_{f}(\bm{q}) {A_f} 
,\qquad  \Delta G^f_{\gamma}(\bm{q}) \equiv   \gamma_f F_{f}(\bm{q}) \Delta A_f 
,\\  G^f_{\delta}(\bm{q}) &\equiv& \delta_f F_{f}(\bm{q}) {A_f}  
, \qquad  \Delta G^f_{\delta}(\bm{q}) \equiv \delta_f F_{f}(\bm{q}) \Delta A_f   
\nonumber .\end{eqnarray} 
Important simplification 
can be obtained if one takes into account that masses of the proton and the neutron are the same
\footnote{Previously it was not considered to be so, although the $f$-index of the mass $m$ and the $f$-functions was omitted.}, 
 $m \equiv m_p=m_n$.
Then the form-factor $f$-functions in (\ref{eq:43RchiA-CrossSection-via-ScalarProducts-WeakCohCS})
will also be independent of the nucleon type, and formulas (\ref{eq:43RchiA-CrossSection-via-ScalarProducts-WeakCohCS})
can be written as  follows:
{\small
\begin{eqnarray}\nonumber
\label{eq:43RchiA-CrossSection-via-ScalarProducts-WeakCohCS-general-mp=mn}
\frac{d\sigma^{\mp\mp}_\text{coh}(\bm{q})}{g_\text{c}\hat c_A  d T_A}&=& 
\cos^2\frac{\theta}{2}  \Big|  G_\alpha(A,\bm{q}) f_{\alpha+}(\theta) \pm \Delta G_\alpha(A,\bm{q})  f_{\alpha-} (\theta)   +  G_\delta(A,\bm{q}) f_{\delta-} (\theta)     \pm  \Delta G_\delta(A,\bm{q}) f_{\delta+} (\theta) 
\\&& \qquad \nonumber
\mp  G_\beta (A,\bm{q}) f_{\beta-} (\theta)-\Delta G_\beta (A,\bm{q})  f_{\beta+} (\theta) 
\mp G_\gamma(A,\bm{q}) f_{\gamma+}(\theta) -\Delta G_\gamma(A,\bm{q})f_{\gamma-} (\theta) \Big|^2 
, \\  \frac{d\sigma^{\mp\pm}_\text{coh}(\bm{q})}{g_\text{c}\hat c_A  d T_A}&=& 
 \sin^2\frac{\theta}{2}\Big| G_\alpha(A,\bm{q})  \hat f_{\alpha}(\theta) - G_\delta (A,\bm{q})  \hat f_{\delta-}(\theta)  \mp \Delta G_\delta (A,\bm{q}) \hat f_{\delta+}(\theta) 
\\&& \nonumber \qquad\qquad  +[\Delta G_\gamma(A,\bm{q}) - \Delta G_\beta(A,\bm{q})] \hat f_{\beta\gamma}(\theta) \Big|^2
\nonumber .\end{eqnarray}}%
In (\ref{eq:43RchiA-CrossSection-via-ScalarProducts-WeakCohCS-general-mp=mn}), 
new $\bm{q}$-dependent effective coupling constants are introduced, 
which take into account the influence of the nuclear and nucleon structure
($\tau$ "runs through"\/  $\alpha, \beta, \gamma, \delta$):
\begin{eqnarray}\nonumber
\label{eq:43RchiA-CrossSection-via-ScalarProducts-BigWeakNuclear-Couplings}
G_{\tau}(A,\bm{q})&\equiv&G^p_{\tau}(\bm{q}) +G^n_{\tau}(\bm{q})=\tau_p A_pF_p(\bm{q}) + \tau_n A_nF_n(\bm{q})  
,\\ \Delta G_{\tau}(A,\bm{q})&\equiv&\Delta G^p_{\tau}(\bm{q}) +\Delta G^n_{\tau}(\bm{q}) 
=\tau_p \Delta A_pF_p(\bm{q}) + \tau_n \Delta A_nF_n(\bm{q})  
. \end{eqnarray} 
One can say that the nuclear form factors "participate"\/ in the weak interaction together
with the corresponding number of (unpaired) protons or neutrons.
  \par
Taking into account expressions for the form-factor $f$-functions, 
(\ref{eq:43RchiA-CrossSection-via-ScalarProducts-WeakCohCS-general-mp=mn}) can be written as
{\small \begin{eqnarray}\nonumber
\label{eq:43RchiA-CrossSection-via-ScalarProducts-Weak-ChiEta-rel-CohCS-mp=mn-general}
\frac{d\sigma^{\mp\mp}_\text{coh}}{g_\text{c} d T_A} &=&\cos^2\frac{\theta}{2} \dfrac{G^2_F  m_A }{4 \pi }
\dfrac{E^2_\chi}{|\bm{k}_\chi|^2} \Big| 
G_\alpha(A,\bm{q})\Big(\sin^2\frac{\theta}{2}+\frac{E_p}{m}\cos^2\frac{\theta}{2}+\frac{|\bm{k}|^2}{m E_\chi}\Big) \pm \Delta G_\alpha(A,\bm{q}) \frac{|\bm{k}|^2}{m E_\chi} \sin^2\frac{\theta}{2}
\\&&\nonumber \quad \qquad +  G_\delta(A,\bm{q})\Big(\frac{E_p}{m}-1\Big)\sin^2\frac{\theta}{2}   
\pm\Delta G_\delta(A,\bm{q})\Big(\frac{E_p}{m} + \frac{|\bm{k}|^2}{m E_\chi}\cos^2\frac{\theta}{2}\Big) 
\\&& \qquad\quad \nonumber
\mp  G_\beta (A,\bm{q}) \frac{|\bm{k}|}{E_\chi} \Big(\frac{E_p}{m}-1\Big)\sin^2\frac{\theta}{2}
- \Delta G_\beta (A,\bm{q})\frac{|\bm{k}|}{m}\Big(\frac{E_p}{E_\chi}+\cos^2\frac{\theta}{2}\Big)
\\&& \qquad\quad  \mp G_\gamma(A,\bm{q}) \frac{|\bm{k}|}{m} 
{\big(1+ \frac{E_p}{ E_\chi} \cos^2\frac{\theta}{2}
+ \frac{m}{ E_\chi} \sin^2\frac{\theta}{2}\big)}
 -\Delta G_\gamma(A,\bm{q}) \frac{|\bm{k}|}{m} \sin^2\frac{\theta}{2}   \Big|^2
,\\   \nonumber
 \frac{d\sigma^{\mp\pm}_\text{coh}}{g_\text{c} d T_A} &=&\sin^2\frac{\theta}{2} \dfrac{G^2_F  m_A}{4 \pi}
 \dfrac{m^2_\chi }{|\bm{k}_\chi|^2 }   \Big|\big(G_\alpha(A,\bm{q}) - G_\delta(A,\bm{q})\big) \frac{E_p}{m} \cos^2\frac{\theta}{2}   + G_\alpha(A,\bm{q})\sin^2\frac{\theta}{2}+ G_\delta(A,\bm{q})\cos^2\frac{\theta}{2}
\\&& \nonumber \qquad  +\big(\Delta G_\gamma(A,\bm{q})  -\Delta G_\beta(A,\bm{q})\big) \frac{|\bm{k}|}{m} \cos^2\frac{\theta}{2}  \mp \Delta G_\delta(A,\bm{q}) \Big|^2
\nonumber .\end{eqnarray}}%
Formulas (\ref{eq:43RchiA-CrossSection-via-ScalarProducts-Weak-ChiEta-rel-CohCS-mp=mn-general})
give the most general set of expressions for relativistic cross sections of coherent scattering of the massive $\chi$ lepton 
on the nucleus due to the weak interaction given the form
(\ref{eq:42RchiA-CrossSection-ScalarProducts-ChiEta-Weak-definition}).
Averaged over the initial projections of $\chi$-lepton spins and summed over the final ones, the coherent cross sections are as follows:\par 
{\small \begin{eqnarray}\nonumber
\label{eq:43RchiA-CrossSection-via-ScalarProducts-ChiEta-rel-CohCS-mp=mn-general-spinful-total}
 \dfrac{d\sigma^{s'=s}_\text{coh}}{d T_A}&=&\cos^2\frac{\theta}{2}\dfrac{G^2_Fm_A}{4\pi}\dfrac{E^2_\chi}{|\bm{k}_\chi|^2}\Big\{\Big[G_\alpha(A,\bm{q})\Big(\sin^2\frac{\theta}{2} + \frac{E_p}{m}\cos^2\frac{\theta}{2}+\frac{|\bm{k}|^2}{mE_\chi}\Big) 
\\&&\nonumber
+G_\delta(A,\bm{q})\Big(\frac{E_p}{m}-1\Big)\sin^2\frac{\theta}{2}
 - \Delta G_\beta (A,\bm{q})\frac{|\bm{k}|}{m}\Big(\frac{E_p}{E_\chi}+\cos^2\frac{\theta}{2}\Big) -  \Delta G_\gamma(A,\bm{q}) \frac{|\bm{k}|}{m}\sin^2\frac{\theta}{2} \Big]^2 
\\&& \nonumber
+\Big[\Delta G_\alpha(A,\bm{q}) \frac{|\bm{k}|^2}{m E_\chi}\sin^2\frac{\theta}{2} 
+ \Delta G_\delta(A,\bm{q}) \Big(\frac{E_p}{m}+ \frac{|\bm{k}|^2}{mE_\chi} \cos^2\frac{\theta}{2}\Big) 
\\&&\quad -G_\beta (A,\bm{q}) \frac{|\bm{k}|}{m} \frac{E_p-m}{E_\chi} \sin^2\frac{\theta}{2}
- G_\gamma(A,\bm{q}) \frac{|\bm{k}|}{m} 
{\big(1+ \frac{E_p}{ E_\chi} \cos^2\frac{\theta}{2}+\frac{m}{ E_\chi} \sin^2\frac{\theta}{2}\big)}\Big]^2\Big\}
;\\ \nonumber
\dfrac{d\sigma^{s'\ne s}_\text{coh}}{d T_A}   &=& \sin^2\frac{\theta}{2} \dfrac{G^2_F  m_A}{4 \pi} \dfrac{m^2_\chi }{|\bm{k}_\chi|^2 }\Big\{
\Big[G_\alpha(A,\bm{q}) \Big(\sin^2\frac{\theta}{2} + \frac{E_p}{m}\cos^2\frac{\theta}{2}\Big)
-G_\delta(A,\bm{q}) \Big(\frac{E_p}{m}-1 \Big)\cos^2\frac{\theta}{2} +
\\&& \qquad\qquad + \big(\Delta G_\gamma(A,\bm{q}) - \Delta G_\beta(A,\bm{q})\big)  \frac{|\bm{k}|}{m}  \cos^2\frac{\theta}{2}\Big]^2 +\big[ \Delta G_\delta (A,\bm{q}) \big]^2\Big\}
\nonumber  .\end{eqnarray}}%
The second formula in 
(\ref{eq:43RchiA-CrossSection-via-ScalarProducts-ChiEta-rel-CohCS-mp=mn-general-spinful-total})
corresponds to the massive $\chi$ lepton coherent scattering on the nucleus with lepton helicity (spin) flip.
The cross section in this case is proportional to the square of the lepton mass $m_\chi$ 
and is strongly suppressed when the lepton escapes into the forward hemisphere.
The cross section without changing the lepton helicity (the first formula) is proportional to the square of the lepton 
energy $E_\chi$ and is maximal at the minimum lepton escape angles.
For spinless nuclei one can obtain "shorter"\/ formulas  from 
(\ref{eq:43RchiA-CrossSection-via-ScalarProducts-ChiEta-rel-CohCS-mp=mn-general-spinful-total})
by simply discarding terms proportional to $\Delta G$, since for $\Delta A_f=0$ all $\Delta G=0$.
\par
The "fully averaged"\/ (averaged over the initial helicities and summed over the final lepton helicities) 
{\em coherent}\/ $\chi A$ cross section is the sum of two expressions from
(\ref{eq:43RchiA-CrossSection-via-ScalarProducts-ChiEta-rel-CohCS-mp=mn-general-spinful-total}).
 For spinless nuclei ($\Delta A_f=0$), this  coherent $\chi A$ cross section has the form
 \begin{eqnarray}\nonumber
\label{eq:43RchiA-CrossSection-via-ScalarProducts-ChiEta-rel-CohCS-mp=mn-general-spinless-total}
\dfrac{d\sigma^{\text{total}}_{\text{coh},w}}{g_\text{c} \hat c_Ad T_A}\!\!&=&\!\!
\cos^2\frac{\theta}{2} \Big\{E^2_\chi  \Big[G_\alpha(A,\bm{q}) \big(m\sin^2\frac{\theta}{2} + E_p\cos^2\frac{\theta}{2}+\frac{|\bm{k}|^2}{E_\chi}\big)+G_\delta(A,\bm{q}) (E_p-m) \sin^2\frac{\theta}{2}  \Big]^2
 \\ \nonumber   \\&&  + |\bm{k}|^2\Big[G_\gamma(A,\bm{q}) 
{[m\sin^2\frac{\theta}{2}+ E_p\cos^2\frac{\theta}{2}+E_\chi ]}
+G_\beta (A,\bm{q}) (E_p-m) \sin^2\frac{\theta}{2}  \Big]^2 \Big\}
\\ &&+ \sin^2\frac{\theta}{2} m^2_\chi\Big[G_\alpha(A,\bm{q})(m\sin^2\frac{\theta}{2} + E_p\cos^2\frac{\theta}{2})-G_\delta(A,\bm{q})(E_p-m)\cos^2\frac{\theta}{2} \Big ]^2
\nonumber.\end{eqnarray}
As a "simplifying example"\/ of general formulas 
(\ref{eq:43RchiA-CrossSection-via-ScalarProducts-WeakCohCS}), 
the nuclear form factors are assumed to be real functions independent of the type of the nucleon, i.e., 
\begin{equation}\label{eq:43RchiA-CrossSections-via-ScalarProducts-Fp=Fn}
F_{}(\bm{q})\equiv F_{p}(\bm{q})=F_{n}(\bm{q}).
\end{equation}
Then they factorize in expressions (\ref{eq:43RchiA-CrossSection-via-ScalarProducts-BigWeakNuclear-Couplings}),
and coherent $\chi A$ cross sections (\ref{eq:43RchiA-CrossSection-via-ScalarProducts-Weak-ChiEta-rel-CohCS-mp=mn-general})
are directly proportional to them
\begin{eqnarray}\nonumber
\label{eq:43RchiA-CrossSections-via-ScalarProducts-ChiEta-rel-CohCSs-mp=mn-Fp=Fn-general} 
\frac{d\sigma^{\mp\mp}_\text{coh}}{g_\text{c} F^2(\bm{q})dT_A}&=&\cos^2\frac{\theta}{2}\dfrac{G^2_Fm_A}{4\pi} \dfrac{E^2_\chi}{|\bm{k}_\chi|^2} \Big|  G_\alpha(A)\Big(\sin^2\frac{\theta}{2}+\frac{E_p}{m}\cos^2\frac{\theta}{2}+\frac{|\bm{k}|^2}{m E_\chi}\Big) \pm \Delta G_\alpha(A) \frac{|\bm{k}|^2}{m E_\chi} \sin^2\frac{\theta}{2}
\\&&\nonumber \quad \qquad +  G_\delta(A)\Big(\frac{E_p}{m}-1\Big)\sin^2\frac{\theta}{2}   
\pm\Delta G_\delta(A)\Big(\frac{E_p}{m} + \frac{|\bm{k}|^2}{m E_\chi}\cos^2\frac{\theta}{2}\Big) 
\\&& \qquad\quad \nonumber
\mp  G_\beta (A) \frac{|\bm{k}|}{E_\chi} \Big(\frac{E_p}{m}-1\Big)\sin^2\frac{\theta}{2}
- \Delta G_\beta (A)\frac{|\bm{k}|}{m}\Big(\frac{E_p}{E_\chi}+\cos^2\frac{\theta}{2}\Big)
\\&& \qquad\quad   \mp G_\gamma(A) \frac{|\bm{k}|}{m}
{\big(1+ \frac{E_p}{ E_\chi} \cos^2\frac{\theta}{2}+\frac{m}{ E_\chi} \sin^2\frac{\theta}{2}\big)}
-\Delta G_\gamma(A) \frac{|\bm{k}|}{m} \sin^2\frac{\theta}{2}   \Big|^2
,\\   \nonumber
 \frac{d\sigma^{\mp\pm}_\text{coh}}{g_\text{c} F^2(\bm{q}) d T_A} &=&\sin^2\frac{\theta}{2} \dfrac{G^2_F  m_A}{4 \pi}  \dfrac{m^2_\chi }{|\bm{k}_\chi|^2 }   \Big|\big(G_\alpha(A)-G_\delta(A)\big) \frac{E_p}{m} \cos^2\frac{\theta}{2}   + G_\alpha(A)\sin^2\frac{\theta}{2}+ G_\delta(A)\cos^2\frac{\theta}{2}
\\&& \nonumber \qquad  +\big(\Delta G_\gamma(A)  -\Delta G_\beta(A)\big) \frac{|\bm{k}|}{m} \cos^2\frac{\theta}{2}  \mp \Delta G_\delta(A) \Big|^2
\nonumber .\end{eqnarray}
In (\ref{eq:43RchiA-CrossSections-via-ScalarProducts-ChiEta-rel-CohCSs-mp=mn-Fp=Fn-general}), 
$\bm{q}$-independent "enlarged"\/ interaction constants are   introduced:
\begin{eqnarray}
\label{eq:43RchiA-CrossSections-via-ScalarProducts-ChiEta-rel-CohCSs-mp=mn-Fp=Fn-couplings} 
G_\tau(A)&=&\tau_p A_p + \tau_n A_n \quad\text{and}\quad \Delta G_\tau(A)=\tau_p \Delta A_p +\tau_n \Delta A_n  , 
\quad \text{where}\quad \tau = \alpha, \beta, \gamma, \delta
.\qquad  \end{eqnarray}

\paragraph{\em Incoherent cross sections.} $\!\!\!\!\!\!$
{\em Relativistic}\/ incoherent cross sections of $\chi A$ scattering due to weak currents 
(\ref{eq:42RchiA-CrossSection-ScalarProducts-ChiEta-Weak-definition}) can be 
obtained from (\ref{eq:43RchiA-CrossSection-via-ScalarProducts-both-CS-main})
after calculating the following values:
$$Q^{s's}_\pm \equiv  S_+^{s's} \pm S_-^{s's}, \quad \text{where} \quad 
S_\pm^{s's} \equiv\sum^{}_{r'=\pm}|(l_{s's}\, h^{f}_{r'\pm})|^2 = |(l_{s's}\, h^{f}_{+\pm})|^2+|(l_{s's}\, h^{f}_{-\pm})|^2 .
$$
With the help of squared scalar products, represented as expansions in effective coupling constants
$\alpha, \beta, \gamma, \delta$
(page~\pageref{PAGE:42RchiA-CrossSection-via-ScalarProducts-via-alpha-beta-gamma-delta-Squared}),
eight $Q^{s's}_{\pm,w}$ values are computed.
They are as follows
\footnote{Hereinafter, for simplicity, the nucleon $f$-index is omitted in $\alpha, \beta, \gamma, \delta$.}:
\begin{eqnarray}\nonumber
 \label{eq:43RchiA-CrossSection-via-ScalarProducts-InCohCS-Q^s's_pm} 
\frac{Q^{\mp\mp}_{+,w}}{16} &=& \alpha^2 F^+_{\alpha^2}(\theta)+\beta^2 F^+_{\beta^2}(\theta) + \gamma^2 F^+_{\gamma^2}(\theta)+ \delta^2 F^+_{\delta^2}(\theta)+ \alpha\delta F^+_{\alpha\delta}(\theta)+  \beta\gamma F^+_{\beta\gamma}(\theta)
\\&& \nonumber \mp\{\alpha\beta F^+_{\alpha\beta}(\theta) +\alpha\gamma F^+_{\alpha\gamma}(\theta)
 + \beta\delta F^+_{\beta\delta}(\theta)+ \gamma \delta F^+_{\gamma\delta}(\theta)\}
, \\ \nonumber
 \frac{ Q^{\mp\mp}_{-,w}}{16} &=& \mp \{+\alpha^2F^-_{\alpha^2}(\theta) +\beta^2 F^-_{\beta^2}(\theta)+\gamma^2 F^-_{\gamma^2}(\theta)+\delta^2 F^-_{\delta^2}(\theta) - \alpha\delta{F^-_{\alpha\delta}(\theta)}-\beta \gamma{F^-_{\beta\gamma}(\theta)}\}
\\&& +  \alpha\gamma {F^-_{\alpha\gamma}(\theta)}- \alpha\beta{F^-_{\alpha\beta}(\theta)} 
+ \beta\delta{F^-_{\beta\delta}(\theta)} - \gamma\delta {F^-_{\gamma\delta}(\theta)}
;\\ \nonumber
\frac{Q^{\mp\pm}_{+,w}}{16 } &=&+\alpha^2 {G^+_{\alpha^2}(\theta)}+ \delta^2 {G^+_{\delta^2}(\theta)}  + \alpha\delta{G^+_{\alpha\delta}(\theta)} + (\gamma-\beta)^2 {G^+_{\beta\gamma}(\theta)}
\mp (\gamma-\beta)\delta {G^+_{\beta\delta}(\theta)}
,\\  \frac{Q^{\mp\pm}_{-,w}}{16 } &=&  \pm \delta^2 {G^-_{\delta^2}(\theta)} \pm  \alpha\delta {G^-_{\alpha\delta}(\theta)} +\alpha(\gamma-\beta){G^-_{\alpha\gamma}(\theta)} +\delta(\beta-\gamma){G^-_{\beta\delta}(\theta)}
. \nonumber \end{eqnarray} 
Substituting $Q^{s's}_{\pm,w}$ into (\ref{eq:43RchiA-CrossSection-via-ScalarProducts-both-CS-main}),
one obtains the incoherent cross sections in the following form:
{\small \begin{eqnarray}\nonumber  
\label{eq:43RchiA-CrossSection-via-ScalarProducts-InCoh-all-viaFG}
\frac{d\sigma^{\mp\mp}_{\text{inc}}}{g_\text{i} \hat c_A d T_A} &=&\frac12\!\! \sum_{f=p,n} A_f  \widehat F_f^2(\bm{q})
\Big[ \dfrac{\Delta A_f}{A_f} \big\{ \alpha\gamma {F^-_{\alpha\gamma}(\theta)}- \alpha\beta{F^-_{\alpha\beta}(\theta)}  + \beta\delta{F^-_{\beta\delta}(\theta)} - \gamma\delta {F^-_{\gamma\delta}(\theta)}\big\}
\\&& \nonumber 
+ \alpha^2 F^+_{\alpha^2}(\theta)+\beta^2 F^+_{\beta^2}(\theta)
+ \gamma^2 F^+_{\gamma^2}(\theta)+ \delta^2 F^+_{\delta^2}(\theta)
+ \alpha\delta F^+_{\alpha\delta}(\theta)+  \beta\gamma F^+_{\beta\gamma}(\theta)
\\&& \nonumber 
  \mp \dfrac{\Delta A_f}{A_f} \{\alpha^2F^-_{\alpha^2}(\theta)
+\beta^2 F^-_{\beta^2}(\theta)+\gamma^2 F^-_{\gamma^2}(\theta)+\delta^2 F^-_{\delta^2}(\theta)
- \alpha\delta{F^-_{\alpha\delta}(\theta)}-\beta \gamma{F^-_{\beta\gamma}(\theta)}\}
\\&& \qquad   \mp\{\alpha\beta F^+_{\alpha\beta}(\theta) +\alpha\gamma F^+_{\alpha\gamma}(\theta)
  + \beta\delta F^+_{\beta\delta}(\theta)+ \gamma \delta F^+_{\gamma\delta}(\theta)\}  \Big]
,\\  \nonumber
\frac{d\sigma^{\mp\pm}_{\text{inc}}}{g_\text{i}\hat c_A d T_A} &=&\frac12\!\! \sum_{f=p,n} A_f \widehat F_f^2(\bm{q}) \Big[\alpha^2 {G^+_{\alpha^2}(\theta)}+ \delta^2 {G^+_{\delta^2}(\theta)}  + \alpha\delta{G^+_{\alpha\delta}(\theta)} + (\gamma-\beta)^2 {G^+_{\beta\gamma}(\theta)}\mp (\gamma-\beta)\delta {G^+_{\beta\delta}(\theta)}
\\&&+\dfrac{ \Delta A_f}{A_f  }\{\pm \delta^2 {G^-_{\delta^2}(\theta)} \pm  \alpha\delta {G^-_{\alpha\delta}(\theta)} +\alpha(\gamma-\beta){G^-_{\alpha\gamma}(\theta)} +\delta(\beta-\gamma){G^-_{\beta\delta}(\theta)}\} \Big]
 \nonumber  .\end{eqnarray}}%
Formulas (\ref{eq:43RchiA-CrossSection-via-ScalarProducts-InCoh-all-viaFG})
are general expressions for the relativistic incoherent cross sections of $\chi A$ scattering due to weak currents.
The nucleon $F$- and $G$-form-factor functions
included in (\ref{eq:43RchiA-CrossSection-via-ScalarProducts-InCoh-all-viaFG}) 
 are obtained using functions from 
(\ref{eq:42RchiA-CrossSection-via-ScalarProducts-f_1--f_17}).
Expressed in terms of the $\chi$-lepton scattering angle and the energy variables
(\ref{eq:ScalarProducts-ChiEta-All-Energies-etc}), they are
{\small \begin{eqnarray} \nonumber
\label{eq:43RchiA-CrossSection-via-ScalarProducts-InCoh-Fs}
\frac{F^-_{\alpha^2}(\theta)}{2}&=&-2\sin^2\frac{\theta}{2}|\bm{k}|^2(|\bm{k}|^2+E_pE_\chi\cos^2\frac{\theta}{2})
,\quad  \frac{F^-_{\delta^2}(\theta)}{2}=-2\sin^2\frac{\theta}{2}E_pE_\chi(E_pE_\chi +|\bm{k}|^2 \cos^2\frac{\theta}{2}) 
,\\ \nonumber
\frac{F^-_{\beta^2}(\theta)}{2}&=&-2\sin^2\frac{\theta}{2}|\bm{k}|^2(E_p^2+E_pE_\chi \cos^2\frac{\theta}{2})  ,\quad \nonumber
\frac{F^-_{\gamma^2}(\theta)}{2}=-2\sin^2\frac{\theta}{2}|\bm{k}|^2(E_\chi^2+E_p E_\chi\cos^2\frac{\theta}{2})
;\\  \frac{F^+_{\alpha^2}(\theta)}{2}&=&
|\bm{k}|^4 (1+\sin^4\frac{\theta}{2})+\cos^2\frac{\theta}{2}[|\bm{k}|^2(E_\chi^2\cos^2\frac{\theta}{2}+ 2 E_pE_\chi  )+ m^2E_\chi^2]
,\\  \nonumber \frac{F^+_{\delta^2}(\theta)}{2}&=&
E_\chi^2 E_p^2 (1+\sin^4\frac{\theta}{2})+\cos^2\frac{\theta}{2}[ |\bm{k}|^2 (|\bm{k}|^2 \cos^2\frac{\theta}{2} +2 E_\chi E_p) + m^2 E_\chi^2 \sin^2\frac{\theta}{2}]
,\\  \nonumber. \frac{F^+_{\beta^2}(\theta)}{2 |\bm{k}|^2}&=&
(E_p+E_\chi \cos^2\frac{\theta}{2})^2+ \sin^2\frac{\theta}{2}[m^2 \cos^2\frac{\theta}{2}+E_p^2(\sin^2\frac{\theta}{2}-\cos^2\frac{\theta}{2}) ]
,\\  \nonumber \frac{F^+_{\gamma^2}(\theta)}{2 |\bm{k}|^2}&=&
(E_p\cos^2\frac{\theta}{2} +E_\chi)^2+\sin^2\frac{\theta}{2} [m^2 \cos^2\frac{\theta}{2}
+ E_\chi^2(\sin^2\frac{\theta}{2}-\cos^2\frac{\theta}{2})]
.\end{eqnarray}}
\label{page:43RchiA-CrossSection-via-ScalarProducts-InCoh-Fs}
{\small \begin{eqnarray*}
\frac{F^+_{\alpha\beta}(\theta)}{4}&=&2 \sin^2\frac{\theta}{2} |\bm{k}|^3 (E_p  + E_\chi  \cos^2\frac{\theta}{2})
,\quad \frac{F^+_{\alpha\delta}(\theta)}{4}=\sin^2\frac{\theta}{2}|\bm{k}|^2[(E_\chi^2+|\bm{k}|^2)\cos^2\frac{\theta}{2}+2E_\chi E_p]
,\\   \frac{F^+_{\gamma\delta}(\theta)}{4}&=&2\sin^2\frac{\theta}{2}E_\chi |\bm{k}|(E_\chi E_p +|\bm{k}|^2 \cos^2\frac{\theta}{2})
,\quad   \frac{F^+_{\beta\gamma}(\theta)}{4}=\sin^2\frac{\theta}{2}|\bm{k}|^2[(E_\chi^2 +|\bm{k}|^2)\cos^2\frac{\theta}{2}  +2E_\chi E_p]
,\\  \frac{F^-_{\alpha\delta}(\theta)}{4}&=&2E_\chi  E_p |\bm{k}|^2 \sin^4\frac{\theta}{2}
+(E_\chi E_p + |\bm{k}|^2)^2  \cos^2\frac{\theta}{2}
,\\  \frac{F^-_{\beta\gamma}(\theta)}{4 }&=&2E_\chi E_p |\bm{k}|^2 \sin^4\frac{\theta}{2}+ 
 |\bm{k}|^2(E_\chi+E_p)^2\cos^2\frac{\theta}{2} 
,\\  \frac{ F^-_{\beta\delta}(\theta)}{4}&=& - \sin^2\frac{\theta}{2}  |\bm{k}| [E_p(E_\chi^2 + |\bm{k}|^2)\cos^2\frac{\theta}{2}+2 E_\chi E_p^2] 
,\\  \frac{F^-_{\alpha\gamma}(\theta)}{4 }&=&- \sin^2\frac{\theta}{2}|\bm{k}|[E_p  (E_\chi^2  + |\bm{k}|^2 ) \cos^2\frac{\theta}{2}+2|\bm{k}|^2  E_\chi] ,
\end{eqnarray*}
\begin{eqnarray*}
\frac{F^-_{\alpha\beta}(\theta)}{4 |\bm{k}| }&=&
  |\bm{k}|^2E_p  (1+\sin^4\frac{\theta}{2}) +\cos^2\frac{\theta}{2} [E_\chi^2E_p\cos^2\frac{\theta}{2}+ E_\chi(|\bm{k}|^2+E_p^2)]
,\\ \frac{F^-_{\gamma\delta}(\theta)}{4|\bm{k}|} &=&E_\chi^2E_p (1+\sin^4\frac{\theta}{2})
+\cos^2\frac{\theta}{2} [ |\bm{k}|^2(E_p\cos^2\frac{\theta}{2}  +E_\chi) +  E_\chi E_p^2]
,\\ \frac{F^+_{\alpha\gamma}(\theta)}{4 |\bm{k}|}&=&
(E_\chi^2   + E_\chi  E_p  +|\bm{k}|^2)E_p \cos^2\frac{\theta}{2}+ |\bm{k}|^2 E_\chi (2\sin^4\frac{\theta}{2}+ \cos^2\frac{\theta}{2})
,\\ \frac{F^+_{\beta\delta}(\theta)}{4 |\bm{k}|}&=&
(E_\chi^2  +E_\chi E_p  + |\bm{k}|^2) E_p  \cos^2\frac{\theta}{2}+ |\bm{k}|^2 E_\chi (2\sin^4\frac{\theta}{2}+ \cos^2\frac{\theta}{2}) +2 m^2 E_\chi \sin^2\frac{\theta}{2}
.\end{eqnarray*}} 
{\small \begin{eqnarray}\nonumber
\label{eq:43RchiA-CrossSection-via-ScalarProducts-InCoh-Gs}
\frac{G^-_{\alpha\delta}(\theta)}{2 m^2_\chi }&=& - 2  m^2 \sin^2\frac{\theta}{2} 
,\qquad  \frac{G^-_{\delta^2}(\theta)}{2 m^2_\chi}= - 2 m^2 \cos^2\frac{\theta}{2}
,\qquad  \frac{G^+_{\beta\gamma}(\theta)}{2 m^2_\chi}=\lambda^2_+ \lambda^2_-  \cos^4\frac{\theta}{2}
=|\bm{k}|^2 \cos^4\frac{\theta}{2}
,\\ \nonumber \frac{G^+_{\alpha^2}(\theta)}{2 m^2_\chi}&=&
 (m^2+|\bm{k}|^2\cos^2\frac{\theta}{2})\sin^2\frac{\theta}{2}
,\qquad \frac{G^+_{\delta^2}(\theta)}{2 m_\chi^2}= 
m^2(1+ \cos^2\frac{\theta}{2}) +|\bm{k}|^2  \sin^2\frac{\theta}{2}\cos^2\frac{\theta}{2}
,\\  \nonumber \frac{G^+_{\beta\delta}(\theta)}{2 m^2_\chi} &=& \frac{G^+_{\gamma\delta}(\theta)}{2 m^2_\chi}=
2 m |\bm{k}| \cos^2\frac{\theta}{2} (\sin^2\frac{\theta}{2}-\cos^2\frac{\theta}{2})
,\qquad   \frac{G^+_{\alpha\delta}(\theta)}{2 m_\chi^2}=   - 2|\bm{k}|^2 \sin^2\frac{\theta}{2}\cos^2\frac{\theta}{2}
,\\ \frac{G^-_{\alpha\gamma}(\theta)}{2 m_\chi ^2}&=&
2 |\bm{k}|\cos^2\frac{\theta}{2}[m + 2(E_p-m) \cos^2\frac{\theta}{2}] \sin^2\frac{\theta}{2}
,\\   \frac{G^-_{\beta\delta}(\theta)}{2 m^2_\chi}&=&
{2 |\bm{k}|\cos^2\frac{\theta}{2}[m + 2(E_p-m)\sin^2\frac{\theta}{2}]\cos^2\frac{\theta}{2}}
\nonumber .\end{eqnarray}}%
All $G$-form-factor functions (corresponding to the lepton spin flip)
are proportional to the square of the lepton mass and disappear at $m_\chi=0$.
Given  expressions (\ref{eq:43RchiA-CrossSection-via-ScalarProducts-InCoh-Gs}), the second formula from
(\ref{eq:43RchiA-CrossSection-via-ScalarProducts-InCoh-all-viaFG})
gives incoherent $\chi A$ cross sections {\em with lepton spin flip}\/ in the general form
\par 
{\small \begin{eqnarray}\nonumber
\label{eq:43RchiA-CrossSection-via-ScalarProducts-InCohSC-spinflip}
\frac{d\sigma^{\mp\pm}_{\text{inc}}}{g_\text{i} d T_A}&=&
\dfrac{G^2_F  m_A}{4 \pi }\dfrac{m^2_\chi}{|\bm{k}_\chi|^2}
\sum_{f=p,n}\!\! A_f \widehat F_f^2(\bm{q})\Big[
 (\alpha^2 \sin^2\frac{\theta}{2}+ \delta^2 (1+ \cos^2\frac{\theta}{2}\sin^2\frac{\theta}{2})) 
+(\alpha- \delta)^2\frac{|\bm{k}|^2}{m^2}\cos^2\frac{\theta}{2} \sin^2\frac{\theta}{2}
\\&+& \nonumber
\Big[ (\gamma-\beta)\frac{|\bm{k}|}{m}\pm\delta\Big]^2 \cos^4\frac{\theta}{2}
 \mp 2 (\gamma-\beta)  \delta \frac{|\bm{k}|}{m}\cos^2\frac{\theta}{2}\sin^2\frac{\theta}{2}
+ \dfrac{2\Delta A_f}{A_f}\big\{\mp(\delta^2\cos^2\frac{\theta}{2} +\alpha\delta\sin^2\frac{\theta}{2})
\\&+& (\gamma-\beta)\frac{|\bm{k}|}{m}[(\alpha\sin^2\frac{\theta}{2} -\delta\cos^2\frac{\theta}{2})\cos^2\frac{\theta}{2}+2(\alpha-\delta)\Big(\frac{E_p}{m}-1\Big)\cos^4\frac{\theta}{2}\sin^2\frac{\theta}{2}] \big\}\Big]
.\end{eqnarray}}%
A similar "direct"\/ expansion of the first formula from (\ref{eq:43RchiA-CrossSection-via-ScalarProducts-InCoh-all-viaFG})
takes up too much space due to more complex $F$-form factors
(\ref{eq:43RchiA-CrossSection-via-ScalarProducts-InCoh-Fs}).
Nevertheless, by introducing generalized incoherent nucleus-nucleon form-factor combinations that accumulate dependence on weak nucleon coupling constants in the form
\begin{eqnarray}
\label{eq:43RchiA-CrossSection-via-ScalarProducts-InCohSC-Phi-definitions}
\Phi^+_{ab}(\bm{q}, A)&=& \sum_{f=p,n} \widehat F_f^2(\bm{q})   A_f  a_f b_f 
\quad\text{and}\quad \Phi^-_{ab}(\bm{q}, A)= \sum_{f=p,n} \widehat F_f^2(\bm{q})  \Delta  A_f a_f b_f
,\end{eqnarray}
where $a$ and $b$ independently "run through"\/ values of $\alpha,\beta,\gamma,\delta$,
one can obtain the {\em most compact}\/ decomposition of the incoherent $\chi A$ cross section
without $\chi$-lepton spin flip in 20 nucleon $F$-form factors corresponding 
to all possible combinations of weak coupling constants.
This decomposition looks like
{\small \begin{eqnarray} 
\label{eq:43RchiA-CrossSections-via-ScalarProducts-ChiEta-rel-InCohCS-viaFG-viaPhi}
\frac{d\sigma^{\mp\mp}_{\text{inc}}}{8 g_\text{i} c_A d T_A} &=& 
  \Phi^+_{\alpha^2}(\bm{q})  F^+_{\alpha^2}(\theta)+\Phi^+_{\beta^2}(\bm{q})  F^+_{\beta^2}(\theta)
+\Phi^+_{\gamma^2}(\bm{q}) F^+_{\gamma^2}(\theta)+\Phi^+_{\delta^2}(\bm{q}) F^+_{\delta^2}(\theta)
+ \Phi^+_{\alpha\delta}(\bm{q}) F^+_{\alpha\delta}(\theta)\qquad\qquad 
\\\nonumber  &&+ \Phi^+_{ \beta\gamma}(\bm{q})  F^+_{\beta\gamma}(\theta)
\mp [\Phi^+_{\alpha\beta}(\bm{q}) F^+_{\alpha\beta}(\theta)+\Phi^+_{\alpha\gamma}(\bm{q}) F^+_{\alpha\gamma}(\theta)+\Phi^+_{\beta\delta}(\bm{q}) F^+_{\beta\delta}(\theta)+\Phi^+_{ \gamma\delta}(\bm{q})F^+_{\gamma\delta}(\theta)] 
\\\nonumber  &+& \Phi^-_{\alpha\gamma}(\bm{q}){F^-_{\alpha\gamma}(\theta)}- \Phi^-_{\alpha\beta}(\bm{q}){F^-_{\alpha\beta}(\theta)}  + \Phi^-_{\beta\delta}(\bm{q})  {F^-_{\beta\delta}(\theta)} - \Phi^-_{\gamma\delta}(\bm{q}) {F^-_{\gamma\delta}(\theta)}
\pm \Phi^-_{\beta\gamma}(\bm{q}) {F^-_{\beta\gamma}(\theta)}
\\&& \nonumber   \mp  \Phi^-_{\alpha^2}(\bm{q}) F^-_{\alpha^2}(\theta)\mp \Phi^-_{\beta^2}(\bm{q}) F^-_{\beta^2}(\theta)  \mp \Phi^-_{\gamma^2}(\bm{q}) F^-_{\gamma^2}(\theta) \mp \Phi^-_{\delta^2}(\bm{q}) F^-_{\delta^2}(\theta)\pm \Phi^-_{\alpha\delta}(\bm{q}) {F^-_{\alpha\delta}(\theta)}
.\end{eqnarray}}%
\par By substituting explicit expressions for the $F$-form factors into this formula and performing simple but somewhat tedious transformations, one can obtain the expansion of the {\em incoherent} $\chi A$ cross section {\em without spin-flip} of the $\chi$-lepton in terms of kinematic structures $|\bm{k}|^2, |\bm{k}|^4, 2 E_pE_\chi$, etc. in the following form:\par 
{\small
 \begin{eqnarray}\nonumber
\label{eq:43RchiA-CrossSections-via-ScalarProducts-ChiEta-rel-InCohCS-viaPhi+Lorentz}
\frac{d\sigma^{\mp\mp}_{\text{inc}}}{g_\text{i}\hat  c_A d T_A}&=&
m^2m_\chi^2[\cos^2\frac{\theta}{2}(\Phi^+_{\alpha^2+\delta^2}(\bm{q})\pm\Phi^-_{2\alpha\delta}(\bm{q}))
+\sin^2\frac{\theta}{2}(\Phi^+_{2\delta^2}(\bm{q})\pm\Phi^-_{2\delta^2}(\bm{q})) ]
\\&+&\nonumber
m^2|\bm{k}|^2 [\cos^2\frac{\theta}{2}(\Phi^+_{\alpha^2+\beta^2+\gamma^2+\delta^2}(\bm{q})
\pm\Phi^-_{2(\alpha\delta+\beta\gamma)}(\bm{q}))
+\sin^2\frac{\theta}{2}(\Phi^+_{2(\beta^2+\delta^2)}(\bm{q})\pm\Phi^-_{2(\beta^2+\delta^2)}(\bm{q}))]
\\&+&\nonumber
m_\chi^2|\bm{k}|^2[ \cos^2\frac{\theta}{2}(\Phi^+_{\alpha^2+\beta^2+\gamma^2+\delta^2}(\bm{q})
\pm \Phi^-_{2(\alpha\delta+\beta\gamma)}(\bm{q}))
+\sin^2\frac{\theta}{2}(\Phi^+_{2(\gamma^2+\delta^2)}(\bm{q}) \pm \Phi^-_{2(\gamma^2+\delta^2)}(\bm{q}))
\\&&\nonumber - \sin^2\frac{\theta}{2}\cos^2\frac{\theta}{2}\Phi^+_{(\alpha-\delta)^2+(\beta-\gamma)^2}(\bm{q})]
\\&\mp&\nonumber m^2_\chi|\bm{k}|E_p 
[\cos^2\frac{\theta}{2}(\Phi^+_{2(\alpha\gamma+\beta\delta)}(\bm{q})\pm \Phi^-_{2(\alpha\beta+\gamma\delta)}(\bm{q})) +\sin^2\frac{\theta}{2}(\Phi^+_{4\gamma\delta}(\bm{q})\pm \Phi^-_{4\gamma\delta}(\bm{q}))
\\&& \pm \sin^2\frac{\theta}{2}\cos^2\frac{\theta}{2}\Phi^-_{2(\alpha-\delta)(\gamma-\beta)}(\bm{q})]
\\&\mp& \nonumber
m^2 |\bm{k}| E_\chi [\cos^2\frac{\theta}{2}(\Phi^+_{2(\alpha\gamma+\beta\delta)}(\bm{q}) 
\pm \Phi^-_{2(\alpha\beta+\gamma\delta)}(\bm{q})) +\sin^2\frac{\theta}{2}(\Phi^+_{4\beta\delta}(\bm{q})
\pm \Phi^-_{4\beta\delta}(\bm{q}))]
\\&+&\nonumber
2E_pE_\chi|\bm{k}|^2[\Phi^+_{\alpha^2+\beta^2+\gamma^2+\delta^2}(\bm{q})\pm\Phi^-_{2(\alpha\delta+\beta\gamma)}(\bm{q}) 
\\&& \nonumber
-\sin^2\frac{\theta}{2}(\Phi^+_{(\alpha-\delta)^2+(\beta-\gamma)^2}(\bm{q}) 
\mp \cos^2\frac{\theta}{2} \Phi^-_{(\alpha-\delta)^2+(\beta-\gamma)^2}(\bm{q}))]
\\&+&\nonumber
2|\bm{k}|^4 [\Phi^+_{\alpha^2+\beta^2+\gamma^2+\delta^2}(\bm{q})\pm \Phi^-_{\alpha^2+\beta^2+\gamma^2+\delta^2}(\bm{q}) 
\\&& \nonumber
-\cos^2\frac{\theta}{2}(\sin^2\frac{\theta}{2}\Phi^+_{(\alpha-\delta)^2+(\beta-\gamma)^2}(\bm{q}) 
 \pm \Phi^-_{(\alpha-\delta)^2+(\beta-\gamma)^2}(\bm{q}))]
\\&\mp&\nonumber
 2 |\bm{k}|^3E_p [\Phi^+_{2(\alpha\gamma+\beta\delta)}(\bm{q})\pm\Phi^-_{2(\alpha\beta+\gamma\delta)}(\bm{q}) +\sin^2\frac{\theta}{2}(\Phi^+_{2(\alpha-\delta)(\beta-\gamma)}(\bm{q}) 
\mp \cos^2\frac{\theta}{2}\Phi^-_{2(\alpha-\delta)(\beta-\gamma)}(\bm{q})) ]
\\&\mp& \nonumber
2|\bm{k}|^3E_\chi [\Phi^+_{2(\alpha\gamma+\beta\delta)}(\bm{q})\pm\Phi^-_{2(\alpha\beta+\gamma\delta)}(\bm{q}) +\sin^2\frac{\theta}{2}(\cos^2\frac{\theta}{2} \Phi^+_{2(\alpha-\delta)(\beta-\gamma)}(\bm{q})
\mp \Phi^-_{2(\alpha-\delta)(\beta-\gamma)}(\bm{q}))]
\nonumber.\end{eqnarray}\par }  
\noindent
To shorten the notations,  sums of the following type from expressions (\ref{eq:43RchiA-CrossSection-via-ScalarProducts-InCohSC-Phi-definitions}) 
are introduced  in (\ref{eq:43RchiA-CrossSections-via-ScalarProducts-ChiEta-rel-InCohCS-viaPhi+Lorentz})
\begin{eqnarray*}
\Phi^+_{\alpha^2\pm \delta^2}(\bm{q})&=&\Phi^+_{\alpha^2}(\bm{q}, A)\pm \Phi^+_{\delta^2}(\bm{q}, A)
=\sum_{f=p,n} \widehat F_f^2(\bm{q})   A_f  (\alpha^2_f\pm \delta^2_f)
,\\\Phi^-_{\alpha\delta\pm \beta\gamma}(\bm{q})&=&\Phi^-_{\alpha\delta}(\bm{q},A)\pm \Phi^-_{\beta\gamma}(\bm{q},A) = \sum_{f=p,n} \widehat F_f^2(\bm{q})  \Delta A_f  (\alpha_f\delta_f\pm \beta_f\gamma_f)
.\end{eqnarray*}
Formulas (\ref{eq:43RchiA-CrossSections-via-ScalarProducts-ChiEta-rel-InCohCS-viaFG-viaPhi}) and
(\ref{eq:43RchiA-CrossSections-via-ScalarProducts-ChiEta-rel-InCohCS-viaPhi+Lorentz}) give 
two different representations of the incoherent cross section of $\chi A$ scattering {\em without lepton spin flip}.
For a spinless nucleus ($\Delta A_f=0$ and all $\Phi^- =0$) formula
(\ref{eq:43RchiA-CrossSections-via-ScalarProducts-ChiEta-rel-InCohCS-viaPhi+Lorentz}) takes the form
\par
{\small \begin{eqnarray}\nonumber
\label{eq:43RchiA-CrossSections-via-ScalarProducts-ChiEta-rel-InCohCS-viaPhi+Lorentz-spinless}
\frac{d\sigma^{\mp\mp}_{\!\text{inc},0}}{g_\text{i}\hat  c_A d T_A}\!\!\!&=&\!\!
m^2m_\chi^2 [\cos^2\frac{\theta}{2}\Phi^+_{\alpha^2+\delta^2}(\bm{q})+2\sin^2\frac{\theta}{2}\Phi^+_{\delta^2}(\bm{q})] +m^2|\bm{k}|^2 [\cos^2\frac{\theta}{2}\Phi^+_{\alpha^2+\beta^2+\gamma^2+\delta^2}(\bm{q})
+2\sin^2\frac{\theta}{2}\Phi^+_{\beta^2+\delta^2}(\bm{q})] 
\\&&\nonumber  +m_\chi^2|\bm{k}|^2[
\cos^2\frac{\theta}{2}\Phi^+_{\alpha^2+\beta^2+\gamma^2+\delta^2}(\bm{q})
+2\sin^2\frac{\theta}{2}\Phi^+_{\gamma^2+\delta^2}(\bm{q}) 
- \sin^2\frac{\theta}{2}\cos^2\frac{\theta}{2} \Phi^+_{(\alpha-\delta)^2+(\beta-\gamma)^2}(\bm{q})] 
\\&& +2E_pE_\chi|\bm{k}|^2[\cos^2\frac{\theta}{2}\Phi^+_{\alpha^2+\beta^2+\gamma^2+\delta^2}(\bm{q}) 
+2\sin^2\frac{\theta}{2}\Phi^+_{\alpha\delta+\beta\gamma}(\bm{q}) ]
\\&&\nonumber  +2|\bm{k}|^4 [\Phi^+_{\alpha^2+\beta^2+\gamma^2+\delta^2}(\bm{q})-\sin^2\frac{\theta}{2}\cos^2\frac{\theta}{2}  \Phi^+_{(\alpha-\delta)^2+(\beta-\gamma)^2}(\bm{q}) ]
\\&&\nonumber \mp 2m^2_\chi|\bm{k}|E_p [\cos^2\frac{\theta}{2}\Phi^+_{\alpha\gamma+\beta\delta}(\bm{q}) +\sin^2\frac{\theta}{2}\Phi^+_{2\gamma\delta}(\bm{q}) ] \mp 2m^2 |\bm{k}| E_\chi[\cos^2\frac{\theta}{2}\Phi^+_{\alpha\gamma+\beta\delta}(\bm{q}) 
+\sin^2\frac{\theta}{2}\Phi^+_{2\beta\delta}(\bm{q})]
\\&&\nonumber  \mp 4|\bm{k}|^3E_\chi [\Phi^+_{\alpha\gamma+\beta\delta}(\bm{q})
+\Phi^+_{(\alpha-\delta)(\beta-\gamma)}(\bm{q})\sin^2\frac{\theta}{2} \cos^2\frac{\theta}{2}]
\\&&\nonumber   \mp  4|\bm{k}|^3E_p [\Phi^+_{\alpha\gamma+\beta\delta}(\bm{q})
+\Phi^+_{(\alpha-\delta)(\beta-\gamma)}(\bm{q})\sin^2\frac{\theta}{2} ]
.\end{eqnarray}}%
From formulas 
(\ref{eq:43RchiA-CrossSections-via-ScalarProducts-ChiEta-rel-InCohCS-viaPhi+Lorentz-spinless})
and (\ref{eq:43RchiA-CrossSections-via-ScalarProducts-ChiEta-rel-InCohCS-viaPhi+Lorentz})
one can get "check limiting cases"\/.
First, when $m_\chi^2\to 0$ (and $E_\chi \to |\bm{k}|$), formula
(\ref{eq:43RchiA-CrossSections-via-ScalarProducts-ChiEta-rel-InCohCS-viaPhi+Lorentz-spinless}) gets the form
\begin{eqnarray*}
\frac{d\sigma^{\mp\mp}_{\text{inc},0}}{|\bm{k}|^2 g_\text{i}\hat c_A d T_A}&=&s\Phi^+_{(\alpha\mp\gamma)^2+(\beta\mp\delta)^2}(\bm{q}) -m^2 \sin^2\frac{\theta}{2}\Phi^+_{(\alpha\mp\gamma)^2-(\beta\mp \delta)^2}(\bm{q})
\\&& \nonumber
-2|\bm{k}| \sin^2\frac{\theta}{2}\Phi^+_{((\alpha-\delta)\pm (\beta-\gamma))^2}(\bm{q}) (E_p  
+ |\bm{k}| \cos^2\frac{\theta}{2})
, \end{eqnarray*}
from which, when $m_\chi^2 =0$ and $\Delta A_f=0$,  
for the incoherent $\chi A$ cross section without spin flip one has the following "explicit expression"\/:
{\small \begin{eqnarray}\nonumber
\label{eq:43RchiA-CrossSections-via-ScalarProducts-ChiEta-rel-InCohCS-viaPhi+Lorentz-spinless-mchi=0}
  \frac{d\sigma^{\mp\mp}_{\text{inc},00}}{d T_A} &=&  \dfrac{G^2_F  m_A}{4 \pi}\dfrac{|\bm{k}|^{2}}{|\bm{k}_\chi|^2}\sum_{f=p,n} \widehat F_f^2(\bm{q})   A_f  \Big[\frac{s}{m^2} [(\alpha\mp\gamma)^2+(\beta\mp\delta)^2]-\sin^2\frac{\theta}{2}[(\alpha\mp\gamma)^2-(\beta\mp \delta)^2]\\&&  -2\frac{|\bm{k}|}{m} \sin^2\frac{\theta}{2}((\alpha\mp\gamma)\pm(\beta\mp\delta))^2 
(\frac{E_p}{m}  + \frac{|\bm{k}|}{m} \cos^2\frac{\theta}{2}) \Big]
.\end{eqnarray}}%
This expression will be used for cross-check.
Next, when $ |\bm{k}|\to 0$, from (\ref{eq:43RchiA-CrossSections-via-ScalarProducts-ChiEta-rel-InCohCS-viaPhi+Lorentz-spinless})
one gets a simple expression
\begin{eqnarray*}
\frac{d\sigma^{\mp\mp}_{\text{inc}}}{d T_A} &=& \dfrac{G^2_F  m_A}{4 \pi }\dfrac{ m_\chi^2 }{|\bm{k}_\chi|^2 }
\big[\Phi^+_{\alpha^2}(\bm{q})\cos^2\frac{\theta}{2}+ \Phi^+_{\delta^2}(\bm{q})(1+\sin^2\frac{\theta}{2})\big]
,\end{eqnarray*}
which coincides with that obtained earlier in the nonrelativistic limit \cite{Bednyakov:2022dmc}.
\par 
In terms of $\Phi$-form-factor functions from (\ref{eq:43RchiA-CrossSection-via-ScalarProducts-InCohSC-Phi-definitions}),
expression (\ref{eq:43RchiA-CrossSection-via-ScalarProducts-InCohSC-spinflip}) for
the incoherent $\chi A$ scattering {\em with lepton spin flip}\/ takes the form
{\small \begin{eqnarray*}
\frac{d\sigma^{\mp\pm}_{\text{inc}}}{8 g_\text{i} c_A d T_A}&=&\Phi^+_{\alpha^2}(\bm{q}) G^+_{\alpha^2}(\theta)
+\Phi^+_{\delta^2}(\bm{q}) G^+_{\delta^2}(\theta)+\Phi^+_{\alpha\delta}(\bm{q}) G^+_{\alpha\delta}(\theta)+\Phi^+_{(\gamma-\beta)^2}(\bm{q}) G^+_{\beta\gamma}(\theta) \mp\Phi^+_{(\gamma-\beta)\delta}(\bm{q})G^+_{\beta\delta}(\theta) \\&&\nonumber \pm \Phi^-_{\delta^2}(\bm{q}) G^-_{\delta^2}(\theta) \pm \Phi^-_{\alpha\delta}(\bm{q}) G^-_{\alpha\delta}(\theta)+\Phi^-_{\alpha(\gamma-\beta)}(\bm{q}) G^-_{\alpha\gamma}(\theta)+\Phi^-_{\delta(\beta-\gamma)}(\bm{q}) G^-_{\beta\delta}(\theta)
.\end{eqnarray*}}
After "expanding"\/ $G$-form-factors, this expression becomes the following:
{\small\begin{eqnarray}\nonumber
\label{eq:43RchiA-CrossSection-via-ScalarProducts-InCohSC-spinflip-viaPhi}
\frac{d\sigma^{\mp\pm}_{\text{inc}}}{g_\text{i}d T_A} &=&\dfrac{G^2_F  m_A}{4 \pi}\dfrac{m^2_\chi }{|\bm{k}_\chi|^2}
\Big\{ \Phi^+_{\alpha^2+\delta^2}(\bm{q})+\cos^2\frac{\theta}{2}\Phi^+_{\delta^2-\alpha^2}(\bm{q})
\\&& \nonumber +\frac{|\bm{k}|}{m}\cos^2\frac{\theta}{2}  
[\frac{|\bm{k}|}{m}|(\sin^2\frac{\theta}{2} \Phi^+_{(\alpha-\delta)^2}(\bm{q})+\cos^2\frac{\theta}{2} \Phi^+_{(\gamma-\beta)^2}(\bm{q})) \pm 2\Phi^+_{(\gamma-\beta)\delta}(\bm{q})(\cos^2\frac{\theta}{2}-\sin^2\frac{\theta}{2})]
\\&& +2\frac{|\bm{k}|}{m}\cos^2\frac{\theta}{2}[ \Phi^-_{(\alpha+\delta)(\beta-\gamma)}(\bm{q})\cos^2\frac{\theta}{2}-\Phi^-_{\alpha(\beta-\gamma)}(\bm{q})] 
\\&& \nonumber
 +4\frac{(E_p-m)}{m}\frac{|\bm{k}|}{m}\cos^4\frac{\theta}{2}\sin^2\frac{\theta}{2}\Phi^-_{(\delta-\alpha)(\beta-\gamma)}(\bm{q})  \mp 2 [\Phi^-_{\alpha\delta}(\bm{q})+\Phi^-_{\delta^2-\alpha\delta}(\bm{q})\cos^2\frac{\theta}{2}] \Big\}
.\end{eqnarray}}%
Fully averaged over the initial and summed over the final
$\chi$-particle helicity, the {\em averaged incoherent} weak $\chi A$ interaction cross section has the general form
\begin{eqnarray}
\label{eq:43RchiA-CrossSections-via-ScalarProducts-ChiEta-rel-InCohCS-total}
\frac{d\sigma^\text{total}_{\text{inc}}}{g_\text{i} c_A d T_A} &\equiv& \dfrac12\sum^{}_{s's} \frac{d\sigma^{s's}_\text{inc}}{ g_\text{i} c_A d T_A}  = \frac12 \sum_{f=p,n} \frac{A_f }{2}  \widehat F^2_f(\bm{q}) 16
\Big[\sum^{}_{s's}\frac{Q^{s's}_{+,w}}{16}+\dfrac{ \Delta A_f}{A_f} \sum^{}_{s's}\frac{Q^{s's}_{-,w}}{16}\Big].
\quad \end{eqnarray}
It can be calculated "directly"\/ by taking the sum of all individual incoherent cross sections from
(\ref{eq:43RchiA-CrossSection-via-ScalarProducts-InCoh-all-viaFG})
over all spin projections $s, s'$ of the initial and final $\chi$ lepton.
For example, when $m_\chi =0$ and $\Delta A_f=0$,  the total $\chi A$ cross section
(\ref{eq:43RchiA-CrossSections-via-ScalarProducts-ChiEta-rel-InCohCS-total})
is the half-sum of only two expressions from
(\ref{eq:43RchiA-CrossSections-via-ScalarProducts-ChiEta-rel-InCohCS-viaPhi+Lorentz-spinless-mchi=0}):
{\small
\begin{eqnarray}
\label{eq:43RchiA-CrossSections-via-ScalarProducts-ChiEta-rel-InCohCS-total-spinless-mchi=0--++in}
&&\!\!\!\!\!\!\!\! \!\!\!\!\! \dfrac12\Big\{\dfrac{d\sigma^{--}_{\!\text{inc},00}}{d T_A}\!+\!\dfrac{d\sigma^{++}_{\!\text{inc},00}}{d T_A}\Big\} 
=\dfrac{G^2_F  m_A}{8 \pi}\dfrac{|\bm{k}|^{2}}{|\bm{k}_\chi|^2}\sum_{f=p,n} \widehat F_f^2(\bm{q}) A_f    \Big[
\\\!\!&+&\!\!\!\!\nonumber
\frac{s}{m^2} [(\alpha\!-\!\gamma)^2+(\beta\!-\!\delta)^2]-\sin^2\frac{\theta}{2}[(\alpha\!-\!\gamma)^2-(\beta\!-\!\delta)^2] -\frac{2|\bm{k}|}{m} \sin^2\frac{\theta}{2}(\alpha\!-\!\gamma\!+\!\beta\!-\!\delta)^2(\frac{E_p}{m}  + \frac{|\bm{k}|}{m} \cos^2\frac{\theta}{2})
\\\!\!&+&\!\!\!\!\nonumber
\frac{s}{m^2} [(\alpha\!+\!\gamma)^2+(\beta\!+\!\delta)^2]-\sin^2\frac{\theta}{2}[(\alpha\!+\!\gamma)^2-(\beta\!+\!\delta)^2]-\frac{2|\bm{k}|}{m} \sin^2\frac{\theta}{2}(\alpha\!+\!\gamma\!-\!\beta\!-\!\delta)^2 (\frac{E_p}{m}  + \frac{|\bm{k}|}{m} \cos^2\frac{\theta}{2})\Big]
.\end{eqnarray}}%
The second line here corresponds to the negative helicity of the massless $\chi$ lepton and completely vanishes 
for the $V\!+\!A$-current.
The bottom line corresponds to the positive helicity of the massless $\chi$ lepton and vanishes for $V\!-\!A$ current
\footnote{For $V\!\mp\!A$ currents one has $(\alpha\mp\gamma)^{}_{V\!\mp\!A}= 2 g_V, (\delta \mp \beta)^{}_{V\!\mp\!A} = \pm 2g_A, (\alpha\mp\gamma)^{}_{V\!\pm\!A} = (\delta\mp\beta)^{}_{V\!\pm \!A} = 0
.$}. 
This "property"\/ of expression (\ref{eq:43RchiA-CrossSections-via-ScalarProducts-ChiEta-rel-InCohCS-viaPhi+Lorentz-spinless-mchi=0}) speaks in favor of its correctness.
After a slight transformation, the formula
(\ref{eq:43RchiA-CrossSections-via-ScalarProducts-ChiEta-rel-InCohCS-total-spinless-mchi=0--++in})
takes the form (recall that $m_\chi =0$ and $\Delta A_f=0$)
{\small \begin{eqnarray}\nonumber
\label{eq:43RchiA-CrossSections-via-ScalarProducts-ChiEta-rel-InCohCS-total-spinless-mchi=0--++}
\frac{d\sigma^\text{total}_{\text{inc},00}}{g_\text{i}d T_A} &=&
\dfrac{G^2_F  m_A}{4\pi}\dfrac{|\bm{k}|^{2}}{|\bm{k}_\chi|^2}\sum_{f=p,n} \widehat F_f^2(\bm{q}) A_f 
\Big[\frac{s}{m^2}(\alpha^2+\delta^2+\beta^2+\gamma^2)- \sin^2\frac{\theta}{2}(\alpha^2-\delta^2+\gamma^2-\beta^2)
\\&& -\frac{2|\bm{k}|}{m} \sin^2\frac{\theta}{2}\Big(\frac{E_p}{m}  + \frac{|\bm{k}|}{m} \cos^2\frac{\theta}{2}\Big)[(\alpha-\delta)^2+(\beta-\gamma)^2] \Big]
.\end{eqnarray}}%
\par
Another way to find the total cross section (\ref{eq:43RchiA-CrossSections-via-ScalarProducts-ChiEta-rel-InCohCS-total})
is to use the results of calculating the sums $\sum Q^{s's}_{\pm,w}$ from the right side of
formula (\ref{eq:43RchiA-CrossSections-via-ScalarProducts-ChiEta-rel-InCohCS-total}), which have the form
\begin{eqnarray}\nonumber
\label{eq:43RchiA-CrossSections-via-ScalarProducts-ChiEta-rel-InCohCS-total-sums}
\sum^{}_{s's}\frac{Q^{s's}_{+,w}}{2^5}&=&
\alpha^2 S^+_{\alpha^2}(\theta)+\beta^2 S^+_{\beta^2}(\theta)+ \gamma^2 S^+_{\gamma^2}(\theta)
+ \delta^2 S^+_{\delta^2}(\theta)+ \alpha\delta S^+_{\alpha\delta}(\theta)+  \beta\gamma S^+_{\beta\gamma}(\theta)
,\\ \sum^{}_{s's}\frac{Q^{s's}_{-,w}}{2^5}&=&
\alpha\gamma S^-_{\alpha\gamma}(\theta) + \beta\delta S^-_{\beta\delta}(\theta) 
- \alpha\beta S^-_{\alpha\beta}(\theta) - \gamma\delta S^-_{\gamma\delta}(\theta)
.\end{eqnarray}
The structural coefficients multiplying combinations of coupling constants are given below:
{\small 
\begin{eqnarray}\nonumber
\label{eq:43RchiA-CrossSections-via-ScalarProducts-ChiEta-rel-FF-Evaluations-InCoh-all-S-plus}
\frac{S^+_{\alpha^2}(\theta)}{2}&=& \frac{F^+_{\alpha^2}(\theta)}{2}+\frac{G^+_{\alpha^2}(\theta)}{2}=
m^2 m^2_\chi + |\bm{k}|^2 [s \cos^2\frac{\theta}{2} +2 |\bm{k}|^2\sin^4\frac{\theta}{2}]
,\\\nonumber
\frac{S^+_{\delta^2}(\theta)}{2}&=& \frac{F^+_{\delta^2}(\theta)}{2}+\frac{G^+_{\delta^2}(\theta)}{2 }=
3 m^2m^2_\chi +|\bm{k}|^2 [s\cos^2\frac{\theta}{2}+ 2|\bm{k}|^2 \sin^4\frac{\theta}{2}+2(m^2 + m^2_\chi ) \sin^2\frac{\theta}{2} ]
,\\  \frac{S^+_{\alpha\delta}(\theta)}{2} &=&\frac{F^+_{\alpha\delta}(\theta)}{2}+\frac{G^+_{\alpha\delta}(\theta)}{2}=2 \sin^2\frac{\theta}{2}|\bm{k}|^2  [s-m^2 - m_\chi^2 - 2|\bm{k}|^2 \sin^2\frac{\theta}{2} ]
,\\ \nonumber
\frac{S^+_{\gamma^2}(\theta)}{2} &=& \frac{F^+_{\gamma^2}(\theta)}{2}+ \frac{G^+_{\beta\gamma}(\theta)}{2}
=|\bm{k}|^2 [(s +m^2_\chi) \cos^2\frac{\theta}{2} +2 \sin^4\frac{\theta}{2} E^2_\chi]
,\\ \nonumber
\frac{S^+_{\beta^2}(\theta)}{2 }&=&\frac{F^+_{\beta^2}(\theta)}{2}+\frac{G^+_{\beta\gamma}(\theta)}{2}
=|\bm{k}|^2[(s +m^2_\chi)\cos^2\frac{\theta}{2}  +2(m^2 -m^2_\chi) \sin^2\frac{\theta}{2}\cos^2\frac{\theta}{2} +2 E_p^2\sin^4\frac{\theta}{2}]
,\\. \frac{S^+_{\beta\gamma}(\theta)}{2 }&=& \frac{F^+_{\beta\gamma}(\theta)}{2}-2\frac{G^+_{\beta\gamma}(\theta)}{2} =2|\bm{k}|^2
[ (s -m^2 +m_\chi^2) \sin^2\frac{\theta}{2} -m_\chi^2 \cos^2\frac{\theta}{2}
-2 E_\chi^2 \sin^4\frac{\theta}{2}]
\nonumber .\end{eqnarray} 
\begin{eqnarray}\nonumber
\label{eq:43RchiA-CrossSections-via-ScalarProducts-ChiEta-rel-FF-Evaluations-InCoh-all-S-minus}
\frac{S^-_{\alpha\gamma}(\theta)}{4  |\bm{k}|} &=&\frac{G^-_{\alpha\gamma}(\theta)}{2}+\frac{F^-_{\alpha\gamma}(\theta)}{2}=-\sin^2\frac{\theta}{2} [m_\chi ^2 (E_p - m)\cos^2\frac{\theta}{2} (1-2\cos^2\frac{\theta}{2}) +2  |\bm{k}|^2 (E_p \cos^2\frac{\theta}{2} +  E_\chi)]
,\\ \nonumber
\frac{S^-_{\beta\delta}(\theta)}{4  |\bm{k}| }&=&\frac{G^-_{\beta\delta}(\theta)}{2}+\frac{F^-_{\beta\delta}(\theta)}{2}= \\&=& 
{  m^2_\chi(m\cos^2\frac{\theta}{2}+E_p\sin^2\frac{\theta}{2})\cos^2\frac{\theta}{2}(1-2\sin^2\frac{\theta}{2}) - 2\sin^2\frac{\theta}{2} E_p (E_\chi E_p+\cos^2\frac{\theta}{2} |\bm{k}|^2)} 
,\\   \nonumber
\frac{S^-_{\gamma\delta}(\theta) }{4 |\bm{k}| } &=&\frac{F^-_{\gamma\delta}(\theta)}{2}+ \frac{G^-_{\beta\delta}(\theta)}{2}= \\&=&  \nonumber  {2E_\chi^2E_p(\sin^2\frac{\theta}{2}+\cos^4\frac{\theta}{2})-m^2_\chi(E_p- m) \cos^4\frac{\theta}{2}(1-2\sin^2\frac{\theta}{2})
+  E_\chi(m^2+2 |\bm{k}|^2) \cos^2\frac{\theta}{2}}
,\\ \nonumber
\frac{S^-_{\alpha\beta}(\theta)}{4|\bm{k}|}  &=&\frac{F^-_{\alpha\beta}(\theta)}{2} + \frac{G^-_{\alpha\gamma}(\theta)}{2}=\\&=&  \nonumber \cos^2\frac{\theta}{2} [m_\chi ^2 \cos^2\frac{\theta}{2}(E_p -m)  (1+ 2 \sin^2\frac{\theta}{2}) 
+ (m_\chi^2   + E_\chi m)  m +  2|\bm{k}|^2 \sqrt{s}]+ 2|\bm{k}|^2 E_p\sin^4\frac{\theta}{2} 
.\nonumber\end{eqnarray}}%
Substituting expressions (\ref{eq:43RchiA-CrossSections-via-ScalarProducts-ChiEta-rel-FF-Evaluations-InCoh-all-S-plus})
and (\ref{eq:43RchiA-CrossSections-via-ScalarProducts-ChiEta-rel-FF-Evaluations-InCoh-all-S-minus})
into formulas (\ref{eq:43RchiA-CrossSections-via-ScalarProducts-ChiEta-rel-InCohCS-total-sums}),
one gets explicit forms
 {\small
\begin{eqnarray}\nonumber
\label{eq:43RchiA-CrossSectionsEvaluations-Weak-ChiEta-rel-InCohCS-total-sumy-both}
\sum\frac{Q^{s's}_{+,w}}{2^6}&=&
(\alpha^2+ 3\delta^2) m^2m^2_\chi +2 |\bm{k}|^4 [(\alpha-\delta)^2+(\beta-\gamma)^2] \sin^4\frac{\theta}{2} 
+ |\bm{k}|^2s [2(\alpha\delta+\beta\gamma)
\\&& \nonumber  +((\alpha-\delta)^2+(\beta-\gamma)^2) \cos^2\frac{\theta}{2}] 
+2|\bm{k}|^2\sin^2\frac{\theta}{2} [m^2(\beta-\gamma)\beta+(m^2+m^2_\chi)(\delta-\alpha)\delta]        
\\&& +|\bm{k}|^2m^2_\chi [(\beta-\gamma)^2(\sin^4\frac{\theta}{2}+\cos^4\frac{\theta}{2})+(\gamma^2-\beta^2) \sin^2\frac{\theta}{2}]
,\\ \nonumber 
\sum\frac{Q^{s's}_{-,w}}{2^7|\bm{k}|}&=&
(\alpha\gamma+\beta\delta)[m^2_\chi E_p\cos^2\frac{\theta}{2}\sin^2\frac{\theta}{2} 
(1-2\sin^2\frac{\theta}{2})-2|\bm{k}|^2(E_p \cos^2\frac{\theta}{2}+ E_\chi)\sin^2\frac{\theta}{2}]
\\&& \nonumber
-(\alpha\beta +\gamma\delta)
 \big[m^2_\chi E_p\cos^4\frac{\theta}{2}(1+2\sin^2\frac{\theta}{2})+2|\bm{k}|^2E_p(1-\cos^2\frac{\theta}{2}\sin^2\frac{\theta}{2})
 \\&& \nonumber
 +(2|\bm{k}|^2+  m^2) E_\chi \cos^2\frac{\theta}{2}\big]
-2\delta (\beta     m^2 E_\chi+\gamma m^2_\chi E_p)\sin^2\frac{\theta}{2} 
\\&& \nonumber
+(\alpha\sin^2\frac{\theta}{2} 
-\delta\cos^2\frac{\theta}{2}) (\gamma-\beta) mm^2_\chi\cos^2\frac{\theta}{2}(1-2\sin^2\frac{\theta}{2}) 
 .\end{eqnarray}}%
Using these sums and formula (\ref{eq:43RchiA-CrossSections-via-ScalarProducts-ChiEta-rel-InCohCS-total}),
one can calculate the fully averaged incoherent $\chi A$ interaction cross section.
By analogy with the coherent cross section
(\ref{eq:43RchiA-CrossSection-via-ScalarProducts-ChiEta-rel-CohCS-mp=mn-general-spinless-total})
we present only the completely averaged {\em incoherent} $\chi A$ cross section for the {\em  spinless}\/ nuclei
\begin{eqnarray}\nonumber  
\label{eq:43RchiA-CrossSections-via-ScalarProducts-ChiEta-rel-InCohCS-total-spinless}
\frac{d\sigma^\text{total}_{\text{inc}}}{g_\text{i} d T_A} &=& 
 \hat c_A  \sum_{f=p,n} {A_f}  \widehat F^2_f(\bm{q})\Big[
(\alpha^2+ 3\delta^2) m^2m^2_\chi +2 |\bm{k}|^4 [(\alpha-\delta)^2+(\beta-\gamma)^2] \sin^4\frac{\theta}{2} 
\\&&\qquad+ |\bm{k}|^2s [((\alpha-\delta)^2+(\beta-\gamma)^2) \cos^2\frac{\theta}{2}+2(\alpha\delta+\beta\gamma)]
\\&&\qquad \nonumber +2|\bm{k}|^2\sin^2\frac{\theta}{2} [m^2(\beta-\gamma)\beta+(m^2+m^2_\chi)(\delta-\alpha)\delta]        
\\&&\qquad +|\bm{k}|^2m^2_\chi [(\beta-\gamma)^2(\sin^4\frac{\theta}{2}+\cos^4\frac{\theta}{2})+(\gamma^2-\beta^2) \sin^2\frac{\theta}{2}] \Big]
\nonumber .\end{eqnarray}
In terms of $\Phi$-functions from (\ref{eq:43RchiA-CrossSection-via-ScalarProducts-InCohSC-Phi-definitions})
this formula looks like
\footnote{Because $\Phi^+_{(\alpha-\delta)^2+(\beta-\gamma)^2}(\bm{q})\cos^2\frac{\theta}{ 2}+\Phi^+_{2(\alpha\delta+\beta\gamma)}(\bm{q}) = \Phi^+_{\alpha^2 +\delta^2+\beta^2+ \gamma^2}(\bm{q})\cos^2\frac{\theta}{2}+ \Phi^+_{2\alpha\delta + 2\beta\gamma}(\bm{q} )\sin^2\frac{\theta}{2}$.}
{\small \begin{eqnarray}\nonumber
\label{eq:43RchiA-CrossSections-via-ScalarProducts-ChiEta-rel-InCohCS-total-spinless-Phi}
\frac{d\sigma^\text{total}_{\text{inc}}}{ g_\text{i} \hat c_A d T_A} &=&
m^2_\chi m^2\Phi^+_{\alpha^2+3\delta^2}(\bm{q})
 +m^2_\chi  |\bm{k}|^2[\Phi^+_{(\beta-\gamma)^2}(\bm{q})(\sin^4\frac{\theta}{2}+\cos^4\frac{\theta}{2})
+\Phi^+_{(\gamma^2-\beta^2)}(\bm{q}) \sin^2\frac{\theta}{2}]
\\&& \nonumber
+2 |\bm{k}|^4 \Phi^+_{(\alpha-\delta)^2+(\beta-\gamma)^2}(\bm{q}) \sin^4\frac{\theta}{2} 
+ |\bm{k}|^2s [ \Phi^+_{(\alpha-\delta)^2+(\beta-\gamma)^2}(\bm{q})\cos^2\frac{\theta}{2}+\Phi^+_{2(\alpha\delta+\beta\gamma)}(\bm{q})]
\\&&  
+2|\bm{k}|^2\sin^2\frac{\theta}{2} [m^2\Phi^+_{(\beta-\gamma)\beta}(\bm{q})
+(m^2+m^2_\chi)\Phi^+_{(\delta-\alpha)\delta}(\bm{q})]        
. \end{eqnarray}}%
Formulas (\ref{eq:43RchiA-CrossSections-via-ScalarProducts-ChiEta-rel-InCohCS-total-spinless})
and (\ref{eq:43RchiA-CrossSections-via-ScalarProducts-ChiEta-rel-InCohCS-total-spinless-Phi}) 
give the {\em incoherent} cross sections of the weak $\chi A$ interaction, 
completely averaged over the initial and summed over the final helicities of the $\chi$ particle, 
in the case of a spinless nucleus.
\par
In the super-relativistic, "neutrino"\/ limit ($m_\chi\to 0$, $|\bm{k}|=E_\chi$), the sums
(\ref{eq:43RchiA-CrossSectionsEvaluations-Weak-ChiEta-rel-InCohCS-total-sumy-both}) are
{\small\begin{eqnarray}\nonumber
\label{eq:43RchiA-CrossSectionsEvaluations-Weak-ChiEta-rel-InCohCS-total-sumy-both=mchi=0}
\sum\frac{Q^{s's}_{+,w}}{2^6|\bm{k}|^2}&=& 2 |\bm{k}|^2 [(\alpha-\delta)^2+(\beta-\gamma)^2] \sin^4\frac{\theta}{2} 
+ s [((\alpha-\delta)^2+(\beta-\gamma)^2) \cos^2\frac{\theta}{2}+2(\alpha\delta+\beta\gamma)]
\\ &&  +2m^2 \sin^2\frac{\theta}{2} [(\beta-\gamma)\beta+(\delta-\alpha)\delta]        
;\\ \nonumber
\sum\frac{Q^{s's}_{-,w}}{2^7|\bm{k}|^2}&=& -2\delta \beta m^2\sin^2\frac{\theta}{2} 
-2(\alpha\gamma+\beta\delta)|\bm{k}|(E_p \cos^2\frac{\theta}{2}+|\bm{k}|)\sin^2\frac{\theta}{2}
\\&& \nonumber -(\alpha\beta +\gamma\delta) \big[
2|\bm{k}|E_p(1-\cos^2\frac{\theta}{2}\sin^2\frac{\theta}{2})+(2|\bm{k}|^2+  m^2) \cos^2\frac{\theta}{2}\big]
\nonumber .\end{eqnarray}}%
Then the fully averaged incoherent $\chi A$ cross section
(\ref{eq:43RchiA-CrossSections-via-ScalarProducts-ChiEta-rel-InCohCS-total})
takes the form
{\small \begin{eqnarray*}
\frac{d\sigma^\text{total}_{\text{inc},0}}{g_\text{i}d T_A} &=&   \dfrac{G^2_F  m_A }{4 \pi }\dfrac{E_\chi^2}{|\bm{k}_\chi|^2}
 \!\!\sum_{f=p,n}\!\! A_f \widehat F^2_f(\bm{q})\Big[ \frac{2E_\chi^2}{m^2}[(\alpha-\delta)^2 +(\beta-\gamma)^2]\sin^4\frac{\theta}{2}+s [(\alpha^2 +\delta^2+\beta^2+\gamma^2)\cos^2\frac{\theta}{2}
\\&&+ 2(\alpha\delta + \beta\gamma) \sin^2\frac{\theta}{2}] +2\sin^2\frac{\theta}{2}[\beta(\beta-\gamma)+\delta (\delta-\alpha)]  
\\ &+& \dfrac{2 \Delta A_f}{A_f}\Big\{  -2(\alpha\gamma+\beta\delta)\frac{E_\chi}{m}\big(\frac{E_p}{m}\cos^2\frac{\theta}{2}+\frac{E_\chi}{m}\big)\sin^2\frac{\theta}{2}
\\ && \nonumber  -(\alpha\beta +\gamma\delta) \big[
\frac{2E_\chi E_p}{m^2}\big(1-\cos^2\frac{\theta}{2}\sin^2\frac{\theta}{2}\big)
+\big(2\frac{E_\chi^2}{m^2}+1\big)\cos^2\frac{\theta}{2}\big]-2\delta \beta \sin^2\frac{\theta}{2}  \Big\}\Big]
\nonumber. \end{eqnarray*}}%
This expression for the spinless nucleus yields
{\small
\begin{eqnarray} 
\label{eq:43RchiA-CrossSections-via-ScalarProducts-ChiEta-rel-InCohCS-total-spinless-mchi=0}
\frac{d\sigma^\text{total}_{\text{inc},00}}{g_\text{i}d T_A} &=&   \dfrac{G^2_F  m_A }{4 \pi }\dfrac{|\bm{k}|^2}{|\bm{k}_\chi|^2}
 \!\!\sum_{f=p,n}\!\! A_f \widehat F^2_f(\bm{q})  \Big[\frac{2|\bm{k}|^2}{m^2}[(\alpha-\delta)^2 +(\beta-\gamma)^2]\sin^4\frac{\theta}{2}
 \\&&  +s [(\alpha^2 +\delta^2+\beta^2+\gamma^2)\cos^2\frac{\theta}{2} + 2(\alpha\delta + \beta\gamma) \sin^2\frac{\theta}{2}] +2\sin^2\frac{\theta}{2}[\beta(\beta-\gamma)+\delta (\delta-\alpha)]  \Big] 
. \nonumber \end{eqnarray}}%
Since $m_\chi= 0$, formula
(\ref{eq:43RchiA-CrossSections-via-ScalarProducts-ChiEta-rel-InCohCS-total-spinless-mchi=0})
for the {\em total}\/ cross section should not have the contribution of the $\chi$-lepton spin-flip cross sections; 
therefore, this expression is the sum of only two terms arising from 
the massless $\chi$ lepton falling on the nucleus with positive and negative (conserved) helicity.
Therefore, expression (\ref{eq:43RchiA-CrossSections-via-ScalarProducts-ChiEta-rel-InCohCS-total-spinless-mchi=0})
must coincide with the direct sum of these contributions, i.e. with formula
(\ref{eq:43RchiA-CrossSections-via-ScalarProducts-ChiEta-rel-InCohCS-total-spinless-mchi=0--++}).
In the nonrelativistic limit, when $|\bm{k}| \ll m_\chi $ (or $|\bm{k}| \to 0$), one gets
\begin{eqnarray*}
\sum\frac{Q^{s's}_{+,w}}{2^6}&=&(\alpha^2 + 3\delta^2) m^2m^2_\chi 
,\qquad \sum\frac{Q^{s's}_{-,w}}{2^5 4}\propto  |\bm{k}| \simeq 0
\nonumber .\end{eqnarray*}
Hence the averaged (total) incoherent $\chi A$ cross section
(\ref{eq:43RchiA-CrossSections-via-ScalarProducts-ChiEta-rel-InCohCS-total}) acquires a "simple form"\:
\begin{eqnarray*}
\frac{d\sigma^\text{total}_{\text{inc}}}{g_\text{i}d T_A} &=&  \dfrac{G^2_F  m_A}{4 \pi }\dfrac{m^2_\chi }{|\bm{k}_\chi|^2 }
\sum_{f=p,n} A_f  \widehat F^2_f(\bm{q}) (\alpha^2 + 3\delta^2) 
, \end{eqnarray*}
which was obtained earlier in \cite{Bednyakov:2022dmc}.

\paragraph{\em The total cross section of the weak $\chi A$ interaction} $\!\!\!\!\!\!$ is
the sum of fully averaged coherent and incoherent terms given in the general form by formulas
(\ref{eq:43RchiA-CrossSection-via-ScalarProducts-ChiEta-rel-CohCS-mp=mn-general-spinful-total}) and
(\ref{eq:43RchiA-CrossSections-via-ScalarProducts-ChiEta-rel-InCohCS-total}).
Let us present, as an example, the expressions for this sum in the case of a
{\em spinless}\/ ($\Delta A_f=0$) nucleus
\footnote{The ratio (\ref{eq:ScalarProducts-Lab-k^2-to-k^2_lab}), the expression for $\hat c_A$
from (\ref{eq:41chiA-CrossSection-Coh-vs-InCoh-GeneralFactor-with-G2_F}) and $g_\text{c} \simeq g_\text{i} \simeq 1$
are taken into account here.}:
{\small \begin{eqnarray}\nonumber
\label{eq:43RchiA-CrossSection-via-ScalarProducts-ChiEta-rel-CohCS-mp=mn-general-spinless-total-weak}
\dfrac{d\sigma^{\text{total}}_{0}}{d T_A}&=&\dfrac{G^2_F  m_A}{4 \pi}
\Bigg[\cos^2\frac{\theta}{2} \dfrac{E^2_\chi}{|\bm{k}_\chi|^2 }
\Big\{ \Big[G_\alpha(A,\bm{q})\big(\frac{E_p}{m}\cos^2\frac{\theta}{2}+\sin^2\frac{\theta}{2}
+\frac{|\bm{k}|^2}{m E_\chi}\big) +  G_\delta(A,\bm{q}) (\frac{E_p}{m} -1) \sin^2\frac{\theta}{2}  \Big]^2
\\&&  \nonumber 
+ \frac{|\bm{k}|^2}{m^2} \Big[G_\beta (A,\bm{q}) \frac{E_p -m}{E_\chi} \sin^2\frac{\theta}{2}
+G_\gamma(A,\bm{q})\big(1+\frac{E_p}{ E_\chi} \cos^2\frac{\theta}{2}+\frac{m}{ E_\chi}\sin^2\frac{\theta}{2}\big) \Big]^2 \Big\}
\\&&+ \nonumber
\sin^2\frac{\theta}{2} \dfrac{m^2_\chi }{|\bm{k}_\chi|^2 }\Big[G_\alpha(A,\bm{q}) (\frac{E_p}{m}\cos^2\frac{\theta}{2}+\sin^2\frac{\theta}{2})-G_\delta(A,\bm{q})(\frac{E_p}{m} -1)\cos^2\frac{\theta}{2} \Big]^2 
+\\&+& \nonumber
\dfrac{m^2_\chi }{|\bm{k}_\chi|^2}\Phi^+_{\alpha^2+3\delta^2}(\bm{q})
+2\frac{|\bm{k}|^2}{s} \Phi^+_{(\alpha-\delta)^2+(\beta-\gamma)^2}(\bm{q}) \sin^4\frac{\theta}{2} 
+\Phi^+_{(\alpha-\delta)^2+(\beta-\gamma)^2}(\bm{q})\cos^2\frac{\theta}{2}+\Phi^+_{2(\alpha\delta+\beta\gamma)}(\bm{q})
\\&& +2 \dfrac{m^2}{s}\sin^2\frac{\theta}{2} [\Phi^+_{(\beta-\gamma)\beta}(\bm{q})
+(1+\frac{m^2_\chi}{m^2})\Phi^+_{(\delta-\alpha)\delta}(\bm{q})]        
\\&& +\dfrac{m^2_\chi }{s} [\Phi^+_{(\beta-\gamma)^2}(\bm{q})(\sin^4\frac{\theta}{2}+\cos^4\frac{\theta}{2})
+\Phi^+_{(\gamma^2-\beta^2)}(\bm{q}) \sin^2\frac{\theta}{2}] \Bigg]
\nonumber. \end{eqnarray}}%
Using this formula, we recalls the algorithm for calculating the cross sections.
The "erxternal fixed"\/ parameters that define the nature of the interaction are the
weak Fermi constant $G_F$, the (common) nucleon mass $m$, and 
two sets of effective proton and neutron weak interaction constants, $\alpha_{p/n}, \beta_{p/n}, \gamma_{p /n}, \delta_{p/n}$.
The "external static"\/ parameters are the mass $m_A$ of the target nucleus and the mass $m_\chi$ of the lepton.
The "external dynamic"\/ parameters are the initial kinetic energy
  $T_0$ (or momentum $\bm{k}_\chi$) of the lepton incident on the nucleus at rest
and the recoil energy of this nucleus $T_A$.
Furthermore, the scattering angle $\theta$ is defined in terms of them 
using expression (\ref{eq:ScalarProducts-Lab-sinTheta-definition}),
and the other quantities $E_p$, $E_\chi$, $\bm{k}$, etc. are defined by means of 
formulas (\ref{eq:ScalarProducts-ChiEta-All-Energies-etc}).

\subsection{\normalsize\em Massive (anti)neutrinos and ${V\!\mp\!A}$ interaction}
Consider an important case of the scattering of the massive neutral $\chi$ lepton on the nucleus
due to the "standard"\/ $V\!\mp\!A$ weak interaction.
\par
Since calculation \cite{Bednyakov:2021pgs} of the scalar products was carried out  when the eigenstates of the $\chi$-lepton spin were the states of its helicity (projections of the spin onto the direction of its momentum), 
the neutrino (whose helicity is negative) corresponds to the cross sections with the "$--$"\/-indices of the $\chi$ lepton, i.e.,  coherent $\dfrac{d\sigma^{--}_\text{coh}}{d T_A}$ and incoherent $\dfrac{d\sigma^{--}_\text{inc}}{d T_A} $ cross sections.
It is only due to the non-zero mass $m_\chi$ that the processes with $\chi$-lepton spin flip whose cross sections are given in the second line of formulas (\ref{eq:43RchiA-CrossSection-via-ScalarProducts-WeakCohCS}) are possible.
\par
First,  from general formulas (\ref{eq:43RchiA-CrossSection-via-ScalarProducts-WeakCohCS}) 
we get expressions for the cross sections for coherent scattering on the nuclei of massive neutrinos from the Standard Model.
Let us use formulas
(\ref{eq:43RchiA-CrossSections-via-ScalarProducts-ChiEta-rel-CohCSs-mp=mn-Fp=Fn-general}),
corresponding to the assumption that the nuclear proton and neutron form factors are equal
(\ref{eq:43RchiA-CrossSections-via-ScalarProducts-Fp=Fn}).
For the $V\!-\!A$ weak interaction of neutrinos with nucleons, 
the latter have the following effective coupling constants
from (\ref{eq:42RchiA-CrossSection-ScalarProducts-ChiEta-Weak-definition}):
\begin{equation}
\label{eq:43RchiA-CrossSections-via-ScalarProducts-Weak-ChiEta-rel-V-A-couplings}
\alpha_f =  + g_V^f ,\quad  \beta_f = - g_A^f , \quad \gamma_f =  - g_V^f ,\quad  \delta_f = + g_A^f .
\end{equation}
Then the $\bm{q}$-independent notations
(\ref{eq:43RchiA-CrossSections-via-ScalarProducts-ChiEta-rel-CohCSs-mp=mn-Fp=Fn-couplings})
are simplified and take the form
\begin{eqnarray*}
G_\alpha(A)&=&g_V^p A_p + g_V^n A_n \equiv G_V(A) ,\qquad \qquad 
G_\gamma(A)= -(g_V^p A_p  + g_V^n A_n) = -  G_V(A)
,\\  G_\delta(A)&=&g_A^p A_p  + g_A^n A_n  \equiv G_A (A) ,\qquad \qquad 
G_\beta(A)=  - (g_A^p A_p + g_A^n  A_n) = - G_A(A) 
;\\ \Delta G_\alpha(A)&=&g_V^p  \Delta A_p + g_V^n \Delta A_n \equiv \Delta G_V(A) 
,\quad \Delta  G_\gamma(A)= -(g_V^p \Delta A_p  + g_V^n \Delta A_n) =-  \Delta G_V(A)
,\\  \Delta  G_\delta(A)&=&g_A^p \Delta A_p  + g_A^n \Delta A_n \equiv \Delta G_A(A)
,\quad \Delta  G_\beta(A)=  - (g_A^p  \Delta A_p + g_A^n \Delta A_n)  = - \Delta  G_A(A)
.\end{eqnarray*}
With their help one obtains 
from (\ref{eq:43RchiA-CrossSections-via-ScalarProducts-ChiEta-rel-CohCSs-mp=mn-Fp=Fn-general})
a general set of $\chi A$ cross sections of coherent scattering
of the massive $\chi$ lepton due to $V$--$A$ interaction in the form
\begin{eqnarray} \nonumber
\label{eq:43RchiA-CrossSections-via-ScalarProducts-Weak-ChiEta-rel-CohCS-Fp=Fn-V-A}
\frac{d\sigma^{\mp\mp}_\text{coh,V-A}}{g_\text{c} \hat c_A  d T_A}&=&
\cos^2\frac{\theta}{2} F^2(\bm{q}) \Big| G_V(A) [f_{\alpha+}(\theta) \pm  f_{\gamma+} (\theta)]  + G_A(A) [f_{\delta-} (\theta)  \pm  f_{\beta-} (\theta)]   \\&& \nonumber  \qquad +\Delta G_V(A) [ f_{\gamma-} (\theta) \pm  f_{\alpha-} (\theta)]  +\Delta G_A(A) [ f_{\beta+} (\theta)  \pm   f_{\delta+} (\theta) ]  \Big|^2
, \\ \frac{d\sigma^{\mp\pm}_\text{coh,V-A}}{g_\text{c} \hat c_A  d T_A}&=&\sin^2\frac{\theta}{2} m^2_\chi F^2(\bm{q}) \Big|G_V(A)  (m + \lambda^2_- \cos^2\frac{\theta}{2}) - G_A(A)  \lambda^2_-\cos^2\frac{\theta}{2} \\&& \nonumber \qquad - \Delta G_V(A) \lambda_+ \lambda_-  \cos^2\frac{\theta}{2} +\Delta G_A(A) [\lambda_+ \lambda_-  \cos^2\frac{\theta}{2} \mp m]\Big|^2
. \nonumber \end{eqnarray}
For negative initial helicity ($s=-1$, upper right index) of the 
massive $\chi$ lepton corresponding to the neutrino with $V$--$A$ weak interaction, 
formulas (\ref{eq:43RchiA-CrossSections-via-ScalarProducts-Weak-ChiEta-rel-CohCS-Fp=Fn-V-A}) take 
the  explicit form
\begin{eqnarray} \nonumber
\label{eq:43RchiA-CrossSections-via-ScalarProducts-Weak-ChiEta-rel-CohCS-Fp=Fn-Nu-massive}
\frac{d\sigma^{--}_\text{coh,V-A}}{g_\text{c}\hat c_A  d T_A}&=& \cos^2\frac{\theta}{2} F^2(\bm{q})  (|\bm{k}|+E_\chi)^2 \Big|  G_V(A)(m\sin^2\frac{\theta}{2} +  E_p\cos^2\frac{\theta}{2} +|\bm{k}|)+\Delta G_V(A) |\bm{k}| \sin^2\frac{\theta}{2}   
\\&& \nonumber  \qquad \qquad \quad + G_A(A)(E_p-m)\sin^2\frac{\theta}{2} 
+ \Delta G_A(A)(E_p + |\bm{k}|\cos^2\frac{\theta}{2})  \Big|^2
, \\  \frac{d\sigma^{+-}_\text{coh,V-A}}{g_\text{c}\hat c_A  d T_A}&=&  m^2_\chi \sin^2\frac{\theta}{2} F^2(\bm{q}) \Big| G_V(A)(m\sin^2\frac{\theta}{2} +  E_p\cos^2\frac{\theta}{2} ) - \Delta G_V(A) |\bm{k}| \cos^2\frac{\theta}{2}
\\&& \nonumber \qquad \qquad  
 - G_A(A) (E_p-m)\cos^2\frac{\theta}{2} + \Delta  G_A(A)(|\bm{k}| \cos^2\frac{\theta}{2}+ m)\Big|^2
. \end{eqnarray} 
It is taken into account that for sums of $f$-form factors from (\ref{eq:43RchiA-CrossSections-via-ScalarProducts-Weak-ChiEta-rel-CohCS-Fp=Fn-V-A}) one has
{\small
\begin{eqnarray*}  f_{\alpha+}(\theta)+f_{\gamma+}(\theta)\!&=&\! 
(|\bm{k}|+E_\chi)(m\sin^2\frac{\theta}{2} +  E_p\cos^2\frac{\theta}{2} +|\bm{k}|) 
,\  f_{\alpha-}(\theta)+f_{\gamma-}(\theta)=(|\bm{k}|+E_\chi) |\bm{k}| \sin^2\frac{\theta}{2} 
,\\ f_{\beta-}(\theta)+f_{\delta-}(\theta)\!&=&\! (|\bm{k}|+E_\chi)(E_p-m)\sin^2\frac{\theta}{2} 
,\  f_{\beta+}(\theta)+f_{\delta+}(\theta)=(|\bm{k}|+E_\chi)(E_p + |\bm{k}|\cos^2\frac{\theta}{2}) 
.\end{eqnarray*}}%
Recall that in the invariant variables the following relations hold
\begin{eqnarray}\nonumber
 E_{\chi/p}=\dfrac{s\pm m^2_\chi \mp m^2}{2\sqrt{s}},&&|\bm{k}|=\dfrac{\lambda(s, m^2, m^2_\chi)}{2\sqrt{s}}
,\quad [\bm{k}^l_\chi]^2=  \dfrac{\lambda^2(s, m^2, m^2_\chi)}{4 m^2},\ \text{~giving} 
\\ \nonumber  4^2  c_A (E_\chi+ |\bm{k}|)^2 &=&
 \dfrac{G^2_F  m_A }{4 \pi} \dfrac{(E_\chi+ |\bm{k}|)^2}{m^2 |\bm{k}^l_\chi|^2 }
 =  \dfrac{G^2_F  m_A }{4 \pi}  \dfrac{\varkappa(s, m^2, m^2_\chi) }{ s }, \text{~~~where}
 \\   \varkappa(s,m^2_\chi)&\equiv& \varkappa(s, m^2, m^2_\chi) = \Big[1+\dfrac{s+m^2_\chi-m^2}{\lambda(s, m^2, m^2_\chi)}\Big]^2, \quad \text{as well}
\label{eq:43RchiA-CrossSections-via-ScalarProducts-Weak-ChiEta-rel-CohCS-Fp=Fn-Nu-massive-kappa}
 \\ \nonumber  4^2 c_Am_\chi^2&=&\dfrac{G^2_Fm_A}{4\pi}\dfrac{m_\chi^2}{m^2|\bm{k}^l_\chi|^2}
=\dfrac{G^2_Fm_A}{\pi}\dfrac{m_\chi^2}{\lambda^2(s, m^2, m^2_\chi)}
.\end{eqnarray}
Then for the massive $\chi$ lepton incident on the nucleus with negative initial helicity 
and interacting with nucleons via the $V$--$A$ weak current, coherent $\chi A$ cross sections
(\ref{eq:43RchiA-CrossSections-via-ScalarProducts-Weak-ChiEta-rel-CohCS-Fp=Fn-Nu-massive})
can be {\em finally}\/ written  in the form
\begin{eqnarray} \nonumber
\label{eq:43RchiA-CrossSections-via-ScalarProducts-Weak-ChiEta-rel-CohCS-Fp=Fn-Nu-massive+}
\frac{d\sigma^{--}_\text{coh,V-A}}{g_\text{c}d T_A}&=&\dfrac{G^2_F  m_A }{4 \pi} \varkappa(s,m^2_\chi)
\cos^2\frac{\theta}{2} F^2(\bm{q}) \Big|  G_V(A)\Big(\frac{m}{\sqrt{s}}\sin^2\frac{\theta}{2} + \frac{E_p}{\sqrt{s}}\cos^2\frac{\theta}{2}+\frac{|\bm{k}|}{\sqrt{s}}\Big)  
\\&& \nonumber 
+ \Delta G_V(A) \frac{|\bm{k}|}{\sqrt{s}}\sin^2\frac{\theta}{2}   + G_A(A)\frac{(E_p-m)}{\sqrt{s}}\sin^2\frac{\theta}{2} + \Delta G_A(A)\Big(\frac{E_p}{\sqrt{s}} +\frac{ |\bm{k}|}{\sqrt{s}} \cos^2\frac{\theta}{2}\Big) \Big|^2
, \\[0pt] \\ \nonumber
\frac{d\sigma^{+-}_\text{coh,V-A}}{g_\text{c}d T_A}&=& \dfrac{G^2_Fm_A}{\pi}\dfrac{s m_\chi^2}{\lambda^2(s,m^2_\chi)}
 \sin^2\frac{\theta}{2} F^2(\bm{q}) \Big| G_V(A)\Big(\frac{m}{\sqrt{s}}\sin^2\frac{\theta}{2} +\frac{E_p}{\sqrt{s}}\cos^2\frac{\theta}{2}\Big) - \Delta G_V(A) \frac{|\bm{k}|}{\sqrt{s}} \cos^2\frac{\theta}{2}
\\&& \nonumber \qquad   - G_A(A) \frac{(E_p-m)}{\sqrt{s}}\cos^2\frac{\theta}{2} 
 + \Delta  G_A(A)\Big(\frac{|\bm{k}|}{\sqrt{s}}\cos^2\frac{\theta}{2}+\frac{m}{\sqrt{s}}\Big)\Big|^2
. \end{eqnarray} 
Formulas (\ref{eq:43RchiA-CrossSections-via-ScalarProducts-Weak-ChiEta-rel-CohCS-Fp=Fn-Nu-massive+})
give the main result for the coherent scattering cross section of the massive neutrino on the nucleus 
due to the weak $V$--$A$ interaction
\footnote{If one wants to "restore"\/ dependence on the nucleon type  in the nuclear form factors $F^2(\bm{q})$, 
one should "hide them back"\/   into the parameters $G_V(A,\bm{q})$ and $G_A(A,\bm{q})$.}. 
In invariant variables, the coherent sections of the $V$--$A$ weak interaction of the massive neutrino incident 
on the {\em spinless}\/ ($\Delta A_f=0$) nucleus take the form
\begin{eqnarray}\nonumber
\label{eq:43RchiA-CrossSections-via-ScalarProducts-Weak-ChiEta-rel-CohCS-Fp=Fn-Nu-massive-spinless-fino}
\dfrac{d\sigma^{--}_\text{coh,V-A}}{g_\text{c}d T_A}&=&\Big[1-\frac{T_A}{T^{\max}_A}\Big] \Big[\dfrac{G^2_F  m_A}{4 \pi }\Big] F^2(\bm{q}) \varkappa(s,m^2_\chi)
\Big\{ G_A(A) \frac{ (\sqrt{s}-m)^2  -m^2_\chi}{2s} \sin^2\frac{\theta}{2} 
\\&&\qquad  + G_V(A)\Big[\dfrac{m}{\sqrt{s}}  + \frac{ (\sqrt{s}-m)^2  -m^2_\chi}{2s} \cos^2\frac{\theta}{2} 
+ \dfrac{\lambda(s, m^2, m^2_\chi)}{2{s} } \Big] \Big\}^2
,\\ \nonumber
\frac{d\sigma^{+-}_\text{coh,V-A}}{g_\text{c}d T_A}&=&\sin^2\frac{\theta}{2}\Big[\dfrac{G^2_F  m_A}{4 \pi }\Big]
 \dfrac{m^2_\chi }{|\bm{k}_\chi|^2 }\dfrac{s} {m^2 } F^2(\bm{q})  
\Big\{- G_A(A) \frac{ (\sqrt{s}-m)^2  -m^2_\chi}{2s}\cos^2\frac{\theta}{2} 
\\&& \qquad +G_V(A)\Big[\dfrac{m}{\sqrt{s}}  + \frac{ (\sqrt{s}-m)^2  -m^2_\chi}{2s} \cos^2\frac{\theta}{2}\Big] \Big\}^2, \quad \text{where}\quad \sin^2\frac{\theta}{2}\simeq \frac{T_A}{T^{\max}_A}
\nonumber . \end{eqnarray}
There are two comments about these formulas.
First, in the case of a massless Standard Model neutrino, the second formula must vanish, which is clearly seen due to its proportionality to the square of the neutrino mass $m^2_\chi$.
Second, for $m_\chi=0$ one gets $\lambda(s, m^2,m^2_\chi=0)= (s-m^2)$. 
Hence it follows that
$$\varkappa(s,m^2_\chi=0) = 4 \text{~~~and~~~} \dfrac{m}{\sqrt{s}}+ \dfrac{(s-m^2)}{2{s} } + \frac{ (\sqrt{s}- m)^2}{2s} \cos^2\frac{\theta}{2} = 1-\dfrac{(\sqrt{s}-m)^2}{2{s}}\sin^2\frac{\theta}{2}.$$
As a result, the first formula goes into the expression
{\small \begin{eqnarray}
\label{eq:43RchiA-CrossSection-via-ScalarProducts-ChiEta-CohCS-neutrino}
\dfrac{d\sigma^{--}_\text{coh}}{d T_A}=\dfrac{G^2_F  m_A}{\pi}\Big[1-\frac{T_A}{T^{\max}_A}\Big]F^2(\bm{q})
\Big[ G_V(A)\big[1-\dfrac{(\sqrt{s}-m)^2}{2{s}}\sin^2\frac{\theta}{2}\big]  + G_A(A) \frac{ (\sqrt{s}-m)^2}{2s} \sin^2\frac{\theta}{2} \Big]^2,\qquad \end{eqnarray}}%
which coincides with the corresponding formula from \cite{Bednyakov:2021ppn}.
In other words, formula
(\ref{eq:43RchiA-CrossSections-via-ScalarProducts-Weak-ChiEta-rel-CohCS-Fp=Fn-Nu-massive-spinless-fino})
for the coherent scattering cross section of massive neutrinos on (spinless) nuclei 
at $m_\chi\to 0$ goes exactly into the well-known formula for the massless neutrino \cite{Bednyakov:2021ppn}.
\par
About the second formula
(\ref{eq:43RchiA-CrossSections-via-ScalarProducts-Weak-ChiEta-rel-CohCS-Fp=Fn-Nu-massive-spinless-fino}),
note the following.
This cross section (with lepton spin flip) has an appreciable value only for "appreciable"\/
mass of this lepton; moreover, the angular distribution proportional to $\sin^2\frac{\theta}{2}$ is very different from the case without the lepton spin flip (the first formula with $\cos^2\frac{\theta}{ 2}$ distribution).
The latter means that the lepton spin flip (due to the conservation of the total angular momentum of the system) is possible only if the spin-changing lepton flies out of the interaction zone almost exactly in the opposite direction.
On the other hand, for (extremely) small momenta $|\bm{k}_\chi|^2 \ll m^2_\chi$ of the lepton incident on the nucleus,
firstly, the momentum $\bm{q}$ transferred to the nucleus is also very small, and hence the nuclear form factor
$F^2(\bm{q}) \simeq 1$ plays no role; second, $\sqrt{s}\simeq m+m_\chi$, and then the coherent
 $\chi A$ cross section with lepton spin flip takes the form
\begin{eqnarray}
\label{eq:43RchiA-CrossSections-via-ScalarProducts-Weak-ChiEta-rel-CohCS-Fp=Fn-Nu-massive-spinflip} 
\frac{d\sigma^{+-}_\text{coh}(\chi A)}{d T_A}&=& 
\sin^2\frac{\theta}{2}\dfrac{G^2_F  m_A}{4 \pi } \dfrac{m^2_\chi }{|\bm{k}_\chi|^2}  G^2_V(A)
.\end{eqnarray}
It can be seen that if $\dfrac{|\bm{k}_\chi|^2}{m^2_\chi} \equiv \Theta \ll 1$,
then the coherent $\chi A$ cross section with spin flip is enhanced not only by the $A^2$ nuclear factor
hidden in the parameter $G^2_V(A)$, but also by a sufficiently large "particle flux" factor\/ $\Theta^{-1}$.
This argument is not invalidated if $|\bm{k}_\chi|^2 \simeq m^2_\chi \ll m^2$, then
$s \simeq m^2$ and the second formula
(\ref{eq:43RchiA-CrossSections-via-ScalarProducts-Weak-ChiEta-rel-CohCS-Fp=Fn-Nu-massive-spinless-fino})
 goes into expression
(\ref{eq:43RchiA-CrossSections-via-ScalarProducts-Weak-ChiEta-rel-CohCS-Fp=Fn-Nu-massive-spinflip}).
\par
Note further that formula for the $V$--$A$ cross sections
(\ref{eq:43RchiA-CrossSections-via-ScalarProducts-Weak-ChiEta-rel-CohCS-Fp=Fn-V-A}), 
while maintaining the positive helicity of the $\chi$ lepton (right superscript $+$), 
includes the following combinations of nucleon $f$-form factors
from (\ref{eq:43RchiA-CrossSection-via-ScalarProducts-WeakCohCS-fs}):
{\small \begin{eqnarray*}
f_{\alpha+}(\theta)-f_{\gamma+}(\theta)\!&=&\!(E_\chi  -|\bm{k}| ) (m\sin^2\frac{\theta}{2} + E_p\cos^2\frac{\theta}{2} - |\bm{k}|)
,\ \ f_{\gamma-} (\theta)-f_{\alpha-} (\theta)=(E_\chi -|\bm{k}|)|\bm{k}|\sin^2\frac{\theta}{2}
, \\ f_{\delta-} (\theta)-f_{\beta-} (\theta)\!&=&\!(E_\chi  -|\bm{k}|)(E_p-m)\sin^2\frac{\theta}{2}
,\quad f_{\beta+} (\theta)- f_{\delta+} (\theta)=  (E_\chi-  |\bm{k}| )( |\bm{k}| \cos^2\frac{\theta}{2}-E_p) 
.\end{eqnarray*}}%
All these quantities for "$++$"\/-helicity are proportional to the difference $(E_\chi-|\bm{k}|)$, and not to the sum $(E_\chi+|\bm{ k}|)$, as for "$--$"\/-helicity.
  This means that $V$--$A$-couplings from
(\ref{eq:43RchiA-CrossSections-via-ScalarProducts-Weak-ChiEta-rel-V-A-couplings})
strongly "suppress"\/ (proportionally to $(E_\chi-|\bm{k}|)^2$) the coherent cross sections of interaction 
between the nucleus and a massive antineutrino-analogue,
and upon passing to the massless limit, these cross sections simply vanish.
Thus, for the "$++$"\/-helicity of the massive analogue of the antineutrino, it is necessary to set the coupling constants according to the $V$+$A$ character of the weak interaction corresponding specifically to the antineutrino.
Indeed, in the case of antineutrinos (due to the $V$+$A$ current), the nucleon weak coupling constants 
look like
\begin{equation}\label{eq:43RchiA-CrossSections-via-ScalarProducts-Weak-ChiEta-rel-V+A-couplings}
\alpha^{V+A}_f \equiv
\alpha^{\bar\nu}_f =  + g_V^f ,\quad  \beta^{\bar\nu}_f = - g_A^f, \quad \gamma^{\bar\nu}_f = + g_V^f ,\quad  \delta^{\bar\nu}_f =  - g_A^f 
.\end{equation}
Then $\bm{q}$-independent factors (\ref{eq:43RchiA-CrossSections-via-ScalarProducts-ChiEta-rel-CohCSs-mp=mn-Fp=Fn-couplings}) for $V$+$A$ cross sections are
\begin{eqnarray*}
G^{\bar\nu}_\alpha(A) &=&g_V^p A_p + g_V^n A_n \equiv G_V(A) 
,\qquad \Delta G^{\bar\nu}_\alpha(A)=g_V^p  \Delta A_p + g_V^n \Delta A_n \equiv \Delta G_V(A) 
,\\ G^{\bar\nu}_\gamma(A) &=& g_V^p A_p  + g_V^n A_n = G_V(A)
,\qquad  \Delta  G^{\bar\nu}_\gamma(A)= g_V^p \Delta A_p  + g_V^n \Delta A_n =  \Delta G_V(A)
,\\   G^{\bar\nu}_\delta(A) &=&- (g_A^p A_p  + g_A^n A_n ) \equiv - G_A (A) 
,\quad \Delta  G^{\bar\nu}_\beta(A)=  - (g_A^p  \Delta A_p + g_A^n \Delta A_n)  = - \Delta  G_A(A)
,\\  G^{\bar\nu}_\beta(A)&=&  - (g_A^p A_p + g_A^n  A_n) = - G_A(A) 
,\quad \Delta  G^{\bar\nu}_\delta(A)=-(g_A^p \Delta A_p  + g_A^n \Delta A_n ) \equiv -\Delta G_A(A)
.\end{eqnarray*}
After substituting these factors into formulas
(\ref{eq:43RchiA-CrossSection-via-ScalarProducts-WeakCohCS-general-mp=mn}), 
expressions are obtained for the coherent cross sections of the massive lepton scattering on the nucleus
due to the $V$+$A$ interaction
\begin{eqnarray} \nonumber
\label{eq:43RchiA-CrossSection-via-ScalarProducts-ChiEta-rel-CohCS-Fp=Fn-V+A}
\frac{d\sigma^{\mp\mp}_\text{coh,V+A}(\bm{q})}{g_\text{c} \hat c_A  d T_A}&=&
\cos^2\frac{\theta}{2} F^2(\bm{q}) \Big|G_V(A) [f_{\alpha+}(\theta) \mp  f_{\gamma+} (\theta)] - G_A(A)[ f_{\delta-} (\theta) \mp  f_{\beta-} (\theta)]  
\\&&\qquad - \Delta G_V(A)[f_{\gamma-} (\theta) \mp f_{\alpha-} (\theta)]+\Delta G_A(A)[f_{\beta+} (\theta)\mp f_{\delta+} (\theta)] \Big|^2
;\\ \nonumber
 \frac{d\sigma^{\mp\pm}_\text{coh,V+A}(\bm{q})}{ g_\text{c} \hat c_A  d T_A} &=&
   \sin^2\frac{\theta}{2} m^2_\chi F^2(\bm{q}) \Big|G_V(A)  (m + \lambda^2_- \cos^2\frac{\theta}{2}) + G_A(A)  \lambda^2_-\cos^2\frac{\theta}{2} \pm \Delta G_A(A) m +\\&& \nonumber \qquad  + [\Delta G_V(A) +\Delta G_A(A)] \lambda_+ \lambda_-  \cos^2\frac{\theta}{2} \Big|^2
\nonumber .\end{eqnarray}
It can be seen that these formulas "correspond"\/ to antineutrinos (helicity is positive
for a weak $V$+$A$-current).
The "$++$"\/-helicity states correspond to the sum $f_{\alpha+}(\theta) + f_{\gamma+} (\theta)\propto (E_\chi +|\bm{k}| )$, which does not vanish in the massless and nonrelativistic cases when $E_\chi \simeq |\bm{k}|$.
If a $\chi$ particle with the "$--$"\/-helicity "wants"\/ to interact via the weak $V$+$A$-current (like a neutrino),
then its coherent cross section 
(first formula (\ref{eq:43RchiA-CrossSection-via-ScalarProducts-ChiEta-rel-CohCS-Fp=Fn-V+A})) will be suppressed
by the  factor $ (E_\chi -|\bm{k}| )^2$ and vanishes when $E_\chi \simeq |\bm{k}|$.
\par 
Therefore, coherent $\chi A$ cross sections for the "antineutrino-like"\/ ($++$, $-+$ and $V\!+\!A$ indices)
and "neu\-trino-like"\/ ($--$, $+-$ and $V\!-\!A$ indices)
interactions of the massive $\chi$ lepton with the nucleus have the form
{\small \begin{eqnarray} \nonumber
\label{eq:43RchiA-CrossSections-via-ScalarProducts-ChiEta-rel-CohCS-Fp=Fn-VpmA-compact}
\frac{d\sigma^{\mp\mp}_\text{coh,V$\mp$A}}{g_\text{c}\hat c_A  d T_A}&=&
\cos^2\frac{\theta}{2}  (|\bm{k}|+E_\chi)^2 F^2(\bm{q}) \Big|G_V(A)(m\sin^2\frac{\theta}{2} +  E_p\cos^2\frac{\theta}{2} +|\bm{k}|)  \pm G_A(A) (E_p-m)\sin^2\frac{\theta}{2}
\\&&
\qquad \pm \Delta G_V(A) |\bm{k}| \sin^2\frac{\theta}{2}  +\Delta G_A(A) (E_p + |\bm{k}|\cos^2\frac{\theta}{2}) \Big|^2
;\\ \nonumber
\frac{d\sigma^{\pm\mp}_\text{coh,V$\mp$A}}{ g_\text{c}\hat  c_A  d T_A}&=&  m^2_\chi \sin^2\frac{\theta}{2} F^2(\bm{q}) \Big| G_V(A)(m\sin^2\frac{\theta}{2} +  E_p\cos^2\frac{\theta}{2} ) 
\mp G_A(A) (E_p-m)\cos^2\frac{\theta}{2} 
 \\&& \nonumber \qquad  
\mp \Delta G_V(A) |\bm{k}| \cos^2\frac{\theta}{2}+ \Delta  G_A(A)( m+|\bm{k}| \cos^2\frac{\theta}{2})\Big|^2
\nonumber .\end{eqnarray}}%
It should be emphasized that each of formulas
(\ref{eq:43RchiA-CrossSections-via-ScalarProducts-ChiEta-rel-CohCS-Fp=Fn-VpmA-compact})
contains only two possible expressions, 
when  only superscripts or only subscripts are {\em simultaneously}\/ taken for the cross section 
(spin indices are $\mp\mp$) 
and for the current ($\mp$).
\par
Note that, as previousely for neutrinos, at (extremely) small  momenta $|\bm{k}_\chi|^2 \ll m^2_\chi$
of the (right-hand) lepton incident on the nucleus, one has  the $F^2(\bm{q}) \simeq 1$ and  $\sqrt{s}\simeq m+m_\chi$,
$|\bm{k}|\simeq 0$, $E_p\simeq m$, 
hence  the coherent $\chi A$ cross section with spin flip of the (left- and right-handed) lepton {\em are the same}:
\begin{eqnarray} 
\label{eq:43RchiA-CrossSections-via-ScalarProducts-ChiEta-rel-CohCS-Fp=Fn-VpmA-k=0}
\frac{d\sigma^{\pm\mp}_\text{coh,V$\mp$A}(\bm{q})}{g_\text{c} d T_A}&=& 
\dfrac{G^2_F m_A }{4 \pi } \dfrac{m^2_\chi }{ |\bm{k}_\chi|^2 } \sin^2\frac{\theta}{2} \Big| G_V(A)+ \Delta  G_A(A)\Big|^2
.\end{eqnarray}
Thus, if $\dfrac{|\bm{k}_\chi|^2}{m^2_\chi} \equiv \Theta \ll 1$,
the coherent $\chi A$ cross sections with spin flip (of both left- and right-handed massive lepton) 
are enhanced not only by the influence of the nucleus,  but also by a sufficiently large "particle flux" factor\/ $\Theta^{-1}$.
These observations {\em look new}\/ and probably are important  for registration of the relic massive (anti)neutrinos
with extremely low energy ($5\div 6 \times 10^{-4}~$eV).
Despite the rather enhanced cross section, the old question of 
{\em what we are going to measure} remains open.
\par
From general formula
(\ref{eq:43RchiA-CrossSection-via-ScalarProducts-ChiEta-rel-CohCS-mp=mn-general-spinless-total})
one can obtain the fully averaged coherent $\chi A$ cross section {\em on the spinless nucleus}\/
due to the  $V\mp A$ interaction, i.e. when the nucleon coupling constants
(\ref{eq:43RchiA-CrossSections-via-ScalarProducts-Weak-ChiEta-rel-V-A-couplings})
and (\ref{eq:43RchiA-CrossSections-via-ScalarProducts-Weak-ChiEta-rel-V+A-couplings}) are
\begin{equation}
\label{eq:43RchiA-CrossSections-via-ScalarProducts-Weak-ChiEta-rel-VmpA-couplings}
\alpha_{V\mp A} =  + g_V ,\  \beta_{V\mp A} = - g_A, \ \gamma_{V\mp A} = \mp g_V ,\ \delta_{V\mp A} =  \pm g_A.
\end{equation}
The $\bm{q}$-dependent effective coupling constants from
(\ref{eq:43RchiA-CrossSection-via-ScalarProducts-BigWeakNuclear-Couplings})
take the "$V\!\mp\!A$"\/ form
{\small \begin{eqnarray}\nonumber
\label{eq:43RchiA-CrossSections-via-ScalarProducts-Weak-ChiEta-rel-VmpA-Eff-couplings}
G_{\alpha_{\mp}}(A,\bm{q})\!\!&=&\!\!\sum (+g_V^f) A_fF_f(\bm{q}) \equiv +G_V(A,\bm{q}) 
,\ \  G_{\gamma_{\mp}}(A,\bm{q})=\sum (\mp g_V^f) A_fF_f(\bm{q}) \equiv \mp G_V(A,\bm{q})
,\qquad\\   \\[-8pt] \nonumber
G_{\beta_{\mp}}(A,\bm{q})\!\!&=&\!\!\sum (-g_A^f) A_f F_f(\bm{q}) = - G_A(A,\bm{q}) 
,\ \ G_{\delta_{\mp}}(A,\bm{q})=\sum (\pm g_A^f) A_f F_f(\bm{q}) \equiv \pm G_A(A,\bm{q})
\nonumber.\qquad \end{eqnarray}}%
Similar to (\ref{eq:43RchiA-CrossSections-via-ScalarProducts-Weak-ChiEta-rel-VmpA-Eff-couplings}), 
spin-dependent factors are
\begin{eqnarray*}
\Delta G_{\alpha_{\mp}} (A,\bm{q})&=&\sum (+g_V^f) \Delta A_fF_f(\bm{q}) \equiv +\Delta G_V(A,\bm{q}) 
,\\ \Delta G_{\gamma_{\mp}}(A,\bm{q}) &=& \sum (\mp g_V^f)  \Delta A_fF_f(\bm{q}) \equiv \mp  \Delta G_V(A,\bm{q})
,\\\Delta G_{\beta_{\mp}}(A,\bm{q})&=&\sum (-g_A^f) \Delta  A_f F_f(\bm{q}) = - \Delta  G_A(A,\bm{q})
,\\ \Delta   G_{\delta_{\mp}}(A,\bm{q})&=&\sum (\pm g_A^f)\Delta  A_f F_f(\bm{q}) \equiv \pm\Delta  G_A(A,\bm{q})
.\end{eqnarray*}
Substitution of (\ref{eq:43RchiA-CrossSections-via-ScalarProducts-Weak-ChiEta-rel-VmpA-Eff-couplings}) in
(\ref{eq:43RchiA-CrossSection-via-ScalarProducts-ChiEta-rel-CohCS-mp=mn-general-spinless-total})
gives {\em the fully averaged coherent $V\!\mp\!A$ cross section}\/ for the
scattering of the massive lepton on the nucleus {\em without spin}\/ ($\Delta A_f=0$):
{\small \begin{eqnarray}\nonumber
\label{eq:43RchiA-CrossSections-via-ScalarProducts-Weak-ChiEta-rel-VmpA-CohCS-spinless}
\dfrac{d\sigma^{\text{total}}_{\text{coh,V$\mp$A},0}}{g_\text{c} \hat c_Ad T_A}&=& 
\cos^2\frac{\theta}{2}\Big[G_V(A,\bm{q})E_\chi \big(E_p\cos^2\frac{\theta}{2}+m\sin^2\frac{\theta}{2}+\frac{|\bm{k}|^2}{E_\chi}\big) \pm G_A(A,\bm{q})E_\chi(E_p -m) \sin^2\frac{\theta}{2}  \Big]^2
\\&+&\cos^2\frac{\theta}{2} \Big[G_V(A,\bm{q})|\bm{k}|(E_p\cos^2\frac{\theta}{2}+m\sin^2\frac{\theta}{2}+E_\chi) 
\pm G_A(A,\bm{q})|\bm{k}|(E_p -m)\sin^2\frac{\theta}{2}\Big]^2
\nonumber
\\&+&m_\chi^2 \sin^2\frac{\theta}{2} \Big[G_V(A,\bm{q}) (E_p\cos^2\frac{\theta}{2}+m\sin^2\frac{\theta}{2})
\mp G_A(A,\bm{q}) (E_p -m)\cos^2\frac{\theta}{2} \Big]^2
.\end{eqnarray}}%
This coherent $V\!\mp\!A$ cross section can be expanded in $\bm{q}$-dependent effective constants
{\small
\begin{eqnarray}
\label{eq:43RchiA-CrossSection-via-ScalarProducts-ChiEta-rel-CohCS-mp=mn-general-spinless-total-viaG}
\dfrac{d\sigma^{\text{total}}_{\text{coh,V$\mp$A},0}}{g_\text{c} d T_A}=
\dfrac{G^2_F  m_A}{4 \pi m^2 |\bm{k}_\chi|^2 } \big[
G_V^2(A,\bm{q})T_V(\theta)+G_A^2(A,\bm{q})T_A(\theta)  \pm 2G_V(A,\bm{q})G_A(A,\bm{q}) T_M(\theta)
\big], \quad 
\end{eqnarray}}%
where the coefficients of this expansion are
{\small \begin{eqnarray*}
T_A&=&(E_p -m)^2\cos^2\frac{\theta}{2}\sin^2\frac{\theta}{2}(2|\bm{k}|^2\sin^2\frac{\theta}{2}+m_\chi^2)
,\\T_M&=&2(E_p -m)\sin^2\frac{\theta}{2}\cos^2\frac{\theta}{2}|\bm{k}|^2 (E_\chi+E_p\cos^2\frac{\theta}{2}+ m\sin^2\frac{\theta}{2})
,\\ T_V&=&(E_p\cos^2\frac{\theta}{2}+m\sin^2\frac{\theta}{2})^2(2 |\bm{k}|^2\cos^2\frac{\theta}{2} +m_\chi^2)+
\cos^2\frac{\theta}{2}|\bm{k}|^2 \big[m_\chi^2+2|\bm{k}|^2+4E_\chi (E_p\cos^2\frac{\theta}{2}+ m \sin^2\frac{\theta}{2})\big]
.\end{eqnarray*}}%
Formula
(\ref{eq:43RchiA-CrossSection-via-ScalarProducts-ChiEta-rel-CohCS-mp=mn-general-spinless-total-viaG})
gives the coherent $\chi A$ cross section ($\Delta A_f=0$) for $V\!\mp\!A$ variants  of the weak interaction of the massive 
$\chi$ lepton with the nucleus, fully summed over the helicities of the final lepton and averaged over the helicities of the initial lepton.
\par
To verify the calculations, let us consider a transition of the general formulas of coherent scattering to the limit 
of $V\!\mp\!A$ currents.
To this end, we substitute 
$V\!\mp\!A$ parameters from
(\ref{eq:43RchiA-CrossSections-via-ScalarProducts-Weak-ChiEta-rel-VmpA-Eff-couplings})
into the general formulas for the cross sections
(\ref{eq:43RchiA-CrossSection-via-ScalarProducts-WeakCohCS-general-mp=mn})
in the following form:
\begin{eqnarray*}\nonumber
G_{\alpha_{\mp}}(A,\bm{q})&=&[+G_V(\bm{q})] ,\qquad  G_{\gamma_{\mp}}(A,\bm{q})=[\bm{\mp} G_V(\bm{q})]
,\\G_{\beta_{\mp}}(A,\bm{q})&=&[\bm{-} G_A(\bm{q})] ,\qquad G_{\delta_{\mp}}(A,\bm{q})=[\bm{\pm} G_A(\bm{q})]
;\\  \Delta G_{\alpha_{\mp}} (A,\bm{q})&=& [\bm{+} \Delta G_V(\bm{q})] ,\quad \Delta G_{\gamma_{\mp}}(A,\bm{q}) =  [\bm{\mp}  \Delta G_V(\bm{q})]
,\\ \Delta G_{\beta_{\mp}}(A,\bm{q})&=&[\bm{-} \Delta  G_A(\bm{q})] ,\quad \Delta G_{\delta_{\mp}}(A,\bm{q})= [\bm{\pm} \Delta  G_A(\bm{q})]
\nonumber .\end{eqnarray*}
The result is {\em all}\/ coherent $\chi A$ cross sections in the $V\!\mp\!A$ approximation
\begin{eqnarray*}\nonumber
\frac{d\sigma^{\mp\mp}_{\text{coh},\bm{V\!-\!A}}}{\cos^2\frac{\theta}{2} g_\text{c}\hat c_A  d T_A}&=& 
\Big|(f_{\alpha+}(\theta) \pm  f_{\gamma+}(\theta)) G_V(\bm{q}) +( f_{\delta-} (\theta) \pm f_{\beta-} (\theta)) G_A(\bm{q}) \\&&+(f_{\gamma-} (\theta)\pm  f_{\alpha-} (\theta)) \Delta G_V(\bm{q}) +(f_{\beta+} (\theta) \pm f_{\delta+} (\theta)) \Delta G_A(\bm{q}) \Big|^2 
, \\  \frac{d\sigma^{\mp\pm}_{\text{coh},\bm{V\!-\!A}}}{\sin^2\frac{\theta}{2} g_\text{c} \hat c_A  d T_A}&=& 
\Big|G_V(\bm{q})  \hat f_{\alpha}(\theta)-G_A(\bm{q})\hat f_{\delta-}(\theta)  -\Delta G_V(\bm{q})  \hat f_{\beta\gamma}(\theta) +(\hat f_{\beta\gamma}(\theta) \mp \hat f_{\delta+}(\theta) ) \Delta  G_A(\bm{q})  \Big|^2
\nonumber ; \\ 
\frac{d\sigma^{\mp\mp}_{\text{coh},\bm{V\!+\!A}}}{\cos^2\frac{\theta}{2} g_\text{c}\hat c_A  d T_A}&=& 
\Big| (f_{\alpha+}(\theta)\mp f_{\gamma+}(\theta)) G_V(\bm{q}) 
\bm{-}(f_{\delta-} (\theta) \mp f_{\beta-} (\theta)) G_A(\bm{q}) 
\\&&-( f_{\gamma-} (\theta) \mp  f_{\alpha-} (\theta)) \Delta G_V(\bm{q})  
+ (f_{\beta+} (\theta) \mp f_{\delta+} (\theta)) \Delta G_A(\bm{q})  \Big|^2 
, \\  \frac{d\sigma^{\mp\pm}_{\text{coh},\bm{V\!+\!A}}}{\sin^2\frac{\theta}{2}g_\text{c} \hat c_A  d T_A}&=& 
\Big|G_V(\bm{q})  \hat f_{\alpha}(\theta) + G_A(\bm{q})\hat f_{\delta-}(\theta)  
+ \Delta G_V(\bm{q})  \hat f_{\beta\gamma}(\theta) 
\bm{+}(\hat f_{\beta\gamma}(\theta) \pm\hat f_{\delta+}(\theta) ) \Delta  G_A(\bm{q}) \Big|^2
\nonumber .\end{eqnarray*}
Since for $f$-form factors from
(\ref{eq:43RchiA-CrossSections-via-ScalarProducts-Weak-ChiEta-rel-CohCS-Fp=Fn-V-A})
for $V\!\mp\!A$ we have
{\small \begin{eqnarray*} 
f_{\alpha+}(\theta)\pm f_{\gamma+}(\theta)\!&=&\!(E_\chi\pm |\bm{k}|)(m\sin^2\frac{\theta}{2} +  E_p\cos^2\frac{\theta}{2} \pm |\bm{k}|) ,\ \ f_{\gamma-}(\theta)\pm f_{\alpha-}(\theta)=(E_\chi\pm |\bm{k}|) |\bm{k}| \sin^2\frac{\theta}{2}  ,\\  f_{\delta-}(\theta)\pm f_{\beta-}(\theta)\!&=& \! (E_\chi\pm |\bm{k}|)(E_p-m)\sin^2\frac{\theta}{2},\quad   f_{\beta+}(\theta)\pm f_{\delta+}(\theta)=(E_\chi\pm |\bm{k}|)(|\bm{k}|\cos^2\frac{\theta}{2}\pm E_p)
,\end{eqnarray*}}%
and according to (\ref{eq:43RchiA-CrossSection-via-ScalarProducts-WeakCohCS-fs})
{\small 
$$\displaystyle 
\hat f_{\alpha}(\theta)= m_\chi (m +(E_p-m) \cos^2\frac{\theta}{2})    
,\ \hat f_{\beta\gamma}(\theta)= m_\chi |\bm{k}| \cos^2\frac{\theta}{2} 
,\  \hat f_{\delta-}(\theta)= m_\chi  (E_p-m) \cos^2\frac{\theta}{2}
,\  \hat f_{\delta+}(\theta)= m_\chi m, $$ }
then {\em the complete set}\/ of coherent $\chi A$ cross sections for $V\!-\!A$- and $V\!+\!A$-weak currents is
\begin{eqnarray}\nonumber
\label{eq:43RchiA-CrossSection-via-ScalarProducts-ChiEta-rel-CohCS-all-V-A-V+A}
\frac{d\sigma^{\mp\mp}_{\text{coh},{V\!-\!A}}}{g_\text{c}\hat c_A  d T_A}&=&\cos^2\frac{\theta}{2} 
(E_\chi\pm |\bm{k}|)^2\Big| (m\sin^2\frac{\theta}{2} +  E_p\cos^2\frac{\theta}{2} \pm |\bm{k}|)G_V(\bm{q}) 
\\\nonumber &&+ (E_p-m)\sin^2\frac{\theta}{2} G_A(\bm{q}) 
+ |\bm{k}| \sin^2\frac{\theta}{2}\Delta G_V(\bm{q}) + (|\bm{k}|\cos^2\frac{\theta}{2}\pm E_p)\Delta G_A(\bm{q}) \Big|^2 
, \\ \nonumber
\frac{d\sigma^{\mp\mp}_{\text{coh},{V\!+\!A}}}{g_\text{c}\hat c_A  d T_A}&=& \cos^2\frac{\theta}{2}
(E_\chi\mp |\bm{k}|)^2\Big| (m\sin^2\frac{\theta}{2} +  E_p\cos^2\frac{\theta}{2} \mp |\bm{k}|)G_V(\bm{q}) 
-(E_p-m)\sin^2\frac{\theta}{2}G_A(\bm{q}) 
\\&&  -  |\bm{k}| \sin^2\frac{\theta}{2}\Delta G_V(\bm{q})  + (|\bm{k}|\cos^2\frac{\theta}{2}\mp E_p) \Delta G_A(\bm{q}) 
\Big|^2 
; \\ \nonumber
\frac{d\sigma^{\mp\pm}_{\text{coh},{V\!-\!A}}}{g_\text{c} \hat c_A  d T_A}&=& \sin^2\frac{\theta}{2}
  m_\chi^2 \Big|m[G_V(\bm{q})\mp\Delta  G_A(\bm{q})]+[G_V(\bm{q})-G_A(\bm{q})](E_p-m) \cos^2\frac{\theta}{2} 
 \\\nonumber &&  +[\Delta  G_A(\bm{q}) -\Delta G_V(\bm{q})]|\bm{k}| \cos^2\frac{\theta}{2} \Big|^2
, \\ \nonumber
\frac{d\sigma^{\mp\pm}_{\text{coh},{V\!+\!A}}}{g_\text{c} \hat c_A  d T_A}&=& \sin^2\frac{\theta}{2}
 m_\chi^2\Big| m [G_V(\bm{q})\pm \Delta  G_A(\bm{q})]  
 +[G_V(\bm{q})  + G_A(\bm{q})]   (E_p-m) \cos^2\frac{\theta}{2}
\\&&+[\Delta G_V(\bm{q})+\Delta  G_A(\bm{q})] |\bm{k}| \cos^2\frac{\theta}{2}  \Big|^2
\nonumber .\end{eqnarray}
The expressions for the cross sections for the spinless nucleus ($\Delta A_f=0$) can be obtained
from (\ref{eq:43RchiA-CrossSection-via-ScalarProducts-ChiEta-rel-CohCS-all-V-A-V+A})
by discarding terms proportional to $\Delta G$.
They can be used, for example, to directly check the expansion 
of the fully averaged coherent
$V\mp A$ cross sections on the  spinless nucleus 
(\ref{eq:43RchiA-CrossSection-via-ScalarProducts-ChiEta-rel-CohCS-mp=mn-general-spinless-total-viaG})
in effective $\bm{q}$-dependent coupling constants.
Indeed, at $\Delta A_f=0$ the $V\!-\!A$ interaction has the following expansions
in effective $\bm{q}$-dependent coupling constants:
\begin{eqnarray*}
\frac{d\sigma^{\mp\mp}_{\text{coh,V-A},0}}{g_\text{c}\hat  c_A  d T_A}&=&\cos^2\frac{\theta}{2} 
(E_\chi\pm |\bm{k}|)^2\Big[ G_V^2(\bm{q})(m\sin^2\frac{\theta}{2} +  E_p\cos^2\frac{\theta}{2}
 \pm |\bm{k}|)^2  + G_A^2(\bm{q}) (E_p-m)^2\sin^4\frac{\theta}{2}  
\\&&   +2G_V(\bm{q})G_A(\bm{q}) (m\sin^2\frac{\theta}{2} +  E_p\cos^2\frac{\theta}{2} \pm |\bm{k}|)
(E_p-m)\sin^2\frac{\theta}{2}\Big]
,\\  \frac{d\sigma^{\mp\pm}_{\text{coh,V-A},0}}{g_\text{c} \hat c_A d T_A}&=&\sin^2\frac{\theta}{2} m_\chi^2  \Big[
G_V^2(\bm{q}) (m\sin^2\frac{\theta}{2}+E_p\cos^2\frac{\theta}{2})^2+G_A^2(\bm{q})(E_p-m)^2 \cos^4\frac{\theta}{2} \\&&-2 G_V(\bm{q})G_A(\bm{q})(m\sin^2\frac{\theta}{2}+E_p\cos^2\frac{\theta}{2})(E_p-m) \cos^2\frac{\theta}{2}\Big]
.\end{eqnarray*}
Then the averaged coherent "$V\!-\!A$"\/ cross section, which is the sum (divided by 2)
of the above-mentioned four terms  can be written as
\begin{eqnarray*}
\dfrac{d\sigma^{\text{total}}_{\text{coh,V-A},0}}{g_\text{c}\hat c_A d T_A}&\equiv&
\frac12 \sum^{}_{s's=\pm} \frac{d\sigma^{s's}_{\text{coh,V-A},0}}{ g_\text{c} \hat c_A d T_A}
\equiv G_V^2(\bm{q})C_V(\theta)+G_A^2(\bm{q})C_A(\theta)+2G_V(\bm{q})G_A(\bm{q}) C_M(\theta)
.\end{eqnarray*}
For the averaged coherent $V\!+\!A$ cross section, exactly the same expansion is obtained
but with the minus sign  in front of the interference term (see formula
(\ref{eq:43RchiA-CrossSection-via-ScalarProducts-ChiEta-rel-CohCS-mp=mn-general-spinless-total-viaG})).
As a result, with (\ref{eq:41chiA-CrossSection-Coh-vs-InCoh-GeneralFactor-with-G2_F}), 
the averaged coherent $V\!\mp\!A$ cross section on the spinless nucleus takes the form
\begin{eqnarray}\label{eq:43RchiA-CrossSection-via-ScalarProducts-CohCS-VmpA-total-viaG}
\dfrac{d\sigma^{\text{total}}_{\text{coh,V$\mp$A},0}}{g_\text{c}d T_A}&=&
\dfrac{G^2_F  m_A}{4 \pi } 
\Big\{G_V^2(\bm{q})\dfrac{C_V(\theta)}{|\bm{k}|^2 s } 
+G_A^2(\bm{q})\dfrac{C_A(\theta)}{|\bm{k}|^2 s}
\pm 2G_V(\bm{q})G_A(\bm{q}) \dfrac{C_M(\theta)}{|\bm{k}|^2 s} \Big\}.\qquad 
\end{eqnarray}
Direct calculations give expressions for the parameters $C_A(\theta)$, $C_V(\theta)$ and $C_M(\theta)$
coinciding with $T(\theta)$-parameters from
(\ref{eq:43RchiA-CrossSection-via-ScalarProducts-ChiEta-rel-CohCS-mp=mn-general-spinless-total-viaG}),
although they were obtained in completely different ways.
\par
Finally, in the {\em massless}\/ $\chi$-lepton limit, when
$m_\chi=0,\ \dfrac{E_p}{\sqrt{s}}=\dfrac{s+m^2}{2{s}} ,\ \dfrac{|\bm{k}|}{ \sqrt{s}}= \dfrac{E_\chi}{\sqrt{s}}=\dfrac{s-m^2}{2{s}}, $
$C_{V,A,M}$-parameters from
(\ref{eq:43RchiA-CrossSection-via-ScalarProducts-CohCS-VmpA-total-viaG})
contain only one term each, which corresponds to the conservation of helicity of the 
neutrino ($V\!-\!A$) or antineutrino ($V\!+\!A$). They are
\begin{eqnarray*}
\dfrac{C_A(\theta)}{|\bm{k}|^2 s} &=&\frac{1}{2}\cos^2\frac{\theta}{2}\hat C_A(\theta)_0=
\frac{1}{2}\cos^2\frac{\theta}{2}\Big(1- \frac{m}{ \sqrt{s}}\Big)^4\sin^4\frac{\theta}{2} 
,\\ \dfrac{C_M(\theta)}{|\bm{k}|^2 s }&=&\frac12 \cos^2\frac{\theta}{2}\hat C_M(\theta)_0=
\frac12 \cos^2\frac{\theta}{2}\Big(1-\frac{m}{ \sqrt{s}}\Big)^2\sin^2\frac{\theta}{2} 
\Big[2-\sin^2\frac{\theta}{2}\Big(1-\frac{m}{ \sqrt{s}}\Big)^2\Big]
,\\ \dfrac{C_V(\theta)}{|\bm{k}|^2 s }&=&\frac12\cos^2\frac{\theta}{2}\hat C_V(\theta)_0
=\frac12\cos^2\frac{\theta}{2}\Big[2-\sin^2\frac{\theta}{2}\Big(1-\frac{m}{ \sqrt{s}}\Big)^2\Big]^2
.\end{eqnarray*}
Then for the spinless nucleus and $m_\chi=0$,  the total coherent cross section
(\ref{eq:43RchiA-CrossSection-via-ScalarProducts-CohCS-VmpA-total-viaG})
can be written as
\begin{eqnarray}
\label{eq:43RchiA-CrossSection-via-ScalarProducts-ChiEta-rel-CohCS-mp=mn-general-spinless-total-viaG-mchi=0}
\dfrac{d\sigma^{\text{total}}_{\text{coh,V$\mp$A},00}}{g_\text{c} d T_A}=
\dfrac{G^2_F  m_A}{8 \pi}\cos^2\frac{\theta}{2}\Big\{G_V^2(\bm{q})\hat C_V(\theta)_0 +G_A^2(\bm{q})\hat  C_A(\theta)_0\pm 2G_V(\bm{q})G_A(\bm{q}) \hat C_M(\theta)_0 \Big\}. \qquad
\end{eqnarray}
In fact, for the massless lepton, the  "completely averaged"\/
coherent cross section contains only one term
from each formula (\ref{eq:43RchiA-CrossSections-via-ScalarProducts-ChiEta-rel-CohCS-Fp=Fn-VpmA-compact}).
For $V\!-\!A$ current, it is {\em half}\/ the cross section with "$--$"\/-helicity ("$-$"\/-superscripts), 
for $ V\!+\!A$ current, it is {\em half}\/ the cross section with "$++$"\/-helicity ("$+$"\/-subscripts) from (\ref{eq:43RchiA-CrossSections-via-ScalarProducts-ChiEta-rel-CohCS-Fp=Fn-VpmA-compact}):
\begin{eqnarray*}
\dfrac{d\sigma^{\text{total}}_{\text{coh,V$\mp$A},00}}{g_\text{c} d T_A}\equiv
\dfrac12\sum^{}_{s's} \frac{d\sigma^{s's}_\text{coh}}{ g_\text{c}  d T_A}  = \frac12 \frac{d\sigma^{\mp\mp}_{\text{coh},{V\!\mp\!A},00}}{g_\text{c}d T_A}=
\frac12 \cos^2\frac{\theta}{2} \dfrac{G^2_F  m_A}{\pi} 
\Big[ G_V(\bm{q}) {T_{V,0}(\theta)} \pm G_A(\bm{q}) {T_{A,0}} \Big]^2
.\end{eqnarray*}
Parameters $T_{V,0}(\theta)$ and $T_{A,0}(\theta)$ for $m_\chi=0$ take the form
\begin{eqnarray*}
T_{V,0}(\theta)&\equiv&\dfrac{m}{\sqrt{s}}\sin^2\frac{\theta}{2} +\dfrac{E_p}{\sqrt{s}}\cos^2\frac{\theta}{2}+\dfrac{|\bm{k}|}{\sqrt{s}}  \to 
1-\Big(1-\dfrac{m}{\sqrt{s}}\Big)^2\frac{\sin^2\frac{\theta}{2}}{2}
,\\  T_{A,0}(\theta)&\equiv& \Big(\dfrac{E_p}{\sqrt{s}}-\dfrac{m}{\sqrt{s}}\Big)\sin^2\frac{\theta}{2}  \to 
\Big(1-\dfrac{m}{\sqrt{s}}\Big)^2\frac{\sin^2\frac{\theta}{2}}{2} 
.\end{eqnarray*}
The parameters $T_{V,0}$ and $T_{A,0}$ completely correspond to the above expression
(\ref{eq:43RchiA-CrossSection-via-ScalarProducts-ChiEta-CohCS-neutrino})
for the coherent cross section of the interaction of the massless neutrino with the nucleus from
\cite{Bednyakov:2021ppn}.
The latter means that the averaged coherent cross section of the $V\!-\!A$ 
interaction of the massless left-handed helicity lepton with the spinless nucleus
(\ref{eq:43RchiA-CrossSection-via-ScalarProducts-ChiEta-rel-CohCS-mp=mn-general-spinless-total-viaG-mchi=0})
is exactly half of the purely "neutrino cross section"\/
(\ref{eq:43RchiA-CrossSection-via-ScalarProducts-ChiEta-CohCS-neutrino}).

\paragraph{Incoherent ${\chi A}$-sections for $\bm{V\!\mp\!A}$ weak currents.}
Recall that the nucleon constants in the case of $V\!\mp\! A$ weak currents have the form
(\ref{eq:43RchiA-CrossSections-via-ScalarProducts-Weak-ChiEta-rel-VmpA-couplings})
$$\alpha_{V\!\mp\!A} \equiv \alpha_{\mp} = + g_V ,\ \beta_{\mp} = - g_A, \ \gamma_{\mp} = \mp g_V ,\ \delta_{\mp} = \pm g_A.
$$
Then the general expansions of the incoherent $\chi A$ cross sections 
(\ref{eq:43RchiA-CrossSection-via-ScalarProducts-InCoh-all-viaFG})
go into the following "$V\!\mp\!A$"\/ expressions for {\em scattering without lepton spin flip}\/:
\begin{eqnarray}\nonumber
\label{eq:43RchiA-CrossSection-via-ScalarProducts-ChiEta-rel-InCohCS-VpmA}
\frac{d\sigma^{\mp\mp}_{\text{inc,V-A}}}{g_\text{i}d T_A} &=&\frac{\hat c_A}{2}  \sum^{}_{f=p,n} A_f \widehat F_f^2 
\Big[  g_V^2 \widetilde V^+_{\pm}(\theta)+ g_A^2 \widetilde A^+_{\pm}(\theta)+ g_V g_A\widetilde M^+_{\pm}(\theta) 
\\  && \qquad \pm \dfrac{\Delta A_f}{A_f}\big\{ g_V g_A\widetilde M^-_{\pm}(\theta)- g_V^2\widetilde V^-_{\pm}(\theta)- g_A^2\widetilde A^-_{\pm}(\theta)\big\}\Big]
,\\ \nonumber
\frac{d\sigma^{\mp\mp}_{\text{inc,V+A}}}{g_\text{i} \hat c_A d T_A} &=&\frac{\hat c_A}{2}  \sum^{}_{f=p,n} A_f \widehat F_f^2 \Big[ 
g_V^2\widetilde V^+_{\mp}(\theta) +g_A^2\widetilde A^+_{\mp}(\theta) - g_Vg_A\widetilde M^+_{\mp}(\theta) 
\\ &&\qquad \mp \dfrac{\Delta A_f}{A_f}\big\{ g_V g_A\widetilde M^-_{\mp}(\theta)+ g_V^2\widetilde V^-_{\mp}(\theta)+ g_A^2\widetilde A^-_{\mp}(\theta)\big\}\Big]
\nonumber.\end{eqnarray}
In (\ref{eq:43RchiA-CrossSection-via-ScalarProducts-ChiEta-rel-InCohCS-VpmA}),  the following notations are introduced:
{\small \begin{eqnarray}\nonumber
\label{eq:43RchiA-CrossSection-via-ScalarProducts-ChiEta-rel-InCohCS-VpmA-Coefficients} 
\frac{\widetilde V^-_{\pm}(\theta)}{2}&=&\frac{F^-_{\alpha^2}(\theta)}{2}+\frac{F^-_{\gamma^2}(\theta)}{2}\pm \frac{F^-_{\alpha\gamma}(\theta)}{2} = -2(E_\chi \pm |\bm{k}|)^2 (|\bm{k}| \pm E_p \cos^2\frac{\theta}{2})|\bm{k}|\sin^2\frac{\theta}{2}
,\\ \nonumber
 \frac{\widetilde A^-_{\pm}(\theta)}{2}&=&\frac{F^-_{\beta^2}(\theta)}{2}+\frac{F^-_{\delta^2}(\theta)}{2}\pm\frac{F^-_{\beta\delta}(\theta)}{2}=  -2(E_\chi \pm |\bm{k}|)^2(E_p  \pm |\bm{k}|\cos^2\frac{\theta}{2}) E_p\sin^2\frac{\theta}{2}
,\\ \nonumber
\frac{\widetilde M^-_{\pm}(\theta)}{4}&=&\frac{F^-_{\alpha\delta}(\theta)}{4}+\frac{F^-_{\beta\gamma}(\theta)}{4} \pm \Big[\frac{F^-_{\alpha\beta}(\theta)}{4}+ \frac{F^-_{\gamma\delta}(\theta)}{4}\Big] 
=(E_\chi\pm |\bm{k}|)^2 \big[(E_p\pm |\bm{k}|)^2 \cos^2\frac{\theta}{2}\pm 2 E_p |\bm{k}| \sin^4\frac{\theta}{2}\big]
,\\  \frac{\widetilde M^+_{\pm}(\theta)}{4}&=&\frac{F^+_{\alpha\delta}(\theta)}{4}+\frac{F^+_{\beta\gamma}(\theta)}{4}\pm \Big[\frac{ F^+_{\alpha\beta}(\theta)}{4}+ \frac{F^+_{\gamma\delta}(\theta)}{4}\Big] 
=2 (E_\chi\pm |\bm{k}|)^2 (|\bm{k}|\cos^2\frac{\theta}{2} \pm E_p) |\bm{k}|\sin^2\frac{\theta}{2}
,\\ \nonumber
\frac{\widetilde A^+_{\pm}(\theta)}{2}&=&\frac{F^+_{\beta^2}(\theta)}{2}+\frac{F^+_{\delta^2}(\theta)}{2}\pm\frac{F^+_{\beta\delta}(\theta)}{2}=(E_\chi \pm |\bm{k}|)^2 [(E_p\pm |\bm{k}|)^2 \cos^2\frac{\theta}{2} +2 (E_p^2 - |\bm{k}|^2  \cos^2\frac{\theta}{2}) \sin^2\frac{\theta}{2}]  
,\\ \nonumber
\frac{\widetilde V^+_{\pm}(\theta)}{2}&=&\frac{F^+_{\alpha^2}(\theta)}{2}+\frac{F^+_{\gamma^2}(\theta)}{2}\pm \frac{F^+_{\alpha\gamma}(\theta)}{2}  =(E_\chi \pm |\bm{k}|)^2 \big[2|\bm{k}|^2  \sin^4\frac{\theta}{2} + (E_p  \pm |\bm{k}|)^2 \cos^2\frac{\theta}{2}\big]
.\end{eqnarray}}%
Form-factor expressions
(\ref{eq:43RchiA-CrossSection-via-ScalarProducts-ChiEta-rel-InCohCS-VpmA-Coefficients})
completely determine the incoherent $\chi A$ cross sections {\em without spin flip}\/ of the massive $\chi$ lepton
(\ref{eq:43RchiA-CrossSection-via-ScalarProducts-ChiEta-rel-InCohCS-VpmA})
in the case of $V\!\mp\!A$ interaction.
It can be seen that all parameters with the subscript "$+$"\/ contain the common factor
$ (E_\chi + |\bm{k}|)^2$, and all parameters with thee subscript "$-$"\/ are proportional to the
 common factor $(E_\chi - |\bm{k}|)^2$, which makes the cross section vanish 
in the relativistic limit when $E_\chi \simeq |\bm{k}|$.
Since 
$$2^4c_A (E_\chi \pm |\bm{k}|)^2 
= \dfrac{G^2_F  m_A}{4 \pi } \dfrac{(E_\chi \pm |\bm{k}|)^2}{m^2 |\bm{k}_\chi|^2} 
= \dfrac{G^2_F  m_A }{4 \pi}  \dfrac{\varkappa_{\pm}(s,m^2_\chi) }{ s }
,$$
where, by analogy with 
(\ref{eq:43RchiA-CrossSections-via-ScalarProducts-Weak-ChiEta-rel-CohCS-Fp=Fn-Nu-massive-kappa}), 
the following notation is introduced:
\begin{equation}
\label{eq:43RchiA-CrossSections-via-ScalarProducts-Weak-ChiEta-rel-kappaPM}
\varkappa(s,m^2_\chi)_{\pm}\equiv \varkappa(s, m^2, m^2_\chi)_{\pm} = 
\Big[1\pm \dfrac{s+m^2_\chi-m^2}{\lambda(s, m^2, m^2_\chi)}\Big]^2
, \end{equation} 
the set of incoherent "$V\!\mp\!A$"\/  cross sections without spin flip of the massive $\chi$ lepton
can {\em finally} be written  as follows:
\begin{eqnarray}\nonumber
\label{eq:43RchiA-CrossSection-via-ScalarProducts-ChiEta-rel-InCohCS-VpmA-fino} 
\frac{d\sigma^{\mp\mp}_{\text{inc,V-A}}}{g_\text{i}d T_A} &=&
\dfrac{G^2_F  m_A }{4 \pi} \varkappa_{\pm}(s,m^2_\chi) \sum A_f \widehat F_f^2 \Big[ 
   g_V^2 {V}^+_{\pm}(\theta)+ g_A^2 {A}^+_{\pm}(\theta)+ g_V g_A {M}^+_{\pm}(\theta)
\\&&\qquad\qquad
\pm \dfrac{\Delta A_f}{A_f}\big\{ 
g_V g_A {M}^-_{\pm}(\theta)- g_V^2 {V}^-_{\pm}(\theta)- g_A^2 {A}^-_{\pm}(\theta) \big\}\Big]
,\\ \nonumber
\frac{d\sigma^{\mp\mp}_{\text{inc,V+A}}}{g_\text{i}d T_A} &=&
\dfrac{G^2_F  m_A }{4 \pi} \varkappa_{\mp}(s,m^2_\chi)\sum A_f \widehat F_f^2 \Big[ 
g_V^2 {V}^+_{\mp}(\theta)+g_A^2 {A}^+_{\mp}(\theta) - g_Vg_A {M}^+_{\mp}(\theta)
\\&&\qquad\qquad
\mp \dfrac{\Delta A_f}{A_f} \big\{ 
g_V g_A {M}^-_{\mp}(\theta) + g_V^2 {V}^-_{\mp}(\theta)+ g_A^2 {A}^-_{\mp}(\theta) \big\}\Big]
\nonumber.\end{eqnarray}
In formulas (\ref{eq:43RchiA-CrossSection-via-ScalarProducts-ChiEta-rel-InCohCS-VpmA-fino})
parameters (\ref{eq:43RchiA-CrossSection-via-ScalarProducts-ChiEta-rel-InCohCS-VpmA-Coefficients})
are redefined as dimensionless quantities
\begin{eqnarray}\nonumber
\label{eq:43RchiA-CrossSection-via-ScalarProducts-ChiEta-rel-InCohCS-VpmA-Coefficients-dimless} 
{V}^-_{\pm}(\theta) \equiv \frac{\widetilde{V}^-_{\pm}(\theta)}{2 s (E_\chi \pm |\bm{k}|)^2}&=&- \frac{|\bm{k}|}{\sqrt{s}}
 \Big(\frac{|\bm{k}|}{\sqrt{s}} \pm \frac{E_p}{\sqrt{s}}\cos^2\frac{\theta}{2}\Big)\sin^2\frac{\theta}{2}
,\\ \nonumber
{A}^-_{\pm}(\theta)=\frac{\widetilde{A}^-_{\pm}(\theta)}{2s (E_\chi \pm |\bm{k}|)^2}&=&   - 2\Big(\frac{E_p^2}{s}  \pm  \frac{|\bm{k}|}{\sqrt{s}}\frac{E_p }{\sqrt{s}}\cos^2\frac{\theta}{2}\Big) \sin^2\frac{\theta}{2}
,\\ \nonumber
{M}^-_{\pm}(\theta)\equiv\frac{\widetilde{M}^-_{\pm}(\theta)}{2s (E_\chi\pm |\bm{k}|)^2 }&=&
 2\frac{(E_p\pm |\bm{k}|)^2}{s} \cos^2\frac{\theta}{2}\pm 4  \frac{|\bm{k}|}{\sqrt{s}}\frac{E_p }{\sqrt{s}} \sin^4\frac{\theta}{2}
,\\  {M}^+_{\pm}(\theta)\equiv\frac{\widetilde{M}^+_{\pm}(\theta)}{2s (E_\chi\pm |\bm{k}|)^2 }&=&
4\frac{|\bm{k}|}{\sqrt{s}} \Big(\frac{|\bm{k}|}{\sqrt{s}} \cos^2\frac{\theta}{2} \pm  \frac{E_p }{\sqrt{s}} \Big)\sin^2\frac{\theta}{2}
,\\ \nonumber
{A}^+_{\pm}(\theta)\equiv\frac{\widetilde{A}^+_{\pm}(\theta)}{2s(E_\chi \pm |\bm{k}|)^2 }&=&
\frac{(E_p\pm |\bm{k}|)^2}{s} \cos^2\frac{\theta}{2}
+ 2 \Big( \frac{E_p^2}{s}  -  \frac{|\bm{k}|^2}{s}  \cos^2\frac{\theta}{2}\Big) \sin^2\frac{\theta}{2}
,\\ \nonumber
{V}^+_{\pm}(\theta)\equiv\frac{\widetilde{V}^+_{\pm}(\theta)}{2 s(E_\chi \pm |\bm{k}|)^2}&=&
2 \frac{|\bm{k}|^2}{s} \sin^4\frac{\theta}{2} + \frac{(E_p  \pm |\bm{k}|)^2 }{s}\cos^2\frac{\theta}{2}
.\end{eqnarray}
Formulas (\ref{eq:43RchiA-CrossSection-via-ScalarProducts-ChiEta-rel-InCohCS-VpmA-fino})
give the incoherent cross sections for massive $\chi$-lepton scattering on the nucleus {\em without spin flip}\/
simultaneously due to $V$--$A$ and $V$+$A$ currents.
Note that  in the first case the cross section is for the lepton with positive helicity (analogue antineutrino),
and in the second case it is  for the lepton with negative helicity (analogue neutrino).
\par
If one 
substitutes parameters from
(\ref{eq:43RchiA-CrossSection-via-ScalarProducts-ChiEta-rel-InCohCS-VpmA-Coefficients-dimless})
in formulas (\ref{eq:43RchiA-CrossSection-via-ScalarProducts-ChiEta-rel-InCohCS-VpmA-fino})
and introduces another notation
$$\xi=\dfrac{|\bm{k}|}{E_p} = \dfrac{\lambda(s, m^2, m^2_\chi)}{s-m^2_\chi+m^2},
$$ 
then one can obtain "more explicit"\/ expressions for "$V\!\mp\!A$"\/ incoherent $\chi A$ cross sections
in the following form:
\begin{eqnarray}\nonumber
\label{eq:43RchiA-CrossSections-via-ScalarProducts-ChiEta-rel-InCohCS-VpmA-spinfixed-fino} 
 \frac{d\sigma^{\mp\mp}_{\text{inc,V-A}}}{g_\text{i} d T_A} &=&\dfrac{G^2_F  m_A }{4 \pi} 
 \dfrac{E_p^2}{{s}} \varkappa_{\pm}(s,m^2_\chi) \sum A_f \widehat F_f^2 \Big[ (g_V^2 + g_A^2) (1\pm \xi )^2 \cos^2\frac{\theta}{2} + 2 (g_V \xi \pm  g_A )^2\sin^2\frac{\theta}{2}
\\&&\nonumber - 2 (g_V-g_A)^2 \xi^2 \cos^2\frac{\theta}{2}\sin^2\frac{\theta}{2} \pm\dfrac{\Delta A_f}{A_f}\big\{2g_V g_A (1\pm \xi )^2\cos^2\frac{\theta}{2} + 2(g_V \xi \pm  g_A )^2\sin^2\frac{\theta}{2} 
\\&& \pm 2(g_V - g_A)^2 \xi \cos^2\frac{\theta}{2}\sin^2\frac{\theta}{2} -g_V^2 \xi \sin^2\frac{\theta}{2} (\xi  \pm \cos^2\frac{\theta}{2}) \big\} \Big]
,\\ \nonumber
\frac{d\sigma^{\mp\mp}_{\text{inc,V+A}}}{g_\text{i} d T_A} &=&\dfrac{G^2_F  m_A }{4 \pi} 
\dfrac{E_p^2}{{s}}  \varkappa_{\mp}(s,m^2_\chi)\sum A_f \widehat F_f^2  \Big[
(g_V^2 + g_A^2)(1\mp \xi)^2\cos^2\frac{\theta}{2} +2 \big(g_V \xi \pm g_A \big)^2 \sin^2\frac{\theta}{2}
\\&&\nonumber 
 -2 (g_V+ g_A)^2 \xi^2 \cos^2\frac{\theta}{2} \sin^2\frac{\theta}{2} 
\mp  \dfrac{\Delta A_f}{A_f}\big\{ 2g_Vg_A(1\mp \xi)^2\cos^2\frac{\theta}{2}
-2 (g_V \xi \pm g_A )^2\sin^2\frac{\theta}{2} 
\\&&\nonumber  \pm 2(g_V+ g_A)^2 \xi\cos^2\frac{\theta}{2}\sin^2\frac{\theta}{2} +g_V^2 \xi \sin^2\frac{\theta}{2} (\xi \mp \cos^2\frac{\theta}{2}) \big\} \Big]
\nonumber.\end{eqnarray}
This completes the discussion of presentations of the incoherent cross sections for 
the weak $V\!\mp\!A$ interaction 
of the massive $\chi$ lepton with the nucleus without lepton spin flip.
\par
In the massless limit, when
\begin{eqnarray*}
 m_\chi =0, \quad  \dfrac{E_p}{\sqrt{s}}=\dfrac{s+m^2}{2{s}} 
, \quad \xi = \dfrac{s-m^2}{s+m^2} ,\quad \varkappa(s,0)_{+}  = 4, \quad \varkappa(s,0)_{-} = 0
, \end{eqnarray*}
cross sections (\ref{eq:43RchiA-CrossSections-via-ScalarProducts-ChiEta-rel-InCohCS-VpmA-spinfixed-fino})
for leptons with "wrong helicity"\/
become zero
$$\frac{d\sigma^{++}_{\text{inc,V-A},0}}{d T_A} \propto \varkappa_{-}(s,0) = 0, \qquad
\frac{d\sigma^{--}_{\text{inc,V+A},0}}{d T_A} \propto  \varkappa_{-}(s,0) =0. 
$$
For massless leptons with "correct helicity"\/,  formulas
(\ref{eq:43RchiA-CrossSections-via-ScalarProducts-ChiEta-rel-InCohCS-VpmA-spinfixed-fino}) turn into
\begin{eqnarray}\nonumber
\label{eq:43RchiA-CrossSections-via-ScalarProducts-ChiEta-rel-InCohCS-VpmA-spinfixed-fino-m_chi=0} 
 \frac{d\sigma^{\mp\mp}_{\text{inc,V$\mp$A},0}}{g_\text{i}d T_A} &=&\dfrac{G^2_F  m_A }{4 \pi} 
 \dfrac{4 E_p^2}{{s}} \sum A_f \widehat F_f^2 \Big[ (g_V^2  +  g_A^2) (1+ \xi )^2 \cos^2\frac{\theta}{2}
+ 2 (g_V\xi  \pm g_A )^2\sin^2\frac{\theta}{2}
\\&& - 2 (g_V \mp g_A)^2 \xi^2 \cos^2\frac{\theta}{2}\sin^2\frac{\theta}{2}
 +\dfrac{\Delta A_f}{A_f}\Big\{2g_V g_A (1+ \xi )^2\cos^2\frac{\theta}{2} 
\\&& \nonumber
 \pm 2(g_V \xi \pm  g_A )^2\sin^2\frac{\theta}{2} 
 \pm 2(g_V \mp g_A)^2 \xi \cos^2\frac{\theta}{2}\sin^2\frac{\theta}{2} 
\mp g_V^2 \xi \sin^2\frac{\theta}{2} (\xi + \cos^2\frac{\theta}{2}) \Big\} \Big]
\nonumber.\end{eqnarray}
This formula corresponds to the incoherent scattering of neutrinos and antineutrinos on the nucleus,
which was discussed in \cite{Bednyakov:2021ppn}.
It is a direct consequence of the general expansion in chiral constants
(\ref{eq:43RchiA-CrossSection-via-ScalarProducts-ChiEta-rel-InCohCS-VpmA-fino}), whose 
"$V$--$A$"\/ formula for a massless neutrino takes the form
\footnote{Here, the $f$-index is omitted in $g_V^f$ and $g_A^f$ and, as in \cite{Bednyakov:2021ppn}, it is assumed that $g_\text{i}\simeq 1$ and $y\equiv \sin^2\dfrac{\theta}{2}\Big(1-\dfrac{m^2}{s}\Big)$.} 
matching the following expression from \cite{Bednyakov:2021ppn}:
\par
{\small 
\begin{eqnarray}\nonumber
\label{eq:43RchiA-CrossSection-via-ScalarProducts-ChiEta-rel-V-A-CS-m_chi=0}
\frac{d\sigma^{--}_{\text{inc,V-A},0}}{d T_A}&=&\!\! \nonumber \dfrac{G^2_Fm_A }{2\pi}\sum_{f=p,n} A_f \widehat F_f^2 \Bigg[ g_A^2\Big[2\Big(1-\dfrac{ys}{s-m^2}\Big) +y^2+y\dfrac{4m^2}{s-m^2}\Big] 
+2  g_V g_A y (2-y) 
\\&&\nonumber + g_V^2 \Big[2 \Big(1-\dfrac{ys}{s-m^2}\Big) +y^2\Big] +\dfrac{2 \Delta A_f}{A_f} 
\Big\{ g_V^2 y\Big(1 - \frac{y}{2}\frac{s+m^2}{s-m^2}\Big) +g_A^2  y \Big (1-\frac{y}{2}\Big) \frac{s+m^2}{s-m^2}
\\&& + g_V g_A \Big[2\Big(1-\frac{y s}{s-m^2}\Big)+y^2 \frac{s+m^2}{s-m^2}\Big]\Big\} \Bigg]
. \end{eqnarray}}%
Finally, let us see what the incoherent $\chi A$ cross section 
(\ref{eq:43RchiA-CrossSections-via-ScalarProducts-ChiEta-rel-InCohCS-viaPhi+Lorentz-spinless-mchi=0}) 
corresponding to the scattering of the massless lepton on the spinless nucleus in the $V\!\pm\!A$ approximation
passes into when
\begin{eqnarray*}
(\alpha\mp\gamma)^{}_{V\!\mp\!A}&=& 2 g_V ,\quad (\beta\mp\delta )^{}_{V\!\mp\!A} = - 2g_A
,\quad (\alpha\mp\gamma)^{}_{V\!\pm\!A} = (\delta\mp\beta)^{}_{V\!\pm \!A} = 0
.\end{eqnarray*}
Substituting these relations into formula (\ref{eq:43RchiA-CrossSections-via-ScalarProducts-ChiEta-rel-InCohCS-viaPhi+Lorentz-spinless-mchi=0}) and taking into account 
(\ref{eq:ScalarProducts-Lab-k^2-to-k^2_lab}), 
one obtains the consequence of formula
(\ref{eq:43RchiA-CrossSections-via-ScalarProducts-ChiEta-rel-InCohCS-viaPhi+Lorentz-spinless-mchi=0}) as follows:
\begin{eqnarray*}\nonumber
\frac{d\sigma^{\mp\mp}_{\text{inc,V$\mp$A},00}}{d T_A} &=& 
\dfrac{G^2_F  m_A}{2\pi}\sum_{f=p,n} \widehat F_f^2(\bm{q}) A_f
\Big[g_A^2\Big(2 +y^2+2y\dfrac{2m^2-s}{s-m^2}\Big)\pm 2g_V g_A y(2-{y})
\\&&  +g_V^2\Big(2+y^2-\dfrac{2ys}{s-m^2}\Big)\Big]
; \\ \frac{d\sigma^{\mp\mp}_{\text{inc,V$\pm$A},00}}{d T_A} &=& 0
.\end{eqnarray*} 
As expected, the first formula, which is 
the incoherent cross section for scattering of the 
massless neutrino (with negative helicity) and massless antineutrino (with positive helicity) on the nucleus, 
coincides (in its spinless part) with the "neutrino formula"\/ (\ref{eq:43RchiA-CrossSection-via-ScalarProducts-ChiEta-rel-V-A-CS-m_chi=0}). 
The second formula, which is the cross section of the massless neutrino scattering  via the $V\!+\!A$ current 
and massless antineutrino scattering via the $V\!-\!A$ current, is equal to zero.

\paragraph{$\bm{V\!\mp\!A}$-interaction with a spin flip of the massive lepton.}
In this case, general expansions of incoherent $\chi A$ cross sections
(\ref{eq:43RchiA-CrossSection-via-ScalarProducts-InCoh-all-viaFG})
go into the following expressions:
\begin{eqnarray}\nonumber 
\label{eq:43RchiA-CrossSectionsEvaluations-Weak-ChiEta-rel-InCoh-VpmA-spinflip} 
\frac{d\sigma^{\mp\pm}_{\text{inc,V-A}}}{g_\text{i} \hat c_A d T_A} &=&\frac12
\sum^{}_{f=p,n} A_f \widehat F_f^2 \Big[g_V^2 \widehat{V}^+_{}(\theta)+g_A^2 \widehat{A}^+_{\mp}(\theta) 
+ g_Vg_A \widehat M^+_{\pm}(\theta)
\\&&\qquad  +\dfrac{ \Delta A_f}{A_f}\big\{g_Vg_A \widehat M^-_{\pm}(\theta)
-g_V^2 \widehat{V}^-_{}(\theta) -g_A^2 \widehat A^-_{\mp}(\theta)\big \}\Big]
,\\ \nonumber 
\frac{d\sigma^{\mp\pm}_{\text{inc,V+A}}}{g_\text{i} \hat c_A d T_A} &=&\frac12
\sum^{}_{f=p,n} A_f \widehat F_f^2  \Big[
  g_V^2 \widehat{V}^+_{}(\theta)+g_A^2 \widehat A^+_{\pm}(\theta)- g_V g_A \widehat M^+_{\mp}(\theta) 
\\&&\qquad +\dfrac{ \Delta A_f}{A_f}\big \{ g_Vg_A  \widehat M^-_{\mp}(\theta) 
+ g_V^2 \widehat{V}^-_{}(\theta) + g_A^2 \widehat A^-_{\pm}(\theta)\big \}\Big]\nonumber
.\end{eqnarray}
Here, all form-factor combinations below, which define defining incoherent $\chi A$ cross sections in the case of 
$V\!\mp\!A$ interaction {\em with the change of the projection of the massive lepton spin}\/,
depend on $G$-form-factors from (\ref{eq:43RchiA-CrossSection-via-ScalarProducts-InCoh-Gs}) and
are proportional to the common factor $2 m_\chi ^2$:
{\small
\begin{eqnarray*}
\frac{\widehat  V^-_{}(\theta)}{2 m_\chi ^2}&=&\frac{G^-_{\alpha\gamma}(\theta)}{2 m_\chi ^2}=
2 |\bm{k}|[m + 2 (E_p-m) \cos^2\frac{\theta}{2}]\cos^2\frac{\theta}{2}\sin^2\frac{\theta}{2}
,\\ \frac{\widehat V^+_{}(\theta)}{2 m^2_\chi}&=&\frac{G^+_{\alpha^2}(\theta)}{2 m^2_\chi}+\frac{G^+_{\beta\gamma}(\theta)}{2 m^2_\chi}=m^2\sin^2\frac{\theta}{2} +|\bm{k}|^2\cos^2\frac{\theta}{2}
= E^2_p \cos^2\frac{\theta}{2} +m^2(\sin^2\frac{\theta}{2}- \cos^2\frac{\theta}{2})
,\\ \frac{\widehat A^-_{\pm}(\theta)}{2 m^2_\chi }&=&\frac{ G^-_{\beta\delta}(\theta)}{2 m^2_\chi }\pm \frac{G^-_{\delta^2}(\theta)}{2 m^2_\chi }
= 2\cos^2\frac{\theta}{2}\{ |\bm{k}|[m + 2(E_p-m)\sin^2\frac{\theta}{2}]\cos^2\frac{\theta}{2}\mp  m^2\}
,\\ \frac{\widehat M^-_{\pm}(\theta)}{2 m^2_\chi }&=&\frac{G^-_{\alpha\gamma}(\theta)}{2 m^2_\chi }+\frac{G^-_{\beta\delta}(\theta)}{2 m^2_\chi }\pm\frac{G^-_{\alpha\delta}(\theta)}{2 m^2_\chi }
=2 |\bm{k}|\cos^2\frac{\theta}{2}\{ m  + 4(E_p-m) \sin^2\frac{\theta}{2}\cos^2\frac{\theta}{2}\}\mp 2  m^2 \sin^2\frac{\theta}{2} 
,\\ \frac{\widehat A^+_{\pm}(\theta)}{2 m^2_\chi }&=&\frac{G^+_{\delta^2}(\theta)}{2 m^2_\chi } +\frac{G^+_{\beta\gamma}(\theta)}{2 m^2_\chi }\pm \frac{ G^+_{\beta\delta}(\theta)}{2 m^2_\chi }
=m^2+ (m- |\bm{k}|)^2 \cos^2\frac{\theta}{2}\mp 4 m |\bm{k}| \cos^4\frac{\theta}{2} 
,\\ \frac{\widehat M^+_{\pm}(\theta)}{2 m^2_\chi }&=&\frac{G^+_{\alpha\delta}(\theta)}{2 m^2_\chi }- 2\frac{G^+_{\beta\gamma}(\theta)}{2 m^2_\chi }\pm \frac{G^+_{\beta\delta}(\theta)}{2 m^2_\chi }
= -2|\bm{k}| \cos^2\frac{\theta}{2}[|\bm{k}|\mp m (\sin^2\frac{\theta}{2}-\cos^2\frac{\theta}{2})]
.\end{eqnarray*}}%
Given (\ref{eq:41chiA-CrossSection-Coh-vs-InCoh-GeneralFactor-with-G2_F}), 
incoherent $\chi A$ cross sections of scattering due to $V\!\mp\!A$ currents with massive lepton spin flip
can be written with dimensionless factors of the chiral constants $g_A$, $g_V$ in the following final form:
\begin{eqnarray}\nonumber 
\label{eq:43RchiA-CrossSectionsEvaluations-Weak-ChiEta-rel-InCoh-VpmA-spinflip-epsilon} 
\frac{d\sigma^{\mp\pm}_{\text{inc,V-A}}}{d T_A} &=&
\dfrac{G^2_F  m_A}{4 \pi } \dfrac{m_\chi^2}{|\bm{k}_\chi|^2}  \sum A_f \widehat F_f^2  \Big[ 
(g_V^2+ g_A^2) + (g_A^2  -g_V^2) \cos^2\frac{\theta}{2} 
\\&&\nonumber + \epsilon^2 (g_V-g_A)^2  \cos^2\frac{\theta}{2} 
-2g_A \epsilon [ ( g_A \mp  g_V)  \mp 2  (g_A - g_V) \cos^2\frac{\theta}{2}] \cos^2\frac{\theta}{2}   
\\&+&\nonumber \dfrac{2 \Delta A_f}{A_f}\Big\{
\epsilon (g_A- g_V) [g_A\cos^2\frac{\theta}{2}(1+2\sin^2\frac{\theta}{2})
+ g_V(1+\cos^2\frac{\theta}{2}-2\cos^2\frac{\theta}{2} \sin^2\frac{\theta}{2})] \cos^2\frac{\theta}{2}
\\&&  \mp g_A  (g_V\sin^2\frac{\theta}{2} +g_A\cos^2\frac{\theta}{2}) - 2 \epsilon \sqrt{1+\epsilon^2} (g_V-g_A)^2\cos^4\frac{\theta}{2} \sin^2\frac{\theta}{2}  \Big\} \Big]
,\\ \nonumber. \frac{d\sigma^{\mp\pm}_{\text{inc,V+A}}}{d T_A} &=&
\dfrac{G^2_F  m_A}{4 \pi } \dfrac{m_\chi^2}{|\bm{k}_\chi|^2} \sum A_f \widehat F_f^2\Big[
(g_V^2+g_A^2)  +  (g_A^2-g_V^2)\cos^2\frac{\theta}{2}
 \\&&\nonumber + \epsilon^2 (g_V+g_A)^2 \cos^2\frac{\theta}{2}
- 2 g_A \epsilon [(g_A \mp g_V) \pm 2 (g_A +g_V)\cos^2\frac{\theta}{2}  ]\cos^2\frac{\theta}{2} 
\\&+&\nonumber \dfrac{ 2\Delta A_f}{A_f}\Big\{
 \epsilon (g_A+  g_V)[g_A\cos^2\frac{\theta}{2} (1- 2\sin^2\frac{\theta}{2})
+g_V(1-\cos^2\frac{\theta}{2}- 2 \sin^2\frac{\theta}{2}\cos^2\frac{\theta}{2}) ] \cos^2\frac{\theta}{2}
\\&& \nonumber
\pm g_A  [g_V  \sin^2\frac{\theta}{2}-g_A \cos^2\frac{\theta}{2}] +2 \epsilon \sqrt{1+\epsilon^2} (g_V+g_A)^2 \cos^4\frac{\theta}{2}\sin^2\frac{\theta}{2} \Big\}\Big]
\nonumber
.\end{eqnarray}
Here the auxiliary notation $\dfrac{|\bm{k}| E_p}{m^2} = \epsilon \sqrt{1+\epsilon^2} $ is introduced, 
where $\epsilon \equiv\dfrac{|\bm{k}|}{m}$.
It is immediately clear that for $m_\chi\to 0$ both formulas
(\ref{eq:43RchiA-CrossSectionsEvaluations-Weak-ChiEta-rel-InCoh-VpmA-spinflip-epsilon})
vanish, as expected.
\par 
The limit $\epsilon \to 0$ corresponds to the case where the momentum of the incident
$\chi$ lepton $|\bm{k}_\chi|\ll m$, including $|\bm{k}_\chi| \le m_\chi \ll m$.
Then in this limit the following approximation is obtained
from formulas
(\ref{eq:43RchiA-CrossSectionsEvaluations-Weak-ChiEta-rel-InCoh-VpmA-spinflip-epsilon}):
{\small
\begin{eqnarray*}
\frac{d\sigma^{\mp\pm}_{\text{inc,V-A}}}{d T_A} = \dfrac{G^2_F  m_A}{4 \pi } \dfrac{m_\chi^2}{|\bm{k}_\chi|^2}  \sum A_f \widehat F_f^2  \Big[  (g_V^2+ g_A^2) + (g_A^2  -g_V^2) \cos^2\frac{\theta}{2} 
\mp \dfrac{2 \Delta A_f}{A_f} g_A \Big\{g_V\sin^2\frac{\theta}{2} +g_A\cos^2\frac{\theta}{2} \Big\} \Big]
,\\  \frac{d\sigma^{\mp\pm}_{\text{inc,V+A}}}{d T_A} =\dfrac{G^2_F  m_A}{4 \pi } \dfrac{m_\chi^2}{|\bm{k}_\chi|^2} \sum A_f \widehat F_f^2\Big[(g_V^2+g_A^2)  +  (g_A^2-g_V^2)\cos^2\frac{\theta}{2}\pm \dfrac{ 2\Delta A_f}{A_f} g_A  \Big\{ g_V  \sin^2\frac{\theta}{2}-g_A \cos^2\frac{\theta}{2}\Big\}\Big]
\nonumber.\end{eqnarray*}}%
It shows that for the nucleus with spin 0 ($\Delta A_f=0$) when $\epsilon \simeq 0$ these cross sections coincide
$$\frac{d\sigma^{\mp\pm}_{\text{inc,V-A}}}{d T_A} =\frac{d\sigma^{\mp\pm}_{\text{inc,V+A}}}{d T_A} =
\dfrac{G^2_F  m_A}{4 \pi } \dfrac{m_\chi^2}{|\bm{k}_\chi|^2} \sum A_f \widehat F_f^2 (\bm{q}) \Big[
(g_V^2+g_A^2)  +  (g_A^2-g_V^2)\cos^2\frac{\theta}{2}\Big].
$$ 
However, this observation is of no practical use, since at such a small momentum of the incident lepton
the value of the transferred momentum is $\bm{q}\simeq 0$, which results in $\widehat F_f^2 (\bm{q}) \simeq 0$.
In other words, for sufficiently small transferred momenta, one can not speak about any 
noticeable role of inelastic (incoherent) scattering processes.

\paragraph{Fully averaged incoherent "$\bm{V\!\mp\!A}$"\/ cross section.}
Recall that the fully averaged (averaged over the initial and summed over the final helicities of the $\chi$ particle)
 {\em incoherent} $\chi A$-interaction cross section is given by
(\ref{eq:43RchiA-CrossSections-via-ScalarProducts-ChiEta-rel-InCohCS-total}).
In the $V\mp A$ approximation, when
$\alpha_{\mp} = + g_V ,\ \beta_{\mp} = - g_A, \ \gamma_{\mp} = \mp g_V ,\ \delta_{\mp} = \pm g_A$
one has
\begin{eqnarray}\nonumber
\label{eq:43RchiA-CrossSectionsEvaluations-Weak-ChiEta-rel-InCoh-w-to-VpmA}
\alpha^2_{\mp}=\gamma^2_{\mp} &=& g_V^2 ,\quad  \beta^2_{\mp} = \delta^2_{\mp} = g_A^2
 ,\quad (\gamma^2-\beta^2)_{\mp}=(g_V^2-g_A^2) 
,\\ \nonumber
(\alpha-\delta)_{\mp}&=&(g_V\mp g_A) ,\quad (\beta-\gamma)_{\mp}= \pm (g_V\mp g_A)
,\\  (\alpha-\delta)\delta_{\mp}&=&\pm (g_V\mp g_A)g_A ,\quad (\beta-\gamma)\beta_{\mp}=\mp (g_V\mp g_A)g_A  ,\\ ( \alpha\gamma+\beta\delta)_{\mp}&=&\mp (g_V^2+g_A^2) ,\quad (\alpha\delta+\beta\gamma)_{\mp}=\pm 2g_Vg_A ,\quad (\alpha\beta +\gamma\delta)_{\mp}=-2g_Vg_A
\nonumber
. \end{eqnarray}
Then sums (\ref{eq:43RchiA-CrossSectionsEvaluations-Weak-ChiEta-rel-InCohCS-total-sumy-both})
defining completely averaged incoherent $\chi A$ cross sections
go into the following "$V\pm A$"\/ expressions:
{\small \begin{eqnarray}\nonumber
\label{eq:43RchiA-CrossSectionsEvaluations-Weak-ChiEta-rel-InCoh-total-suma+VpmA}
\sum_{s',s}\frac{Q^{s's}_{+,\text{V$\mp$A}}}{2^6}&=&
(g_V^2+3g_A^2) m^2m^2_\chi \pm 4 g_Vg_A |\bm{k}|^2 s
 +4g_A(g_A \mp g_V)  m^2 |\bm{k}|^2\sin^2\frac{\theta}{2}
\\&&  +(g_V\mp g_A)^2  |\bm{k}|^2 [4 |\bm{k}|^2  \sin^4\frac{\theta}{2}+2s \cos^2\frac{\theta}{2}
+ m^2_\chi(\sin^2\frac{\theta}{2}+\sin^4\frac{\theta}{2}+\cos^4\frac{\theta}{2})]
,\\ \nonumber \sum_{s',s}\frac{Q^{s's}_{-,\text{V$\mp$A}}}{2^7 |\bm{k}|}&=&
\mp (g_V\mp g_A)^2  [m^2_\chi E_p \cos^2\frac{\theta}{2}\sin^2\frac{\theta}{2}(1-2\sin^2\frac{\theta}{2})-2|\bm{k}|^2(E_p\cos^2\frac{\theta}{2}+E_\chi) \sin^2\frac{\theta}{2}]
\\&& \nonumber \mp(g_V^2 - g_A^2)m^2_\chi m\cos^2\frac{\theta}{2}\sin^2\frac{\theta}{2}(1-2\sin^2\frac{\theta}{2}) 
+ g_A^2 [mm^2_\chi\cos^2\frac{\theta}{2}(1-2\sin^2\frac{\theta}{2}) 
 \pm 2 m^2 E_\chi \sin^2\frac{\theta}{2}]
\\&& \nonumber + g_Vg_A [mm^2_\chi\cos^2\frac{\theta}{2}(1-2\sin^2\frac{\theta}{2}) 
+2m^2_\chi E_p+4|\bm{k}|^2(E_\chi +E_p)+2m^2 E_\chi \cos^2\frac{\theta}{2}]
\nonumber.\end{eqnarray}}%
As a result, the fully averaged {\em incoherent} $\chi A$ interaction cross section
(\ref{eq:43RchiA-CrossSections-via-ScalarProducts-ChiEta-rel-InCohCS-total})
with allowance for sums from 
(\ref{eq:43RchiA-CrossSectionsEvaluations-Weak-ChiEta-rel-InCoh-total-suma+VpmA})
takes the "$V\!\mp\!A$"\/ form
\begin{eqnarray*}
\frac{d\sigma^\text{total}_{\text{inc,V$\mp$A}}}{g_\text{i}d T_A} &=& 
\dfrac{G^2_F  m_A}{4\pi }\dfrac{1}{m^2 |\bm{k}_\chi|^2 }\sum_{f=p,n} A_f  \widehat F^2_f(\bm{q})
\Big[\sum^{}_{s's}\frac{Q^{s's}_{+,\text{V$\mp$A}}}{2^6}+\dfrac{2 \Delta A_f}{A_f} |\bm{k}| \sum^{}_{s's}\frac{Q^{s's}_{-,\text{V$\mp$A}}}{2^7 |\bm{k}|}\Big]
. \end{eqnarray*}
Further, since the $V\!\mp\!A$ limit satisfies the relations
(\ref{eq:43RchiA-CrossSectionsEvaluations-Weak-ChiEta-rel-InCoh-w-to-VpmA}),
there are the following expressions 
then for parameters from (\ref{eq:43RchiA-CrossSections-via-ScalarProducts-ChiEta-rel-InCohCS-total-spinless-Phi}): 
{\small \begin{eqnarray*}\nonumber
\Phi^+_{\alpha^2+3\delta^2}(\bm{q})&=&\Phi^+_{g_V^2}+3\Phi^+_{g_A^2}(\bm{q})
,\quad \Phi^+_{\gamma^2-\beta^2}(\bm{q})=\Phi^+_{g_V^2}(\bm{q})-\Phi^+_{g_A^2}(\bm{q})
,\\\Phi^+_{(\beta-\gamma)^2}(\bm{q})&=& \Phi^+_{(g_V\mp g_A)^2}(\bm{q})
,\qquad \Phi^+_{(\alpha-\delta)^2+(\beta-\gamma)^2}(\bm{q})= 2\Phi^+_{(g_V\mp g_A)^2}(\bm{q})
,\\\Phi^+_{2(\alpha\delta+\beta\gamma)}(\bm{q})&=& \pm 4 \Phi^+_{g_Vg_A}(\bm{q})
,\quad \Phi^+_{(\beta-\gamma)\beta}(\bm{q}) =\Phi^+_{(\delta-\alpha)\delta}(\bm{q})=\mp\Phi^+_{g_Vg_A \mp g_A^2 }(\bm{q})
. \end{eqnarray*}}%
After their substituting into general formula
(\ref{eq:43RchiA-CrossSections-via-ScalarProducts-ChiEta-rel-InCohCS-total-spinless-Phi}), 
for the spinless nucleus there is
\begin{eqnarray*}
\frac{d\sigma^\text{total}_{\text{inc,V$\mp$A},0}}{g_\text{i}\hat c_A d T_A} &=&
[\Phi^+_{g_V^2} (\bm{q})+3\Phi^+_{g_A^2}(\bm{q})]m^2_\chi m^2 \pm 4 \Phi^+_{g_Vg_A}(\bm{q}) |\bm{k}|^2s 
+4 \Phi^+_{ (g_A\mp g_V)g_A }(\bm{q}) m^2 |\bm{k}|^2\sin^2\frac{\theta}{2} 
\\&&.  +\Phi^+_{(g_V\mp g_A)^2} (\bm{q}) |\bm{k}|^2
 [4 |\bm{k}|^2\sin^4\frac{\theta}{2} +  2s  \cos^2\frac{\theta}{2} +m^2_\chi(\cos^2\frac{\theta}{2}+2\sin^4\frac{\theta}{2})]
. \end{eqnarray*}
Here, by analogy with the definitions from (\ref{eq:43RchiA-CrossSection-via-ScalarProducts-InCohSC-Phi-definitions}), the following notations are given:
\begin{eqnarray}\nonumber
\label{eq:43RchiA-CrossSection-via-ScalarProducts-InCohSC-Phi-definitions-VmpA}
\Phi^+_{V^2}(\bm{q})\equiv \Phi^+_{g_V^2} (\bm{q}) &=& \sum_{f=p,n} \widehat F_f^2(\bm{q})   A_f [g_V^f]^2 
,\quad  \Phi^+_{VA}(\bm{q})\equiv \Phi^+_{g_Vg_A}(\bm{q})= \sum_{f=p,n} \widehat F_f^2(\bm{q})   A_f  g^f_A g^f_V 
,\\ \Phi^+_{A^2}(\bm{q})\equiv \Phi^+_{g_A^2} (\bm{q})&=& \sum_{f=p,n} \widehat F_f^2(\bm{q})   A_f [g_A^f]^2 
.\end{eqnarray}
With these notations, the fully averaged incoherent $\chi A$ cross section on the {\em spinless}\/ nucleus
in the form of an expansion in chiral coupling constants $g_V$ and $g_A$ takes the form
{\small \begin{eqnarray}
\label{eq:43RchiA-CrossSection-via-ScalarProducts-InCohCS-VmpA-total-viaPhi-spinnless}
\frac{d\sigma^\text{total}_{\text{inc,V$\mp$A},0}}{g_\text{i} d T_A}&\equiv&
\dfrac{G^2_F  m_A}{4\pi m^2 |\bm{k}_\chi|^2}\Big\{\Phi^+_{V^2} (\bm{q})S_{V^2} (\theta) 
+\Phi^+_{A^2} (\bm{q}) S_{A^2} (\theta)\pm 2\Phi^+_{VA}(\bm{q})S_{VA} (\theta)\Big\},\quad \text{where}
\\\nonumber 
S_{V^2}(\theta)&\equiv&m^2_\chi m^2 +4 |\bm{k}|^4\sin^4\frac{\theta}{2} +  2s  |\bm{k}|^2 \cos^2\frac{\theta}{2} +m^2_\chi |\bm{k}|^2 (\cos^2\frac{\theta}{2}+2\sin^4\frac{\theta}{2})
,\\\nonumber S_{A^2} (\theta)&\equiv& 3m^2_\chi m^2 +4m^2 |\bm{k}|^2\sin^2\frac{\theta}{2} 
 +|\bm{k}|^2[4 |\bm{k}|^2\sin^4\frac{\theta}{2} +  2s  \cos^2\frac{\theta}{2} +m^2_\chi
 (\cos^2\frac{\theta}{2}+2\sin^4\frac{\theta}{2})]
,\\\nonumber S_{VA} (\theta)&\equiv&  |\bm{k}|^2\big[ 2s -2m^2  \sin^2\frac{\theta}{2} 
- [4 |\bm{k}|^2\sin^4\frac{\theta}{2} +  2s  \cos^2\frac{\theta}{2} +m^2_\chi
(\cos^2\frac{\theta}{2}+2\sin^4\frac{\theta}{2})]\big]
.\end{eqnarray}}
For the {\em massless}\/ $\chi$ lepton ($m_\chi=0$), summation over all lepton helicities
(\ref{eq:43RchiA-CrossSectionsEvaluations-Weak-ChiEta-rel-InCoh-total-suma+VpmA})
gives the following result:
\begin{eqnarray}\nonumber
\label{eq:43RchiA-CrossSectionsEvaluations-Weak-ChiEta-rel-InCoh-total-suma+VpmA-mchi=0}
\sum\frac{Q^{s's}_{+,\text{V$\mp$A},0}}{2^7|\bm{k}|^2 }&=&+2g_A^2 m^2\sin^2\frac{\theta}{2}
\pm 2 g_Vg_A (s-m^2\sin^2\frac{\theta}{2}) +(g_V\mp g_A)^2 (2|\bm{k}|^2  \sin^4\frac{\theta}{2}+s \cos^2\frac{\theta}{2})
, \\ \sum\frac{Q^{s's}_{-,\text{V$\mp$A},0}}{2^7 |\bm{k}|^2}&=& \pm  2g_A^2  m^2 \sin^2\frac{\theta}{2}
+2 g_Vg_A  (s-m^2\sin^2\frac{\theta}{2}) \\&& \nonumber \pm(g_V\mp g_A)^2  
 ((s-m^2)\cos^2\frac{\theta}{2}+2|\bm{k}|^2\sin^2\frac{\theta}{2})\sin^2\frac{\theta}{2}
\nonumber.\end{eqnarray}
Then, for $m_\chi=0$ in the $V\!\mp\!A$ variant, the fully averaged 
incoherent $\chi A$ cross section is
\begin{eqnarray}\nonumber
\label{eq:43RchiA-CrossSectionsEvaluations-Weak-ChiEta-rel-InCoh-total-VpmA-mchi=0}
\frac{d\sigma^\text{total}_{\text{inc,V$\mp$A},0}}{g_\text{i}d T_A}\!&=&\! 
\dfrac{G^2_F  m_A}{2\pi} \sum_{f=p,n}A_f  \widehat F^2_f(\bm{q})\Big\{2g_A^2 \frac{m^2}{s}\sin^2\frac{\theta}{2}
+(g_V\mp g_A)^2\big[\dfrac12 \big(1-\dfrac{m^2}{s} \big)^2\sin^4\frac{\theta}{2}+\cos^2\frac{\theta}{2}\big]
\\\nonumber &&\pm 2 g_Vg_A \big(1-\frac{m^2}{s}\sin^2\frac{\theta}{2}\big) 
+\dfrac{\Delta A_f}{A_f}\Big[2 g_Vg_A\big(1-\frac{m^2}{s}\sin^2\frac{\theta}{2}\big) \pm 2g_A^2\frac{m^2}{s}\sin^2\frac{\theta}{2} 
\\&& \qquad  \pm(g_V\mp g_A)^2 \big(1-\frac{m^2}{s}\big)\sin^2\frac{\theta}{2}
 \big[\dfrac12\big(1-\dfrac{m^2}{s} \big)\sin^2\frac{\theta}{2}+\cos^2\frac{\theta}{2} \big]\Big]\Big\}
. \end{eqnarray}
This $\chi A$ cross section, like each expression 
(\ref{eq:43RchiA-CrossSectionsEvaluations-Weak-ChiEta-rel-InCoh-total-suma+VpmA-mchi=0})
corresponding to the {\em massless}\/ lepton, despite the presence of a sum over
lepton helicities, generally speaking, contains only one term.
For the $V\!-\!A$ variant, it is the only contribution of the lepton with the negative {\em conserved}\/ helicity
(neutrino, same as formula (\ref{eq:43RchiA-CrossSection-via-ScalarProducts-ChiEta-rel-V-A-CS-m_chi=0})).
For the $V\!+\!A$ variant, it is the only contribution of the lepton with positive
{\em conserved}\/ helicity (antineutrino).
This expression coincides with the formulas for (anti)neutrinos from \cite{Bednyakov:2021ppn}.

\paragraph{Total averaged ${\chi A}$ cross section for the massive lepton and $\bm{V\!\mp\!A}$ currents.}
It is the sum of two terms: 
the fully averaged coherent (\ref{eq:43RchiA-CrossSection-via-ScalarProducts-CohCS-VmpA-total-viaG})
and fully averaged incoherent (\ref{eq:43RchiA-CrossSection-via-ScalarProducts-InCohCS-VmpA-total-viaPhi-spinnless})
$\chi A$ interaction cross sections due to $V\!\mp\!A$ currents.
Let us present this expression for the spinless nucleus, taking into account (\ref{eq:ScalarProducts-Lab-k^2-to-k^2_lab})
in the form of expansion in the effective $V\!\mp\!A$ constants
(assuming that $g_\text{i}\simeq g_\text{c} \simeq 1$)
\begin{eqnarray}\label{eq:CrossSections-VmpA-Coh+InCoh-total-spinless}
\dfrac{d\sigma^{\text{total}}_{V\!\mp\!A,0}}{d T_A}&=&
\dfrac{G^2_F  m_A}{4 \pi}\Big\{\Phi^+_{V^2} (\bm{q}) S_{V^2} (\theta)+\Phi^+_{A^2} (\bm{q})S_{A^2} (\theta)
\pm 2\Phi^+_{VA}(\bm{q}) S_{VA} (\theta) \\&&\qquad + G_V^2(\bm{q})C_V(\theta) +G_A^2(\bm{q})C_A(\theta)
\pm 2G_V(\bm{q})G_A(\bm{q}) C_M(\theta) \Big\}
.\nonumber
\end{eqnarray}
Here the coefficients in front of the effective $V\!\mp\!A$ constants are redefined by including into them 
the common factor $(|\bm{k}|^2 s)^{-1}$, i.e., 
$S_{V^2} (\theta)\equiv \dfrac{S_{V^2} (\theta)}{|\bm{k}|^2 s}$, etc. 
They have the following form:
\begin{eqnarray*}
S_{V^2} (\theta)&=& \dfrac{m^2_\chi m^2 }{|\bm{k}|^2 s}  +  2 \cos^2\frac{\theta}{2}  +\dfrac{4|\bm{k}|^2}{s} \sin^4\frac{\theta}{2} 
  +\dfrac{m^2_\chi}{ s}\Big(\cos^2\frac{\theta}{2}+2\sin^4\frac{\theta}{2}\Big)
,\\ S_{A^2} (\theta)&=& \dfrac{3 m^2_\chi m^2 }{|\bm{k}|^2 s}+\dfrac{4m^2 }{s}\sin^2\frac{\theta}{2}
+  2\cos^2\frac{\theta}{2}  +\dfrac{4 |\bm{k}|^2}{s} \sin^4\frac{\theta}{2} 
 +\dfrac{m^2_\chi }{s}\Big(\cos^2\frac{\theta}{2}+2\sin^4\frac{\theta}{2}\Big) 
 ,\\ S_{VA} (\theta)&=& 2\sin^2\frac{\theta}{2}\Big(1-\dfrac{m^2}{s}\Big)
- \dfrac{4|\bm{k}|^2}{s}\sin^4\frac{\theta}{2}
-\dfrac{m^2_\chi}{ s}\Big(\cos^2\frac{\theta}{2}+2\sin^4\frac{\theta}{2}\Big)
;\\ C_A(\theta)&=&\Big(\frac{E_p}{\sqrt{s}}-\frac{m}{\sqrt{s}}\Big)^2
\sin^2\frac{\theta}{2}\cos^2\frac{\theta}{2} \Big(\frac{m_\chi^2}{|\bm{k}|^2}+2\sin^2\frac{\theta}{2}\Big)
,\\ C_M(\theta)&=&2\Big(\frac{E_p}{\sqrt{s}}-\frac{m}{\sqrt{s}}\Big)\cos^2\frac{\theta}{2}\sin^2\frac{\theta}{2} 
\Big(\frac{E_\chi}{\sqrt{s}} + \frac{m}{\sqrt{s}}\sin^2\frac{\theta}{2} +\frac{E_p}{\sqrt{s}}\cos^2\frac{\theta}{2}\Big)
,\\ C_V(\theta)&=& \Big(\frac{m}{\sqrt{s}}\sin^2\frac{\theta}{2}+\frac{E_p}{\sqrt{s}}\cos^2\frac{\theta}{2}\Big)^2
\Big(\frac{m_\chi^2}{|\bm{k}|^2}+2\cos^2\frac{\theta}{2}\Big) 
\\&& +\cos^2\frac{\theta}{2} \Big(\dfrac{m_\chi^2}{s} +\dfrac{2|\bm{k}|^2}{s} 
+ \dfrac{4 E_\chi}{\sqrt{s}} \Big(\dfrac{m}{\sqrt{s}} \sin^2\frac{\theta}{2} +\dfrac{E_p}{\sqrt{s}}\cos^2\frac{\theta}{2}\Big)\Big)
.\end{eqnarray*}
Formula (\ref{eq:CrossSections-VmpA-Coh+InCoh-total-spinless}) can also be obtained
by the corresponding substitutions into the general expression
(\ref{eq:43RchiA-CrossSection-via-ScalarProducts-ChiEta-rel-CohCS-mp=mn-general-spinless-total-weak}), which
holds  for the total cross section of the weak interaction of the massive lepton and the spinless nucleus.
The other variables included in expression
(\ref{eq:CrossSections-VmpA-Coh+InCoh-total-spinless})
are defined in the text of this chapter by formulas
(\ref{eq:43RchiA-CrossSections-via-ScalarProducts-Weak-ChiEta-rel-VmpA-Eff-couplings})
and (\ref{eq:43RchiA-CrossSection-via-ScalarProducts-InCohSC-Phi-definitions-VmpA}).
\par
{\em Note}\/ that expression (\ref{eq:CrossSections-VmpA-Coh+InCoh-total-spinless})
can be considered {\em as a new prescription}\/ for computing the
cross sections of the (massive) lepton scattering  on the nucleus via weak $V\!\mp\!A$ interaction 
at energies below hundreds of MeV.
For example, such modern software products as {\sf Achilles} from \cite{Isaacson:2022cwh}
claiming to allow precision description of weak processes including those considered in this paper
should apparently take into account the results obtained here.

\section{\large Conclusions} \label{60chiA-Conclusions} 
This paper outlines all elements of the theoretical approach to the description of the scattering of a {\em massive}\/ (neutral) lepton on a nucleus as an object,  the internal structure of which is due to 
mutually interacting nucleons.
This is a natural generalization of the new concept in the theory of 
(anti)neutrino scattering on the nucleus as a composite system proposed by D.V. Naumov and the author
in 
\cite{Bednyakov:2018mjd,Bednyakov:2019dbl,Bednyakov:2021ppn,Bednyakov:2021bty}.
The boundary condition for the applicability of the approach is the requirement to preserve
integrity of the nucleus after interaction. 
The nucleus as a complex quantum system can be "excited"\/, can pass from one of its quantum states to another,
but in the canonical version of this approach, the nucleus should not "fall apart".
This requirement imposes restrictions on the magnitude of the momentum transferred by the lepton to the nucleus,
which should not be so large as to completely "destroy the nucleus"\/.
In addition, the applicability of the approach is, strictly speaking,
also limited by the approximations made, which, in turn, are
explicitly formulated in the text, look quite natural, and, 
if necessary, can be specially investigated and taken into account as,
for example, correction factors or contributions.
\par
The approach constructively uses the so-called condition for the completeness of the quantum state of the nucleus (or the probability conservation condition), which, being based on the summation over all possible initial and final states of the nucleus,
allows one to explicitly separate the elastic process from all other inelastic processes
capable of contributing to the total observable $\chi A$ scattering cross section.
The approach further relies on the microscopic description of the nucleus as a bound state
of its constituent nucleons on the basis of the many-particle wave function of the nucleus.
The {\em weak}\/ interaction of a point-like $\chi$ lepton is considered to be between structureless protons and neutrons of the nucleus.
This interaction is parameterized
in the form of an  expansion of the scalar product of the 
lepton and nucleon currents in four effective coupling constants
reflecting the (axial) vector nature of the weak interaction
  $$ \displaystyle (l^w_{s's}, h^{w,f}_{r'r})=\alpha_f (l^v_{s's}\, h^v_{r'r}) + \beta_f (l^v_{s's}\, h^a_{r'r}) + \gamma_f (l^a_{s's}\, h^v_{r'r}) + \delta_f (l^a_{s's}\ , h^a_{r'r}).$$
\par
{\em Relativistic}\/ expressions for the cross sections of the massive (neutral) $\chi$ lepton scattering 
on the nucleus are obtained in a general form.
It is shown that the {\em observable}\/ cross section of the process $\chi A\to \chi A^{(*)}$
(\ref{eq:43RchiA-CrossSection-via-ScalarProducts-ChiEta-rel-CohCS-mp=mn-general-spinless-total-weak})
includes the elastic (or coherent) contribution
(\ref{eq:43RchiA-CrossSection-via-ScalarProducts-Weak-ChiEta-rel-CohCS-mp=mn-general}) when the nucleus
remains in its original quantum state and the inelastic (incoherent) contribution
(\ref{eq:43RchiA-CrossSection-via-ScalarProducts-InCoh-all-viaFG})
when the nucleus goes into another (excited) quantum state.
Transition from the elastic scattering regime to the inelastic scattering regime
 is {\em automatically}\/ determined by the dependence of the nucleon-nucleus form factors
$F_{p/n}(\bm{q})$ on the momentum $\bm{q}$ transferred to the nucleus.
At small $\bm{q}$ the elastic scattering dominates, and as $\bm{q}$ increases
the contribution of the inelastic scattering increases, 
and the latter dominates at sufficiently large $\bm{q}$.
This {\em automatic behavior}\/ makes this approach fundamentally different
from Friedmann's concept of coherence
\cite{Freedman:1973yd,Freedman:1977xn},
within which, before using formulas for the coherent scattering of the (anti)neutrino on the nucleus, 
one has to make sure in advance that this can be done, i.e., 
that the product of the characteristic radius of the target nucleus
and some characteristic momentum transferred to the nucleus is noticeably less than unity ($|\bm{q}|R\ll 1$).
In our approach, this question does not arise at all.
\par
As an important application of the general formulas, the scattering of {\em massive}\/ (anti)neu\-trinos 
interacting with nucleons through the $V\!\mp\!A$ current of the Standard Model is considered in detail.
The weak $V\!-\!A$ interaction with the nucleus corresponds to the massive analogue of the neutrino (negative helicity),
and the case of the weak $V\!+\!A$ interaction is the massive analog of the antineutrino (positive helicity).
A complete set of expressions for the corresponding cross sections is obtained
(e.g. formulas
(\ref{eq:43RchiA-CrossSection-via-ScalarProducts-ChiEta-rel-CohCS-all-V-A-V+A})
and (\ref{eq:43RchiA-CrossSections-via-ScalarProducts-ChiEta-rel-InCohCS-VpmA-spinfixed-fino})).
The transition of these formulas  to the formulas from \cite{Bednyakov:2021ppn}
corresponding to the scattering of {\em massless}\/ (anti)neutrinos of the Standard Model 
on the nucleus is demonstrated.
\par
Owing to the nonzero mass of the (anti)neutrino, an additional channel for the elastic
(coherent) and inelastic (incoherent) scattering of (anti)neutrinos on the nuclei arises from 
the possibility of a change in helicity in massive (anti)\-neutrinos.
For example, despite the smallness of the mass of neutrinos at kinetic energies of (anti)neutrinos much lower than the mass of neutrinos (for example, relict ones), the cross section of their interaction with the nucleus turns out to be multiply enhanced by the
"nucleus coherence effect"\/.
\par
The resulting expressions can be used, for example, in the analysis of 
results of experiments on the direct detection of neutral massive weakly interacting
relativistic and non-relativistic particles of dark matter,
since, unlike the generally accepted case,  
both the elastic and inelastic interactions of such particles with the target nucleus
are simultaneously taken into account.
In this case, the presence of the "inelastic signal"\/ with its characteristic signature in the form of
$\gamma$-quanta from deexcitation of the nucleus may be the only
registered evidence 
\cite{Bednyakov:2021ppn,Bednyakov:2021bty,Bednyakov:2022dmc}
of the interaction of the dark matter particle.

\section*{Acknowledgments}
The author is grateful to V.A. Kuzmin, D.V. Naumov, E.A. Yakushev
and other colleagues for the important comments and discussions.

\section{\large Appendix} \label{ARchiA-Appendix} 
\paragraph{\em Wave functions.}
Recall \cite{Bednyakov:2018mjd,Bednyakov:2019dbl,Bednyakov:2021ppn} that
the state of the nucleus is denoted by the symbol $|P_{l}\rangle$, which means that the nucleus has a 4-momentum $P_{l}$, is in some $l$th internal quantum state ($l=n,m$) and the $|P_{l}\rangle$ 
is a superposition of free nucleons $|\{p\}\rangle$ multiplied by the wave function of the bound state $\tilde{\psi}'_n(\{p\})$.
The latter is a product of the wave function $\widetilde{\psi}_n(\{p{^\star}\})$, which describes the internal structure of the nucleus in its rest frame (the corresponding momenta are marked with index $\star$), and the wave function  $\Phi_n(p)$  responsible for the motion of the nucleus as a whole with the momentum $\bm{p}=\sum^A_{i=1}\bm{p}_i$ and the projection of the nuclear spin $s$
  \begin{equation}\label{A:ScatteringAmplitude-TildePrimePsi}
  \widetilde{\psi}'_n(\{p\}) = \widetilde{\psi}_n(\{p{^\star}\})\Phi_n(p), \quad \text{where} \quad p=(\bm{p},s).
  \end{equation}
 The function $\Phi_n(p)$ depends on $A-1$ terms of 3-momenta, since one combination of $A$ 3-momenta is used to describe the motion of the nucleus as a whole. Taking into account (\ref{A:ScatteringAmplitude-TildePrimePsi}) for the state $|P_n\rangle$, which completely characterizes the nucleus $A$, we will use the (antisymmetrized) expression
  (\ref{eq:ScatteringAmplitude-P_n}) from the main text of the paper.
\par
For nuclear states $|n\rangle$ describing the nucleus at rest in the $n$th internal quantum state ($n$th level), the normalization condition is adopted in the form
\begin{equation} \label{A:ScatteringAmplitude-mn-norm}
\langle m|n\rangle \equiv\int\Big(\prod^{A}_{i}\frac{d\bm{p}^\star_i}{(2\pi)^3}\Big)\widetilde{\psi}_n(\{p^\star\})\widetilde{\psi}^*_m(\{p^\star\})(2\pi)^3 \delta^3(\sum^ A_{i=1} \bm{p}^\star_i) =\delta_{mn}.
\end{equation}
For nuclear states $|P_n\rangle$ from (\ref{eq:ScatteringAmplitude-P_n}) it gives the normalization condition
\begin{equation}
\label{A:ScatteringAmplitude-PmPn-norm}
\langle P'_m|P_n\rangle = (2\pi)^3 2P^0_n \delta^3(\bm{P}-\bm{P}')\delta_{nm}.
\end{equation}
A Formal definition of the $|n\rangle$-state satisfying
(\ref{A:ScatteringAmplitude-mn-norm}) can be given as follows:
\begin{equation}\label{A:ScatteringAmplitude-n_state}
|n\rangle = \int \Big(\prod_{i=1}^{A} d\widetilde{\bm{p}}_i^\star\Big) \frac{\widetilde\psi_n(\{p{^\star}\}) }{\sqrt{A!}} \Big[(2\pi)^3\delta^3\Big(\sum_{i=1}^A\bm{p}_i^\star\Big)\Big]^{1/2}|\{p^\star\}\rangle.
\end{equation}

\paragraph{\em Formula for the hadronic current $\bm{h^\mu_{mn}(q)}$.}
The definition of the matrix element in terms of the effective Lagrangian 
$\mathcal{L}_{\rm int}(x) = \dfrac{G_{\rm F}}{\sqrt{2}} H^\mu (x)\, L_\mu(x)$   looks as follows:
\begin{eqnarray} \label{A-ScatteringAmplitude-S-Matrix-and-M-Element}
i (2\pi)^4\delta^4(\sum p_i-\!\sum p_f) \mathcal{M}\equiv\langle{\rm f} | i \!\int \! d^4 x \mathcal{L}_{\rm int}(x)| {\rm i}\rangle 
=\langle{\rm f}| \dfrac{iG_{\rm F}}{\sqrt{2}} \!\int\!\! d^4 x \!:\!H^\mu (x) L_\mu(x)\!:\! |{\rm i}\rangle. 
\end{eqnarray}
In the case of $\chi A\to \chi A^{(*)}$ scattering, the initial and final states are
\begin{eqnarray}\nonumber \label{A-ScatteringAmplitude-In-Out-States}
|{\rm i}\rangle &=&|\chi(k,s), A(P_n)\rangle = (2\pi)^{3/2}\sqrt{2E_\chi }\, a^+_{\chi}(\bm{k},s) |0\rangle | P_n\rangle,  \\[-15pt]\\
\langle{\rm f}|&=&\langle \chi(k',s'),  A^{(*)}(P'_m)|= (2\pi)^{3/2}\sqrt{2E_{\chi'}}\, \langle 0| a_{\chi}(\bm {k}',s')  \langle P'_m|
.\nonumber \end{eqnarray}
Therefore,  expression (\ref{A-ScatteringAmplitude-S-Matrix-and-M-Element})
for the process $\chi A\to \chi A^{(*)}$ becomes
\begin{eqnarray}  \label{A-ScatteringAmplitude-S-Matrix-and-M-Element-via-H-and-L}
(2\pi)^4\delta^4(k + P_n -k'- P'_m) i \mathcal{M}_{mn} &=&
\dfrac{i G_{\rm F}}{\sqrt{2}} \int d^4 x \, H^\mu_{nm}(x) \,  L_\mu^\chi(x),\quad \text{where}
\\ \nonumber H^\mu_{nm}(x) &\equiv&  \langle P'_m|\normord{\hat{H}^\mu (x)}| P_n\rangle
, \\ \nonumber  L_\mu^\chi(x) &\equiv& (2\pi)^{3}\sqrt{2E_{\chi'} 2E_\chi }\langle 0| a_{\chi}(\bm {k}',s') \normord{\hat{L}_\mu(x)}  a^+_{\chi}(\bm{k},s) |0\rangle  
.\end{eqnarray}
Using the fermionic quantum field operators of the $\chi$ particle and the nucleon
\begin{eqnarray} 
\label{A-ScatteringAmplitude-Psi-operators}
\overline{\hat{\psi}}(x)=\int\frac{d\bm{y}' e^{i y' x } }{\sqrt{(2\pi)^{3} 2E_{\bm{y}'}}}\sum_{s=1,2} a^+_s(\bm{y}')\overline{u}(\bm{y}',s),  \ \ 
\hat{\psi}(x)=\int\frac{d\bm{y} e^{-i y x } }{\sqrt{(2\pi)^{3}2E_{\bm{y}}}}\sum_{r=1,2}a_r(\bm{y})u(\bm{y},r)
, \quad \end{eqnarray}
one can write the nucleon $\hat{H}^\mu (x)$ and the lepton $\hat{L}_\mu(x)$ operators
in the following way:
$$ \hat{H}^\mu (x) \equiv \sum^A_{k} \overline{\hat{\psi}}_k(x)\, O^{\mu}_k\, \hat{\psi}_{k}(x) \quad \text{and}\quad
\hat{L}_\mu (x) \equiv \overline{\hat{\psi}}_\chi(x)\, O_{\mu}\, \hat{\psi}_{\chi}(x)
, $$
where $O^{\mu}$ is some combination of $\gamma$-matrices corresponding to the 
specific Lorentz interaction structure.
Then the lepton element from formula (\ref{A-ScatteringAmplitude-S-Matrix-and-M-Element-via-H-and-L})
is transformed  as follows:
\begin{eqnarray*}
L_\mu^\chi(x) &=&  
\sqrt{E_{k'} E_k} \int\frac{d\bm{y}' d\bm{y} e^{i x (y' - y) }}{\sqrt{E_{\bm{y}'} E_{\bm{y}}} }\sum_{r,r'=1,2}
\langle 0| a_{s'}(\bm {k}')\!:\!a^+_{r'}(\bm{y}')\overline{u}(\bm{y}',r') O_{\mu} a_r(\bm{y})u(\bm{y},r)\!:\!a^+_{s}(\bm{k}) |0\rangle.
\end{eqnarray*}
Further, since $a_{s'}(\bm {k}')a^+_{r'}(\bm{y}') =\delta_{s',r'}\delta(\bm {k }'- \bm{y}')\text{~~and~~}
a_r(\bm{y}) a^+_{s}(\bm{k}) = \delta_{s,r}\delta(\bm {k}- \bm{y})$, one gets
\begin{equation}
\label{A-ScatteringAmplitude-Lepton-PreCurrent}
L_\mu^\chi(x) = e^{i x (k' - k) } \overline{u}_\chi(\bm{k}',s') O_{\mu} u_\chi (\bm{k},s)
\equiv e^{- i x q } \, \overline{u}_\chi(\bm{k}',s') O_{\mu} u_\chi (\bm{k},s)
.  \end{equation}
According to (\ref{eq:ScatteringAmplitude-P_n}), 
the initial and final wave functions of the nucleus are given by the formulas
\begin{equation}
\label{A-ScatteringAmplitude-P_m-and-P_n}
\langle P'_m|=\!\int\!\Big(\!\prod^{A}_{j}d\tilde{\bm{p}'}^\star_j\Big)\frac{\tilde{\psi}^{*}_m(\{{p'}^\star\})}{\sqrt{{A!}}}\! \Phi^*_m(p')\langle\{{p'}^\star\}|,   \ \
|P_n\rangle =\!\int\!\Big(\!\prod^{A}_{i}d\tilde{\bm{p}}^\star_i\Big)\frac{\tilde{\psi}_n(\{p^\star\})}{\sqrt{{A!}}} \Phi_n(p)|\{p^\star\}\rangle
. \qquad \end{equation}
In this case, the product of the "external"\/ wave functions is
\begin{eqnarray}  \label{A-ScatteringAmplitude-Phi_m-times Phi_n}
\Phi^*_m(p') \Phi_n(p) &=& (2\pi)^3 \sqrt{2P^{0'}_m}\delta^3(\bm{p}'-\bm{P}'_m)\cdot
(2\pi)^3 \sqrt{2P^0_n}\delta^3(\bm{p}-\bm{P}_n),
\end{eqnarray} 
where $\bm{P}_n$ and $\bm{P}'_m$ are the 3-momenta of motion of the entire nucleus as a whole, and
$\bm{p}$ and $\bm{p}'$ are the arguments of the wave functions that have the meaning
$ \sum^A_i \bm{p}_i$ and $ \sum^A_i \bm{p}'_i$, respectively.
By virtue of (\ref{A-ScatteringAmplitude-P_m-and-P_n}), the nucleon matrix element from formula
(\ref{A-ScatteringAmplitude-S-Matrix-and-M-Element-via-H-and-L})
with allowance for the summation over all $A$ nucleons takes the form
\begin{eqnarray}
\label{A-ScatteringAmplitude-Nucleus-current-1}
H^\mu_{mn}(x) &=&
\sum^A_{k}\sum^{1,2}_{s_k,r_k}\int\frac{d\bm{y}'_k d\bm{y}_k  e^{i x (y'_k-y_k) } }{(2\pi)^{3} \sqrt{2E_{\bm{y}_k}  2E_{\bm{y}'_k}}}\int\Big(\prod^{A}_{j}d\tilde{\bm{p}'}^\star_j\Big)\frac{\tilde{\psi}^{*}_m(\{{p'}^\star\})}{\sqrt{{A!}}} \Phi^*_m(p')
\times \\&& \times \nonumber
\int\Big(\prod^{A}_{i}d\tilde{\bm{p}}^\star_i\Big) \frac{\tilde{\psi}_n(\{p^\star\})}{\sqrt{{A!}}} \Phi_n(p)
\langle\{{p'}^\star\}| a^+_{s_k}(\bm{y}'_k) \big\{\overline{u}(\bm{y}'_k,s_k)O^\mu_k u(\bm{y}_k,r_k)\big\} a_{r_k}(\bm{y}_k) 
|\{p^\star\}\rangle.
\end{eqnarray}
To further transform of expression (\ref{A-ScatteringAmplitude-Nucleus-current-1}), 
it is necessary to calculate the matrix element of the single-particle ($k$th) nucleon current
\begin{equation}	
\label{A-ScatteringAmplitude-k-Nucleon-current}
h^\mu_k(\bm{y}'_k,r'_k, \bm{y}_k,r_k)\equiv \overline{u}(\bm{y}'_k,r'_k)O^\mu_k u(\bm{y}_k,r_k)
\end{equation}
from  the many-particle state of $A$ free nucleons
\begin{equation}	
\label{A-ScatteringAmplitude-weakk_A}
w^k_A\equiv \langle\{{p'}^\star\}|a^+_{r'_k}(\bm{y}'_k) 
\big\{ h^\mu_k(\bm{y}'_k, r'_k, \bm{y}_k,r_k) \big\} a_{r_k}(\bm{y}_k)|\{p^\star\}\rangle 
.\end{equation}
Here and below, the renaming is done: $r'_k \equiv s_k$.
The states $\langle\{{p'}^\star\}|$ and $|\{p^\star\}\rangle$ can be written by isolating the active $k$th nucleon, for example, as
$${}_A\langle\{{p'}^\star\}|=\langle0|(..,p^{'\star}_k,..);..,a_{s_{p'}}(\bm{p^{'\star}}_k),..|, \quad
|\{p^\star\}\rangle_A = |..,a^+_{r_p}(\bm{p^\star}_k),..;(..,p^\star_k,..)|0\rangle.
$$
Then expression (\ref{A-ScatteringAmplitude-weakk_A}) can be represented as follows:
\begin{eqnarray} 
\label{A-ScatteringAmplitude-weakk_A-via-weakk_1} 
w^k_A&=& {{}_{A-1}\langle\{{p'}^\star\}||\{p^\star\}\rangle_{A-1}} \times w^k_1,
\quad \text{where} \quad 
\end{eqnarray}
\begin{equation} \label{A-ScatteringAmplitude-weakk_1}
w^k_1\equiv \langle\bm{p^{'\star}}_k,s_{p'}|a^+_{r'_k}(\bm{y}'_k)\big\{\overline{u}(\bm{y}'_k,r'_k)O^\mu_k u(\bm{y}_k,r_k)\big\} a_{r_k}(\bm{y}_k)|\bm{p^\star}_k,r_p\rangle
\end{equation}
is the matrix element of the one-particle current (\ref{A-ScatteringAmplitude-k-Nucleon-current}) in the one-nucleon ($k$th) state,  and
\begin{eqnarray*}
{}_{A-1}\langle\{{p'}^\star\}||\{p^\star\}\rangle_{A-1}
&\equiv&\langle0|(..,p^{'\star}_{k-1},p^{'\star}_{k+1},..)|(..,p^\star_{k-1},p^\star_{k+1},..)|0\rangle = 
\\&=& (A-1)!\Big(\prod^{A-1}_{l\ne k}(2\pi)^{3}2E_{\bm{p}^\star_l}\delta^3(\bm{p}^{\star}_l-\bm{p}^{'\star}_l)\delta_{r_l,r'_l}\Big)
\end{eqnarray*}
is the normalization condition for the state of $(A-1)$ free nucleons.
\par
Considering the normalization factor
$(2\pi)^{3}\sqrt{2E_{p^{\star}_k}2E_{p^{'\star}_k}}$ 
for the single-particle
initial and final $k$th nucleon state, commutation conditions for creation and annihilation operators
$$
a_{s_{p'}}(\bm{p^{'\star}}_k) a^+_{r'_k}(\bm{y}'_k) =\delta^3(\bm{p^{'\star}}_k-\bm{y}'_k)\delta_{r'_k,s_{p'}},
\quad   a_{r_k}(\bm{y}_k) a^+_{r_p}(\bm{p^\star}_k)=\delta^3(\bm{p^{\star}}_k-\bm{y}_k)\delta_{r_k,r_p}
$$ 
as well as the vacuum normalization $\langle0| 0\rangle=1$,
the single-particle matrix element $w^k_1$ (\ref{A-ScatteringAmplitude-weakk_1})
 is transformed as follows
\begin{eqnarray*}
w^k_1 &=& (2\pi)^{3}\sqrt{2E_{p^{\star}_k}2E_{p^{'\star}_k}} \delta^3(\bm{p^{'\star}}_k-\bm{y}'_k))\delta_{r'_k,s_{p'}} \delta^3(\bm{p^{\star}}_k-\bm{y}_k))\delta_{r_k,r_p} \big\{ \overline{u}(\bm{y}'_k,r'_k)O^\mu_k u(\bm{y}_k,r_k)\big\} .
\end{eqnarray*} 
With this expression for $w^k_1$, the matrix element of the ($k$th) nucleon current in 
the state of $A$ free nucleons (\ref{A-ScatteringAmplitude-weakk_A-via-weakk_1}) takes the form
\begin{eqnarray*}
w^k_A &=& (A-1)!\Big(\prod^{A-1}_{l\ne k}(2\pi)^{3}2E_{\bm{p}^\star_l}\delta^3(\bm{p}^{\star}_l-\bm{p}^{'\star}_l)\delta_{r_l,r'_l}\Big)
\big\{\overline{u}(\bm{y}'_k,r'_k)O^\mu_k u(\bm{y}_k,r_k)\big\} 
\\&&\times  (2\pi)^{3}\sqrt{2E_{p^{\star}_k}2E_{p^{'\star}_k}} 
\delta^3(\bm{p^{'\star}}_k-\bm{y}'_k))\delta_{r'_k,s_{p'}}\delta^3(\bm{p^{\star}}_k-\bm{y}_k))\delta_{r_k,r_p}.
\end{eqnarray*}
Substituting this expression into the formula for the hadronic current (\ref{A-ScatteringAmplitude-Nucleus-current-1}), and taking into account that the active $k$th nucleon can be in any of the $A$ places in the nucleus, which gives the factor $A$, one gets
{\small
\begin{eqnarray*}
H^\mu_{mn}(x)&\!\!=\!\!&\sum^A_{k}
\int\Big[\prod^{A}_{j,i}\frac{d\bm{p}^{'\star}_j\ d\bm{p}^\star_i}{(2\pi)^6 \sqrt{4E_{\bm{p'}^\star_j}E_{\bm{p}^\star_i}}}\Big]
\Big[\prod^{A-1}_{l\ne k}(2\pi)^{3}2E_{\bm{p}^\star_l}\delta^3(\bm{p}^{\star}_l-\bm{p}^{'\star}_l)\delta_{r^{}_l,r'_l}\Big]
\tilde{\psi}^{*}_m(\{{p'}^\star\}) \tilde{\psi}_n(\{p^\star\})\Phi^*_m(p') \Phi_n(p)
\\&&\times
\int\frac{d\bm{y}'_k d\bm{y}^{}_k e^{i x (y'_k-y^{}_k) }
\delta^3(\bm{p^{'\star}}_k-\bm{y}'_k)\delta^3(\bm{p^{\star}}_k-\bm{y}^{}_k)
}{\sqrt{4E_{\bm{y}^{}_k} E_{\bm{y}'_k}}} 
\sqrt{4E_{p^{\star}_k}E_{p^{'\star}_k}} 
 [\overline{u}(\bm{y}'_k,k'_k)O^\mu_k u(\bm{y}_k,r_k)]  
  .\end{eqnarray*}}%
Integration in this expression over $d\bm{y}'_k d\bm{y}^{}_k$, summation over
coinciding spin indices of spectator nucleons
and separation integration over $\bm{p}^{\star}_k$ and $\bm{p}^{'\star}_k$, give the  expression
\begin{eqnarray*}
H^\mu_{mn}(x)&=& 
\sum^A_{k}\int\Big[\prod^{A}_{j,i\ne k} \frac{d\bm{p}^{'\star}_j d\bm{p}^\star_i 
e^{i x (p^{'\star}_j-p^\star_i) } \delta^3(\bm{p}^{\star}_i-\bm{p}^{'\star}_j)} 
{(2\pi)^3\sqrt{E_{\bm{p'}^\star_j}E_{\bm{p}^\star_i}}}  E_{\bm{p}^\star_i} \Big]
\big\{\overline{u}(\bm{p^{'\star}}_k,r_{p'_k})O^\mu_k u(\bm{p^{\star}}_k,r_{p_k})\big\}
\times \\&&\times
\tilde{\psi}^{*}_m(\{{p'}^\star\}) \tilde{\psi}_n(\{p^\star\})
[\Phi^*_m(p') \Phi_n(p)]
\frac{d\bm{p}^{'\star}_k d\bm{p}^\star_k } {(2\pi)^{6}\sqrt{4E_{\bm{p'}^\star_k}E_{\bm{p}^\star_k}}} 
, \end{eqnarray*}
where after integration over $d\bm{p}^{'\star}_j$ (except for $\bm{p}^{'\star}_k$), one gets
\begin{eqnarray*}
H^\mu_{mn}(x)&=&
\big[\sqrt{4P^{0'}_m P^0_n}\delta^3(\bm{q}+\bm{P}_n-\bm{P}'_m) \delta^3(\sum^A_{i=1}\bm{p}^\star_i)
\big] \sum^A_{k} \int\Big[\prod^{A}_{i\ne k} \frac{ E_{\bm{p}^\star_i}\ 
d\bm{p}^\star_i } {(2\pi)^3\sqrt{E_{\bm{p}^\star_i}E_{\bm{p}^\star_i}}}  \Big]
\times  \\&&\times~~
\widetilde{\psi}^{*}_m(\{p^{(k)}_\star\}) \widetilde{\psi}_n(\{p^\star\})
\big\{ \overline{u}(\bm{p^{'\star}}_k,r_{p'_k})O^\mu_k u(\bm{p^{\star}}_k,r_{p_k}) \big\}
\frac{d\bm{p}^{'\star}_k d\bm{p}^\star_k 
e^{i x (p^{'\star}_k-p^\star_k) } } {\sqrt{ 4E_{\bm{p'}^\star_k}E_{\bm{p}^\star_k}}}
.\end{eqnarray*}
It is used here that the product of functions (\ref{A-ScatteringAmplitude-Phi_m-times Phi_n}) can be written as
  \cite{Bednyakov:2018mjd,Bednyakov:2021ppn}
\begin{equation} \label{A3DM-ScatteringAmplitude-Phi*(p')*Phi(p)}
\Phi^*_m(p') \Phi_n(p)=(2\pi)^6 \sqrt{4P^{0'}_mP^0_n}
\delta^3(\bm{q}+\bm{P}_n-\bm{P}'_m) \delta^3(\sum^A_{i=1}\bm{p}^\star_i),
\end{equation}
and the notation
$\widetilde{\psi}^{*}_m(\{p^{(k)}_\star\})\equiv \widetilde{\psi}^{*}_m(\{p^\star\},  \bm{p^{'\star}}_k\ne \bm{p^{\star}}_k, r_{p'_k} \ne r_{p_k})$, is introduced,  
where the argument $\{p^{(k)}_\star\}$ is the same as $\{p^\star\}$, except that at the $k$th place in $\{p^{ (k)}_\star\}$ there is not the pair $(\bm{p^{\star}}_k, r_{p_k})$ but another pair $(\bm{p^{'\star}} _k, r_{p'_k})\ne (\bm{p^{\star}}_k, r_{p_k})$.
\par
The next step is the integration of expression
(\ref{A-ScatteringAmplitude-S-Matrix-and-M-Element-via-H-and-L}) over $x$,
which
\footnote{Relation $\displaystyle \int d^4 x \, e^{i x (p^{'\star}_k-p^\star_k - q )} =
(2\pi)^4\delta^4(p^{\star}_k+q-p^{'\star}_k)$ was used.}
after one more integration over $ d\bm{p}^{'\star}_k$ due to 
$\delta^3(\bm{p^{\star}}_k+\bm{q}-\bm{p^ {'\star}}_k)$
reduces the integral $\int d^4 x \, H^\mu_{nm}(x) \, L_\mu^\chi(x)$ to the following form:
{\small \begin{eqnarray*} 
&&\!\!\!\!\!\!\!\!\!\!\!\! (2\pi)^4 \big[\sqrt{4P^{0'}_m P^0_n}\delta^3(\bm{q}+\bm{P}_n-\bm{P}'_m) \delta^3(\sum^A_{i=1}\bm{p}^\star_i)\big] 
\overline{u}_\chi(\bm{k}',s') O_{\mu} u_\chi (\bm{k},s) (2\pi)^3\sum^A_{k} \int
\Big(\prod^{A}_{i} \frac{ d\bm{p}^\star_i} {(2\pi)^3 }\Big) 
\times \\&& \times
 \delta(p^{\star}_{0,k}+q_0-p^{'\star}_{0,k})
\frac{\overline{u}(\bm{p^{'\star}}_k\!=\!\bm{p^{\star}}_k+\bm{q},r_{p'_k})O^\mu_k u(\bm{p^{\star}}_k,r_{p_k})} {\sqrt{4E_{\bm{p'}^\star_k=\bm{p^{\star}}_k+\bm{q}} E_{\bm{p}^\star_k}}} 
\widetilde{\psi}^{*}_m(\{p^{(k)}_\star\}, \bm{p^{'\star}}_k\! =\! \bm{p^{\star}}_k+\bm{q} ) 
\widetilde{\psi}_n(\{p^\star\})
. \end{eqnarray*} }%
Equating this integral to the definition of matrix element
(\ref{A-ScatteringAmplitude-S-Matrix-and-M-Element-via-H-and-L}),
 using the equality $\delta^4(k + P_n -k'- P'_m) = \delta^3(\bm{q}+\bm{P}_n-\bm{P}'_m)\delta(q_0 + P_{0,n}- P'_{0,m})$ and $q=k-k'$, 
 introducing notation for the lepton current 
 $l_\mu(k',k,s',s') \equiv \overline{u}_\chi(\bm{k}',s') O_{\mu} u_\chi (\bm{k},s) $, one gets the matrix element in the form
 {\small
 \begin{eqnarray*} 
i \mathcal{M}_{mn}\delta(q_0\!+\!P_{0,n}\!-\!P'_{0,m})
\!\!&=&\!\! \dfrac{i G_{\rm F}}{\sqrt{2}}  l_\mu(k',k,s',s) \sqrt{4P^{0'}_m P^0_n}
\sum^A_{k} \int\Big[\prod^{A}_{i} \frac{ d\bm{p}^\star_i} {(2\pi)^3}\Big] (2\pi)^3\delta^3(\sum^A_{i=1}\bm{p}^\star_i)
\times \\&&\!\!\!\!\!\!\!\!\!\!\!\!\!\!\!\!\!\!\!\!\times 
\delta(q_0+p^{\star}_{0,k}-p^{'\star}_{0,k}) 
\frac{\overline{u}(\bm{p^{\star}}_k+\bm{q},r_{p'_k})O^\mu_k u(\bm{p^{\star}}_k,r_{p_k})}{\sqrt{4E_{
\bm{p^{\star}}_k+\bm{q}} E_{\bm{p}^\star_k}}} 
\widetilde{\psi}^{*}_m(\{p^{(k)}_\star\},\bm{p^{\star}}_k+\bm{q} ) \widetilde{\psi}_n(\{p^\star\})
.\end{eqnarray*}}%
The delta functions (with common argument $q_0$) on the right and left sides of this relation   
cancel each other only  if the nuclear integrity condition 
(\ref{eq:ScatteringAmplitude-Energy-and-Identity-Conservation})  is met
$$ P_{0,n}\!-\!P'_{0,m} \equiv - T_A - \Delta\varepsilon_{mn}= p^{\star}_{0,k}-p^{ '\star}_{0,k} \equiv \sqrt{m^2+ {\bm{p^{\star}_k}}^2} - \sqrt{m^2+ (\bm{p^{ \star}}_k+\bm{q})^2}.$$
Then integration over $\bm{p}^\star_k$ in the hadronic current $h^\mu_{mn}$ 
(see (\ref{eq:ScatteringAmplitude-M-Element-via-H-and-L}))
can also be removed using the delta function of the form
\begin{equation}
\label{A-ScatteringAmplitude-Energy-and-Identity-Conservation}
\delta( - T_A - \Delta\varepsilon_{mn}+\sqrt{m^2+ {\bm{p^{\star}_k}}^2} -  \sqrt{m^2+ (\bm{p^{\star}}_k+\bm{q})^2})
\equiv  \delta\big(f(\bm{p^{\star}_k})\big) = 1 .
\end{equation}
This allows one to take the single-particle hadronic current at $\bm{\bar{p}^{\star}_k}(\bm{q})$ 
outside the integral sign, 
where $\bm{\bar{p}^{\star}_k}(\bm{q})$ is the {\em $\bm{q}$-dependent}
solution of the equation $ f(\bm{\bar{p}^{\star}_k}) = 0$, 
and finally get
{\small
\begin{eqnarray*}
h^\mu_{mn}=\sum^A_{k} 
\frac{ \overline{u}(\bm{\bar{p}}^\star_k+\bm{q},s_{p_k})O^\mu_k u(\bm{\bar{p}^{\star}}_k,r_{p_k})}
{\sqrt{4E_{\bm{\bar{p}}^\star_k}E_{\bm{\bar{p}}^\star_k+\bm{q}}}} 
\int \prod^{A}_{i\ne k}\frac{d\bm{p}^\star_i  }{(2\pi)^3}
\widetilde{\psi}^{*}_m(\{p^{(k)}_\star\},\bm{\bar{p}^{\star}}_k+\bm{q} )
\widetilde{\psi}_n(\{p^{(k)}_\star\}, \bm{\bar{p}^{\star}}_k) 
(2\pi)^{3} \delta^3(\sum^A_{i=1}\bm{p}^\star_i)
.\end{eqnarray*}}%
Formally, one can restore the integration over $A$ momenta in the form
\begin{eqnarray}\nonumber
\label{A-ScatteringAmplitude-h-mu-mn-with-delta-p_k}
h^\mu_{mn}&=& \sum^A_{k} \frac{ \overline{u}(\bm{\bar{p}}^\star_k+\bm{q},s_{p_k})O^\mu_k u(\bm{\bar{p}^{\star}}_k,r_{p_k})} {\sqrt{4E_{\bm{\bar{p}}^\star_k}E_{\bm{\bar{p}}^\star_k+\bm{q}}}} 
\times\\&\times& \int \prod^{A}_{i=1}\frac{d\bm{p}^\star_i   \delta\big(f(\bm{p^{\star}_k})\big)}{(2\pi)^3}
\widetilde{\psi}^{*}_m(\{p^{(k)}_\star\},\bm{p^{\star}}_k\!+\!\bm{q} )\widetilde{\psi}_n(\{p^\star\}) 
(2\pi)^{3} \delta^3(\sum^A_{i=1}\bm{p}^\star_i)
.\end{eqnarray} 

\paragraph{\em Formula for the hadronic structure ${f^k_{mn}(\bm{q})}$.}
Let us demonstrate that the multidimensional integral in formula 
(\ref{A-ScatteringAmplitude-h-mu-mn-with-delta-p_k})  is 
the matrix element $ \langle m|e^{i \bm{q}\hat{\bm{X}}_k} |n\rangle$.
To this end, we note that the operator $\displaystyle e^{i\bm{q}\hat{\bm{X}}}$ increases the 3-momentum $\bm{p}$ of the state $|\bm{p}\rangle $ by momentum $\bm{q}$, transforming the state 
$|\bm{p}\rangle$ into the state $|\bm{p}+\bm{q}\rangle$ by the operation
\footnote{Justification of formula (\ref{A-ScatteringAmplitude-shift_q}) can be found, for example, in
\cite{Bednyakov:2021ppn}.}
\begin{equation}
\label{A-ScatteringAmplitude-shift_q}
e^{i\bm{q}\hat{\bm{X}}}|\bm{p}\rangle =\frac{\sqrt{2E_{\bm{p}}}}{\sqrt{2E_{\bm{p}+\bm{q}}}}|\bm{p}+\bm{q}\rangle.
\end{equation}
From (\ref{A-ScatteringAmplitude-shift_q}) and the normalization condition of the single-particle states
$\langle \bm{k}|\bm{p} \rangle = (2\pi)^32E_{\bm{p}}\delta^3(\bm{p}-\bm{k})$ one gets
\begin{equation}\label{A-ScatteringAmplitude-one-nucleion-shift}
\langle \bm{k}| e^{i\bm{q}\hat{\bm{X}}}|\bm{p}\rangle 
=\frac{\sqrt{2E_{\bm{p}}}}{\sqrt{2E_{\bm{p}+\bm{q}}}} \langle \bm{k}| \bm{p}+\bm{q}\rangle
=(2\pi)^3  \sqrt{2E_{\bm{p}} 2E_{\bm{p}+\bm{q}}}\delta^3(\bm{p}+\bm{q}-\bm{k}).
\end{equation}
By definition, the matrix element $\langle m| e^{i \bm{q}\hat{\bm{X}}_k} |n\rangle$ should be calculated in the nuclear rest frame, 
where the nuclear state wave function $|n\rangle$ is given by (\ref{A:ScatteringAmplitude-n_state}).
Let us write them in the form
\begin{eqnarray}     \label{A-ScatteringAmplitude-NuclearStates-in-cms}
\langle m|=\!\int\!\phi^*_m(0)\Big(\prod^{A}_{j}d\tilde{\bm{p}'}^\star_j\Big)\frac{\tilde{\psi}^{*}_m(\{p'_\star\})}{\sqrt{{A!}}}\! \langle\{p'_\star\}| \text{~~and~~}   |n\rangle = \int\!\phi_n(0)\Big(\prod^{A}_{i}d\tilde{\bm{p}}^\star_i\Big)\frac{\tilde{\psi}_n(\{p_\star\})}{\sqrt{{A!}}} |\{p_\star\}\rangle. \qquad 
\end{eqnarray}
With (\ref{A-ScatteringAmplitude-NuclearStates-in-cms}),
the matrix element $f^k_{mn}\equiv\langle m| e^{i \bm{q}\hat{\bm{X}}_k} |n\rangle$ becomes
\begin{eqnarray}
\label{A-ScatteringAmplitude-f^k_mn-via-Q_k}
f^k_{mn}( \bm{q})&=& \int C\ \Big(\prod^{A}_{j,i}\frac{d\bm{p}^{'\star}_j d\bm{p}^\star_i}{(2\pi)^6 \sqrt{2E_{\bm{p}'^\star_j}2E_{\bm{p}^\star_i}}}\Big)\frac{\tilde{\psi}^{*}_m(\{p'_\star\})\tilde{\psi}_n(\{p^\star\})}{A!} \ {Q_k}( \bm{q}), 
\end{eqnarray}
where $C\equiv \phi^*_m(0)\phi_n(0)\equiv (2\pi)^3 \delta^3(\sum^A_j\bm{p}^{\star}_j)$,
and the matrix element of the $k$th nucleon shift operator $e^{i \bm{q}\hat{\bm{X}}_k}$ 
with respect to the many-particle state of free nucleons is introduced
\begin{eqnarray}
\label{A-ScatteringAmplitude-Operator-Q_k}
Q_k( \bm{q})&\equiv&  \langle\{p'_\star\}|e^{i \bm{q}\hat{\bm{X}}_k}|\{p_\star\}\rangle
= \langle...,p^{'\star}_k,...|e^{i \bm{q}\hat{\bm{X}}_k}|...,p^\star_k,...\rangle
. \end{eqnarray}
It can be presented as a product of $(A-1)$ non-interacting nucleons,
 $$\displaystyle 
 \langle...,p^{'\star}_{k-1},p^{'\star}_{k+1},...|...,p^\star_{k-1},p^\star_{k+1},...\rangle= 
 (A-1)!\Big[\prod^{A-1}_{l\ne k}(2\pi)^{3}2E_{\bm{p}^\star_l}\delta^3(\bm{p}^{\star}_l-\bm{p}^{'\star}_l)\delta_{r_l,r'_l}\Big], $$
 and the matrix element of the action of the operator $e^{i \bm{q}\hat{\bm{X}}_k}$ on the $k$th nucleon
\begin{eqnarray}\label{A-ScatteringAmplitude-Operator-Q_k-01}
Q_k( \bm{q}) &=& \langle...,p^{'\star}_{k-1},p^{'\star}_{k+1},...|...,p^\star_{k-1},p^\star_{k+1},...\rangle
\langle \bm{p}^{'\star}_k|e^{i \bm{q}\hat{\bm{X}}_k}|\bm{p}^\star_k\rangle \delta_{r_k,r'_k}.
\end{eqnarray}
From formula (\ref{A-ScatteringAmplitude-one-nucleion-shift}) 
for the single-particle nucleon state shift operator one obtains
\begin{equation}
\label{A-ScatteringAmplitude-Shift-One-particle}
\langle \bm{p}^{'\star}_k|e^{i \bm{q}\hat{\bm{X}}_k}|\bm{p}^\star_k\rangle =(2\pi)^3 \sqrt{2E_{\bm{p}^\star_k} 2E_{\bm{p}^\star_k+\bm{q}}} \delta^3(\bm{p}^\star_k+\bm{q}-\bm{p}^{'\star}_k).
\end{equation}
 Then formula (\ref{A-ScatteringAmplitude-Operator-Q_k-01}) becomes
\begin{eqnarray}
\label{A-ScatteringAmplitude-Shift-k-4-N-Q_k}
Q_k( \bm{q}) =A!\Big[\prod^{A-1}_{l\ne k}(2\pi)^{3}2E_{\bm{p}^\star_l}\delta^3(\bm{p}^{\star}_l-\bm{p}^{'\star}_l)\delta_{r_l,r'_l}\Big] \delta_{r_k,r'_k}\sqrt{2E_{\bm{p}^\star_k} 2E_{\bm{p}^\star_k+\bm{q}}}(2\pi)^{3}\delta^3(\bm{p}^\star_k+\bm{q}-\bm{p}^{'\star}_k)
, \qquad \end{eqnarray} 
where one takes into account that the $k$th nucleon can be located at any of the $A$ possible positions in the $A$-nucleus, 
which gives an additional permutation factor $A$.
From formula (\ref{A-ScatteringAmplitude-Shift-k-4-N-Q_k}) it follows that
the matrix element $Q_k(\bm{q})$ does not depend on the spin indices.
\par
Further, from the factorization of the spin and momentum dependences
(\ref{eq:ScatteringAmplitude-factorize_spin}) and
normalization conditions (\ref{eq:ScatteringAmplitude-spin_functions_norm})
it follows that the value (\ref{A-ScatteringAmplitude-f^k_mn-via-Q_k})
is nonzero only when the spin wave functions of the initial and final states are the same,
The spin structure of the nucleus does not change under the action of the shift operator.
As a result, the product of nuclear wave functions enters into formula
(\ref{A-ScatteringAmplitude-f^k_mn-via-Q_k}) only via its "impulse component"\/
\begin{equation}\label{A-ScatteringAmplitude-f^k_mn-Spin-cancellation}
\widetilde{\psi}^{*}_m(\{\bm{p}'_\star\})\chi^*_m(\{r'\})\, 
\widetilde{\psi}_n(\{\bm{p}_\star\})\chi_n(\{r\})\prod^{A}_{l}\delta_{r_l,r'_l} = 
\widetilde{\psi}^{*}_m(\{\bm{p}'_\star\})  \widetilde{\psi}_n(\{\bm{p}_\star\})
. \end{equation}
Substituting expression (\ref{A-ScatteringAmplitude-Operator-Q_k-01}) 
into (\ref{A-ScatteringAmplitude-f^k_mn-via-Q_k})
 with allowance for (\ref{A-ScatteringAmplitude-f^k_mn-Spin-cancellation}),
after integration over $d\bm{p}^{'\star}_j$ (without $d\bm{p}^{'\star}_k$) due to $\delta^3(\bm{p}^ {\star}_l-\bm{p}^{'\star}_l)$ 
one arrives at the following:
\begin{eqnarray}\label{A-ScatteringAmplitude-f^k_mn(q)}
f^k_{mn}( \bm{q})= 
\int\Bigg[ \dfrac{d\bm{p}^\star_k}{(2\pi)^3} \prod^{A}_{i\ne k}\frac{C\, d\bm{p}^\star_i}{(2\pi)^3 }\Bigg]
\frac{\langle \bm{p}^{'\star}_k|e^{i \bm{q}\hat{\bm{X}}_k}|\bm{p}^\star_k\rangle  }{(2\pi)^3 \sqrt{4E_{\bm{p'}^\star_k}E_{\bm{p}^\star_k}}} \widetilde{\psi}^{*}_m(\{\bm{p}_\star\},\bm{p}^{'\star}_k \ne \bm{p}^\star_k)  
\widetilde{\psi}_n(\{\bm{p}_\star\})
d\bm{p}^{'\star}_k.\qquad 
\end{eqnarray}
Taking into account the explicit expression for the one-particle matrix element 
(\ref{A-ScatteringAmplitude-Shift-One-particle}), which is proportional to 
$\delta(\bm{p}^{\star}_k- \bm{p}'^\star_k+\bm{q})$, 
and performing subsequent integration over $d\bm{p}^{'\star}_k$, 
which leads to cancelation of normalization factors $(2\pi)^3 \sqrt{4E_{\bm{p'}^\star_k}E_{\bm{p}^\star_k}}$ 
and to the "shift"\/ of the argument of the wave function of the final state of the nucleus, $\bm{p}^{\star}_k\to\bm{p}^\star_k+\bm{q} $, one gets
\begin{eqnarray}\label{A-ScatteringAmplitude-f^k_mn(q)-fino}
f^k_{mn}(\bm{q})=\int\Big[\prod^{A}_{i}\frac{d\bm{p}^\star_i}{(2\pi)^3}\Big]
\widetilde{\psi}^{*}_m(\{\bm{p}^{(k)}_\star\}, \bm{p}^\star_k+\bm{q})\widetilde{\psi}_n(\{\bm{p}_\star\}) 
(2\pi)^3\delta^3(\sum^A_{i=1}\bm{p}^\star_i)
.\quad \end{eqnarray}
Let us return back to the condition of simultaneous conservation of energy and integrity of the nucleus.
The need to apply this condition to the expression $\langle m| e^{i \bm{q}\hat{\bm{X}}_k} |n\rangle$ follows from the implicit assumption that the action of the operator $e^{i \bm{q}\hat{\bm {X}}_k}$, implemented as a momentum shift of the $k$th nucleon in the initial $|n\rangle$ state of the nucleus by $\bm{q}$, does not lead to the disintegration of the nucleus (the nucleus retains its integrity).
Otherwise, the use of the final state of the nucleus in the form of the wave function $\langle m|$ becomes meaningless.
Roughly speaking, something from the outside strikes the $k$th nucleon of the nucleus in the $|n\rangle$ state,
increasing the momenta of both this nucleon and the entire nucleus by $\bm{q}$, and
as a result, the nucleus, while maintaining its integrity, passes into the $\langle m|$ state.
The condition for just such a "course of things"\/ is relation
(\ref{A-ScatteringAmplitude-Energy-and-Identity-Conservation}).
Then integration over $\bm{p}^\star_k$ in expression (\ref{A-ScatteringAmplitude-f^k_mn(q)-fino})
can be removed by the delta function $ \delta\big(f(\bm{p^{\star}_k})\big)$, and it can be rewritten as
\begin{eqnarray}\label{A-ScatteringAmplitude-f^k_mn(q)-with-delta-p-star-k-01}
\bar{f}^k_{mn}( \bm{q})=
\int\Big[ \prod^{A}_{i }\frac{d\bm{p}^\star_i}{(2\pi)^3 }\delta\big(f(\bm{p^{\star}_k})\big)\Big]
\widetilde{\psi}^{*}_m(\{\bm{p}^{(k)}_\star\},\bm{p^{\star}}_k+\bm{q} )\widetilde{\psi}_n(\{\bm{p}^\star\})
 (2\pi)^{3} \delta^3(\sum^A_{i=1}\bm{p}^\star_i)  
.\quad \end{eqnarray}
Note that expression (\ref{A-ScatteringAmplitude-f^k_mn(q)-with-delta-p-star-k-01}) for $\bm{q}=0$
turns into the condition for the normalization of nuclear states in the rest frame of the nucleus
(\ref{A:ScatteringAmplitude-mn-norm}).
Indeed, at $\bm{q}=0$ one has $q_0=0$, and from the condition $\delta(q_0+P_{0,n}-P'_{0,m})=1$
follows $P_{0,n}-P'_{0,m} = - T_A - \Delta\varepsilon_{mn} =0$, i.e., 
$$f(\bm{p^{\star}_k})=  - T_A - \Delta\varepsilon_{mn}+\sqrt{m^2+ {\bm{p^{\star}_k}}^2} -  \sqrt{m^2+ {\bm{p^{\star}_k}}^2} \equiv - T_A - \Delta\varepsilon_{mn} = 0 \quad \text{for any}\ \ \bm{p^{\star}_k}.
$$ 
So $\delta(f(\bm{p^{\star}_k}))\equiv 1$ can be taken out of  the integral over the variable 
$\bm{p^{\star}_k}$, and formula (\ref{A-ScatteringAmplitude-f^k_mn(q)-with-delta-p-star-k-01}) becomes
\begin{eqnarray*} 
\bar{f}^k_{mn}( \bm{0})=\langle m| e^{i \bm{0}\hat{\bm{X}}_k} |n\rangle=
\int\Big[ \prod^{A}_{i }\frac{d\bm{p}^\star_i}{(2\pi)^3 }\Big]
\widetilde{\psi}^{*}_m(\{\bm{p}^{(k)}_\star\},\bm{p^{\star}}_k)\widetilde{\psi}_n(\{\bm{p}^\star\})
 (2\pi)^{3} \delta^3(\sum^A_{i=1}\bm{p}^\star_i) 
 \equiv  \langle m|n\rangle =\delta_{mn} .
 \end{eqnarray*}
Therefore, formula (\ref{A-ScatteringAmplitude-f^k_mn(q)-with-delta-p-star-k-01}) has the same meaning as expression (\ref{A-ScatteringAmplitude-f^k_mn(q)-fino}), being without the delta function.
  \par
Using formula (\ref{A-ScatteringAmplitude-f^k_mn(q)-with-delta-p-star-k-01}) for $\bar{f}^k_{mn}(\bm{q})$,  
one can write the hadronic current (\ref{A-ScatteringAmplitude-h-mu-mn-with-delta-p_k}) as
\begin{eqnarray}\label{A-ScatteringAmplitude-bar-h^k_mn(q)-with-delta-p-star-k}
\bar{h}^\mu_{mn}(\bm{q})=\sum^A_{k} \frac{ \overline{u}(\bm{\bar{p}^{\star}_k}+\bm{q},r'_{k})O^\mu_k\, u(\bm{\bar{p}^{\star}_k},r_{k})}{\sqrt{4E_{\bm{\bar{p}^{\star}_k}}E_{\bm{\bar{p}^{\star}_k}+\bm{q}}}} 
\bar{f}^k_{mn}( \bm{q}) \, \lambda^{mn}(r',r)
,\end{eqnarray}
where $\bm{\bar{p}^{\star}_k}$ is the solution of equation (\ref{A-ScatteringAmplitude-Energy-and-Identity-Conservation}).
It is worth comparing expression (\ref{A-ScatteringAmplitude-bar-h^k_mn(q)-with-delta-p-star-k}) 
with a similar formula from \cite{Bednyakov:2018mjd,Bednyakov:2021ppn} that
has the form
$$  h^\mu_{mn}(\bm{q})=  \sum_{k=1}^{A} \frac{\bar{u}(\bar{\bm{p}}+\bm{q},r'_k) O^\mu_k u(\bar{\bm{p}},r_k)}{\sqrt{4E_{\bar{\bm{p}}}E_{\bar{\bm{p}}+\bm{q}}}} {f}^k_{mn}( \bm{q})   \lambda^{mn}(r',r), 
 \quad \text{where}  $$
\begin{equation}\label{A-ScatteringAmplitude-f^k_mn(q)-OLD}
{f}^k_{mn}( \bm{q}) \equiv    
 \int \Big[\prod_{j=1}^{A}\frac{d\bm{p}^\star_j}{(2\pi)^3}\Big]  \widetilde{\psi}_m^*(\{\bm{p}^{(k)}_\star\} ) \widetilde{\psi}_n(\{\bm{p}_\star\})(2\pi)^3 \delta^3(\sum_{l=1}^A \bm{p}^\star_l)
.\end{equation}
To obtain {\em this} expression,   a {\em special assumption}\/ was made in 
\cite{Bednyakov:2018mjd,Bednyakov:2021ppn} about the possibility of  taking 
 the single-particle matrix element, 
$\dfrac{\bar{u}(\bar{\bm{p}}+\bm{q},r'_k) O^\mu_k u(\bar{\bm{p}},r_k)}{ \sqrt{4E_{\bar{\bm{p}}}E_{\bar{\bm{p}}+\bm{q}}}}$, out of the integration right at the momentum $\bar{\bm{p}}$, 
being the solution of equation (\ref{A-ScatteringAmplitude-Energy-and-Identity-Conservation}).
In formula (\ref{A-ScatteringAmplitude-bar-h^k_mn(q)-with-delta-p-star-k})
this factor is obtained automatically, but one has to slightly redefine the form factor function, 
i.e.,  instead of ${f}^k_{mn}( \bm{q})$ from (\ref{A-ScatteringAmplitude-f^k_mn(q)})
one takes $\bar{f}^k_{mn}( \bm{q}) $ from (\ref{A-ScatteringAmplitude-f^k_mn(q)-with-delta-p-star-k-01}).
Nevertheless, this "redefinition"\/ does not play any role,
since the "physical"\/nuclear form factors of protons and neutrons
$F_{p/n}(\bm{q})$ are defined in terms of the functions ${f}^k_{mn}( \bm{q})$ only  formally
(see,  for example formula (\ref{eq:41chiA-CrossSection-Coh-vs-InCoh-form-factors-with-s-sprime})),
i.e., {\em without explicit calculation}\/  of the above-mentioned multidimensional integrals.
\par
Based on the definition of $f^k_{mn}(\bm{q})$, for example in the
form of (\ref{A-ScatteringAmplitude-f^k_mn(q)-OLD}),
and on the symmetry properties of the nuclear wave function, one can conclude
\cite{Bednyakov:2018mjd,Bednyakov:2021ppn}
that the structure factor $f^k_{mn}$ does not depend on the number $k$, 
and depends only on whether this $k$ corresponds to the proton or the neutron.
This property is used below.

\paragraph{\em Coherent and incoherent terms of the cross section ${\chi A\to \chi A^{(*)}}$.} 
The measurable differential cross section can be obtained by averaging (over all possible initial nucleus states
$|n\rangle$) and summing (over all possible final states $|m\rangle$)
of the differential cross section, which corresponds to the nuclear transition from the initial state $|n\rangle$ 
to the final state $|m\rangle$ due to the interaction with the $\chi$ particle (\ref{eq:Kinematics-dSigma-po-dT-A-kin-rel})
\begin{equation}
\label{A41RchiA-CrossSection-Coh-vs-InCoh-CrossSection-definition}
\frac{d\sigma}{dT_A}(\chi A\to \chi A^{(*)}) = \sum_{n,m}\omega_n \dfrac{|{i\cal M}_{mn}|^2}{2^5\pi  |\bm{k^l_\chi}|^2 m_A} C_{mn} , \quad \text{where}\quad  \sum_n \omega_n=1 
.\end{equation}
With the matrix element from (\ref{eq:Kinematics-dSigma-po-dT-A-kin-rel})
summed over the nucleon spin indices \cite{Bednyakov:2018mjd,Bednyakov:2019dbl,Bednyakov:2021ppn,Bednyakov:2022dmc}
\begin{eqnarray}
\label{A41RchiA-CrossSection-Coh-vs-InCoh-MatrixElement-with-s-sprime}
i\mathcal{M}_{mn}^{s's}  = i\frac{G_F}{\sqrt{2}} \frac{m_A}{m_n}C_{1,mn}^{1/2} \sum_{k=1}^{A} \sum_{r'r} f^k_{mn}\lambda^{mn}(r',r) (l_{s's}\,h^k_{r'r})
, \end{eqnarray}
the differential cross section (\ref{A41RchiA-CrossSection-Coh-vs-InCoh-CrossSection-definition})
can be written as a sum of two terms from
(\ref{eq:41chiA-CrossSection-Coh-vs-InCoh-CrossSection-with-s-sprime-and-Tmns})
\begin{eqnarray}
\label{A41RchiA-CrossSection-Coh-vs-InCoh-Term-nn-definition-with-s-sprime}
T^{s's}_{m=n} &\equiv& g^{}_\text{c}\sum^A_{k,j}\sum_{n}\omega_n\Big[  f^k_{nn}f^{j*}_{nn}
\sum_r  (l_{s's}\,h^k_{rr})\sum_{x} (l_{s's}\,h^j_{xx})^{*}\Big]\text{~~and~~}\\
\label{A41RchiA-CrossSection-Coh-vs-InCoh-Term-mn-definition-with-s-sprime}
T^{s's}_{m\ne n}&\equiv&g^{}_\text{i}  \sum^A_{k,j}\sum_{n}\omega_n  \Big[ \sum_{m\ne n}  f^k_{mn} f^{j*}_{mn}  \sum_{r'r}\lambda^{mn}_{r'r}(l_{s's}\, h^k_{r'r}) \Big(\sum_{x'x}\lambda^{mn}_{x'x}(l_{s's}\, h^j_{x'x}) \Big)^{\dag} \Big].  \qquad 
\end{eqnarray} 
Here, correction factors are introduced  in the form
\begin{eqnarray}\label{A41RchiA-CrossSection-Coh-vs-InCoh-g-coh-not-m-n-depended}
 g^{}_\text{i/c}\equiv  C^{}_{1,mn}C^{}_{mn}   \simeq 
 \dfrac {\Big(1+\dfrac{\varepsilon_n}{m_A}\Big)\Big(1+\dfrac{\varepsilon_m + T_A}{m_A}\Big)} {\sqrt{1+\dfrac{\bar{\bm{p}}^2}{m^2}}\Big(\sqrt{1+\dfrac{\bar{\bm{p}}^2}{m^2}} + \dfrac{T_A+\Delta\varepsilon_{mn}}{m}\Big) }   \dfrac{\Big(1+\dfrac{T_A}{m_A}\Big)/\Big(1+\dfrac{T_A+\varepsilon_m}{m_A}\Big)}{1+ \dfrac{\varepsilon_n }{m_A}\Big(1+ \dfrac{ m^2_\chi }{|\bm{k^l_\chi}|^2 }\Big)}. \qquad  
 \end{eqnarray}
The factors differ from 1 at a level of $O(10^{-3})$ and with the same accuracy 
they do not depend on the nuclear indices $m, n$ and the recoil energy $T_A$
\footnote{Since in the accepted approximation
one has  $\ \dfrac{\varepsilon_m + T_A}{m_A}\le 10^{-3}$, 
$\dfrac{\Delta\varepsilon_{mn}+T_A}{m}\le 10^{-3}$ and $\dfrac{|\bar{\bm{p}}|^2}{m^2} \le 0.01$.
}. 
For this reason, the summation over the index $n$ in formula
(\ref{A41RchiA-CrossSection-Coh-vs-InCoh-Term-nn-definition-with-s-sprime})
can be interpreted as the appearance of the form factors averaged over all possible initial states of the nucleus, 
which are given by formula (\ref{eq:41chiA-CrossSection-Coh-vs-InCoh-form-factors-with-s-sprime})
from the main text.
This allows expression (\ref{A41RchiA-CrossSection-Coh-vs-InCoh-Term-nn-definition-with-s-sprime})
to be separated into contributions of protons and neutrons as follows:
\begin{eqnarray*}
\dfrac{T^{s's}_{m=n}}{g_{\text{c}}}&=&\sum^Z_{k,j}F_p\sum^2_{r=1}(l_{s's}\,h^p_{rr}) {F^*_p} \sum^2_{x=1}(l_{s's}\,h^p_{xx})^{*}
+\sum^N_{k,j} F_n \sum^2_{r=1}(l_{s's}\,h^n_{rr}) {F^*_n} \sum^2_{x=1} (l_{s's}\,h^n_{xx})^{*}
+\\&&+\sum^Z_{k}\sum^N_{j} F_{p}\sum^2_{r=1} (l_{s's}\,h^p_{rr}) F_{n}^* \sum^2_{x=1}(l_{s's}\,h^n_{xx})^{*}
+ \sum^N_{k}\sum^Z_{j}F_{n}  \sum^2_{r=1} (l_{s's}\,h^n_{rr}) F_{p}^* \sum^2_{x=1}(l_{s's}\,h^p_{xx})^{*} 
. \end{eqnarray*}
The formula can be presented as the squared modulus of the sum over proton and neutron contributions:
\begin{eqnarray}
\label{A41RchiA-CrossSection-Coh-vs-InCoh-coherent_term-1-with-s-sprime}
T^{s's}_{m=n}=g_{\text{c}} \Big|\sum^Z_{k}\sum_{r}  (l_{s's}\,h^p_{rr})F_p+\sum^N_{j}\sum_{r}  (l_{s's}\,h^n_{rr})F_n\Big|^2 =g_{\text{c}} \Big|\sum_{f=p,n}\sum^{A_f}_{k=1}\sum_{r} (l_{s's}\,h^f_{rr})F_f\Big|^2.
\qquad \end{eqnarray} 
The second term from (\ref{eq:41chiA-CrossSection-Coh-vs-InCoh-CrossSection-with-s-sprime-and-Tmns}), or
expression (\ref{A41RchiA-CrossSection-Coh-vs-InCoh-Term-mn-definition-with-s-sprime}),
contains summation over both indices $m,n$.
To carry out this summation and to continue transformations of the term $T_{m\ne n}$, 
certain assumptions should be made about the unknown behavior of  $\lambda^{mn}_{r'r}$.
Recall that the spin functions of the nucleus are normalized by condition
(\ref{eq:ScatteringAmplitude-spin_functions_norm})
$$\chi_m^{*}(\{r\})\chi_n(\{r\}) = \delta_{nm}, \quad \chi_n^{*}(\{r'\})\chi_n(\{r\}) =\delta_{\{r'\}\{r'\}}.
$$
The definition (\ref{eq:ScatteringAmplitude-lambda_def}) was adopted 
for the products of the spin functions.
Since formula (\ref{eq:ScatteringAmplitude-lambda_def}) is valid for any value of $k$, 
the index $k$ of $r'_k$ and $r_k$ was omitted.
If one keeps the index $k$, expression (\ref{eq:ScatteringAmplitude-lambda_def}) takes the form
$$ \chi^*_m(\{r^{(k)}\},r'_k)\chi_n(\{r\},r^{}_k) \equiv \lambda^{mn}(r'_k, r^{}_k)
= \delta_{mn}\delta_{r'_kr^{}_k}+(1-\delta_{mn})\lambda^{mn}_{r'_kr^{}_k}.$$
Here
$\{r\}\equiv \{r_1, r_2, .., r_k,.., r_A\}$ is the set of all nucleon spin projections in the initial nucleus $A$, 
and $\{r^{(k)}\}$ is the similar set for the final nucleus, 
differing only by that  the $k$th place in the set $\{r^{(k)}\}$ is occupied by another number $r'_k$ 
not equal to $r_k$ from $\{r\}$.
 All other projection values are the same.
In other words, $\{r^{(k)}\}\equiv \{r_1, r_2, .., r'_k,.., r_A\}$.
It can be seen that for $m=n$ one has
\begin{equation} \label{A41RchiA-CrossSection-Coh-vs-InCoh-spin-n-n-normalization}
\chi^*_n(\{r^{(k)}\},r'_k)\chi_n(\{r\},r^{}_k) = \delta_{r'_kr^{}_k},
\end{equation}
i.e.,  all values in two sets of spin index variables $\{r^{(k)}\}$ and $\{r\}$ must fully coincide 
if the spin functions refer to the same ($n$th) internal quantum state of the nucleus.
In other words, if the nucleus has not changed, then all values of the spin indices of the nucleus
$\{r\}$ should also remain unchanged.
\par
Next, consider the product of spin functions 
included in the squared matrix element
(\ref{A41RchiA-CrossSection-Coh-vs-InCoh-Term-mn-definition-with-s-sprime})
{\em for the same}\/ active $k$th nucleon when $m\ne n$. 
It is
$$ \lambda^{mn}_{r'_kr^{}_k} [\lambda^{mn}_{x'_kx^{}_k}]^* =\chi^*_m(\{r^{(k)}\},r'_k)\chi_n(\{r\},r^{}_k)\cdot [ \chi^*_m(\{x^{(k)}\},x'_k)\chi_n(\{x\},x^{}_k)]^*
.$$ 
Performing complex conjugation and rearranging the functions, one gets
$$ \lambda^{mn}_{r'_k,r_k} [\lambda^{mn}_{x'_k,x_k}]^* =
 \big[ \chi^*_m(\{r^{(k)}\},r'_k)  \chi_m(\{x^{(k)}\},x'_k)\big]  \big[ \chi_n(\{r\},r_k) \chi^*_n(\{x\},x_k) \big]
 .$$
Then, according to (\ref{A41RchiA-CrossSection-Coh-vs-InCoh-spin-n-n-normalization}),
for each of these two products corresponding to the same
state of the nucleus ($m$ and $n$, respectively) to be different from zero, 
the sets of spin indices in each pair of products must completely coincide, i.e., 
$\{r^{(k)}\}=\{x^{(k)}\}$ and $\{r\}=\{x\}$.
 In other words, {\em for any}\/ $k$ there should be the following:
$$   \chi^*_m(\{r^{(k)}\},r'_k)  \chi_m(\{x^{(k)}\},x'_k)=  \delta_{r'_k, x'_k} \quad \text{and}\quad  
 \chi_n(\{r\},r_k) \chi^*_n(\{x\},x_k) = \delta_{r_k, x_k}. $$
As a result, it turns out that
\begin{equation}
\label{A41RchiA-CrossSection-Coh-vs-InCoh-NuclearSpinAmplitudeProduct}
[\lambda^{mn}_{r'_kr_k}][\lambda^{mn}_{x'_kx_k}]^*  
 = \delta_{r'_k, x'_k} \delta_{r_k, x_k}, \quad \text{and}\quad  |\lambda^{mn}_{r'_kr_k}|^2=1
.\end{equation}
Here the index $k$ explicitly takes into account the invariance of the type and number of the active nucleon.
The last equality in (\ref{A41RchiA-CrossSection-Coh-vs-InCoh-NuclearSpinAmplitudeProduct})
means that in the case of scattering {\em on the same active nucleon}, regardless of its number $k$ and 
of the change in the state of the nucleus $|n\rangle\to|m\rangle$,
any final orientation of the spin of this nucleon (index $r'_k$) is equally possible 
for any initial orientation of the spin of the active nucleon (index $r_k$).
Formally, this result is a consequence of the
normalization conditions for the spin wave functions
(\ref{A41RchiA-CrossSection-Coh-vs-InCoh-spin-n-n-normalization}).
There is also no operator that changes the spin of the nucleon and {\em remains standing}\/ between these spin functions, preventing their normalization condition  "to work"\/.
Indeed, if the {\em interaction did not occur}\/, then the initial $|n\rangle$ and final $|m\rangle$ states are completely independent. 
In these states there should be no correlation between the spin directions
of the "active"\/ nucleon (the concept of which is undefined in such a situation).
When {\em the interaction}\/ occured, transferring the nucleus from the $|n\rangle$ state to 
the $|m\rangle$ state, a correlation arises between the $r$- and $r'$-spins of the active nucleon, 
{\em the intensity}\/ of which (up to zero) is regulated by the scalar product $(l_{s's}\, h^{}_{r'r})$.
\par
Thus, {\em it can be accepted}\/ that the products of $\lambda^{mn}_{r'r}$
corresponding to the scattering on the same active nucleon
do not depend on the indices $m$ and $n$ and are equal to some averaged values for protons and neutrons
\footnote{When $x_k=r_k$ and $x'_k=r'_k$, the physical meaning of the square $|\lambda^{mn}_{r'_kr_k}|^2\equiv \lambda^{mn}_{ r'_kr_k} [\lambda^{mn}_{x'_k=r'_k, x_k=r_k}]^*$ is clear. It is the  probability for the $k$th nucleon to meet the lepton with its spin projection $r_k$ and sent the lepton away with the spin projection $r'_k$, provided that the nucleus passes from the $|n\rangle$ state to the $|m\rangle$ state. The meaning of the product $\lambda^{mn}_{+-} [\lambda^{mn}_{++}]^*$, where the initial spin index $r_k=-1$ does not coincide with the initial spin index $x_k =+1$, is unclear, {\em if one considers the same active nucleon}. Such a situation can apparently arise when $\lambda^{mn}_{+-}$ refers to one nucleon, and $[\lambda^{mn}_{++}]^*$ refers {\em to another}\/ one, i.e.,  when {\em two different active nucleons}\/ participate in the scattering process.
As shown below, such a scenario contributes negligibly.} 
\begin{equation}
\label{A41RchiA-CrossSection-Coh-vs-InCoh-spin-amplitudes-p-and-n} 
\lambda^{mn}_{r'r} \simeq \lambda^{p/n}_{r'r}.
\end{equation}
Therefore, due to (\ref{A41RchiA-CrossSection-Coh-vs-InCoh-spin-amplitudes-p-and-n}), the
spin factors $\lambda^{mn}_{r'r}\equiv \lambda^{f}_{r'r}$ (together with scalar products) can be taken out of 
the sum over $m,n$.
\par
Using conditions (\ref{A41RchiA-CrossSection-Coh-vs-InCoh-NuclearSpinAmplitudeProduct})
and (\ref{A41RchiA-CrossSection-Coh-vs-InCoh-spin-amplitudes-p-and-n}), 
one can remove  part of the summation over the indices $x',x$ (when $k=j$)
in formula (\ref{A41RchiA-CrossSection-Coh-vs-InCoh-Term-mn-definition-with-s-sprime}),
"separate"\/ protons from neutrons, and get the expression
\begin{eqnarray}\nonumber
\label{A41RchiA-CrossSection-Coh-vs-InCoh-Tssprime-mnen-with-lambdas_f}
\dfrac{T^{s's}_{m\ne n}}{g^{}_{\text{i}}}  &=&   \nonumber
\sum^Z_{k=j}\sum_{n} \omega_n\sum_{m\ne n} f^p_{mn} f^{p*}_{mn} \sum_{r'_kr_k} |(l_{s's}, h^p_{r'_kr_k}) |^2 
+\sum^N_{k=j} \sum_{n} \omega_n\sum_{m\ne n} f^n_{mn} f^{n*}_{mn}\sum_{r'_kr_k} |(l_{s's}, h^n_{r'_kr_k}) |^2
\\&+& \sum^Z_{k\ne j}{\sum_{n} \omega_n\sum_{m\ne n} f^{k}_{mn} f^{j*}_{mn}}
\sum_{r'_kr_k} \sum_{x'_jx_j}  \lambda^{p}_{r'_kr_k} [\lambda^{p}_{x'_jx_j}]^*  (l_{s's}, h^p_{r'_kr_k}) (l_{s's}, h^p_{x'_jx_j})^{\dag}
+\\&+& \nonumber \sum^N_{k\ne j}{\sum_{n} \omega_n\sum_{m\ne n} f^{k}_{mn} f^{j*}_{mn}}
\sum_{r'_kr_k} \sum_{x'_jx_j} \lambda^{n}_{r'_kr_k} [ \lambda^{n}_{x'_jx_j}]^*(l_{s's}, h^n_{r'_kr_k})(l_{s's}, h^n_{x'_jx_j})^{\dag}
+ \\&+&  \nonumber \sum^Z_{k}  \sum^N_{j}{\sum_{n} \omega_n\sum_{m\ne n}  f^p_{mn} f^{n*}_{mn}}
\sum_{r'r}\sum_{x'x} \lambda^{p}_{r'r} [\lambda^{n}_{x'x}]^* (l_{s's}, h^p_{r'r})(l_{s's}, h^n_{x'x})^{\dag}
,\end{eqnarray}
where the summation over only one index remains in the first two terms,
since both indices $k$ and $j=k$ "point"\/ to the same nucleon.
In the first term it is the proton, and in the second it is the neutron.
In the third and fourth terms, the indices $k$ and $j$ also point to
nucleons of the same type, but these nucleons are different, $k\ne j$.
In the last line, the index $k$ indicates a proton, and the index $j$ indicates a neutron.
\par
Given condition (\ref{A41RchiA-CrossSection-Coh-vs-InCoh-spin-amplitudes-p-and-n}),
calculation of  sums in (\ref{A41RchiA-CrossSection-Coh-vs-InCoh-Tssprime-mnen-with-lambdas_f})
should start with the first line, where the indices $k$ and $j$ "point"\/ to the same proton.
Since $k=j$, then, taking into account the condition of completeness of nuclear states
$\displaystyle \sum_m|m\rangle \langle m|=\hat{I},$
the condition for conservation of probabilities $\displaystyle \sum_n \omega_n=1,$
the condition for normalizing nuclear states $ \displaystyle \langle m|n\rangle =\delta_{mn},$
function definitions $\displaystyle f^k_{mn}(\bm{q})=\langle m|e^{i\bm{q}\hat{\bm{X}}_{k}}|n\rangle $ and
expressions for their squares (\ref{eq:41chiA-CrossSection-Coh-vs-InCoh-form-factors-with-s-sprime}),
one can sum over $n$ and $m$
in the first term of formula (\ref{A41RchiA-CrossSection-Coh-vs-InCoh-Tssprime-mnen-with-lambdas_f})
 and obtain the following sequence of expressions:
\begin{eqnarray} 
\label{A41RchiA-CrossSection-Coh-vs-InCoh-incoherent_term1}
\sum_n \omega_n\sum_{m\ne n} f_{mn}^k f_{mn}^{k*}  &=&\sum_n \omega_n \Big[\sum_{m} f_{mn}^k f_{mn}^{k*} -f_{nn}^k f_{nn}^{k*}\Big]=\\ &=&\sum_n \omega_n \Big[\langle n|e^{i\bm{q}\bm{X}_k}\sum_{m}|m\rangle\langle m|e^{-i\bm{q}\bm{X}_k}|n\rangle\Big] -|F_{p}(\bm{q})|^2 =1-|F_{p}(\bm{q})|^2.
\nonumber
\end{eqnarray}
For the second term in formula (\ref{A41RchiA-CrossSection-Coh-vs-InCoh-Tssprime-mnen-with-lambdas_f}),  
the similarly obtained expression is $1-|F_{n}(\bm{q})|^2$.
For the second (proton) line in (\ref{A41RchiA-CrossSection-Coh-vs-InCoh-Tssprime-mnen-with-lambdas_f}) 
one has $k\ne j$, i.e.,  these indices indicate {\em different}\/ protons.
Then,  taking into account all above-mentioned conditions as in the case with the derivation of expression
(\ref{A41RchiA-CrossSection-Coh-vs-InCoh-incoherent_term1}),
for protons (index $p$) one can write the following:
 \begin{equation}		
 \label{A41RchiA-CrossSection-Coh-vs-InCoh-incoherent_term2}
 \sum_n \omega_n\sum_{m\ne n} f_{mn}^k f_{mn}^{j*}  =  \langle\text{cov}(e^{i\bm{q}\hat{\bm{X}}_k}, e^{-i\bm{q}\hat{\bm{X}}_j})\rangle_p
 .\end{equation}
Here the covariance of the shift operators
  $e^{-i\bm{q}\hat{\bm{X}}_j}$ and $e^{i\bm{q}\hat{\bm{X}}_k}$ over $| n\rangle$ 
  is introduced   in the form
 \begin{equation}
 \label{A41RchiA-CrossSection-Coh-vs-InCoh-nn-Covariance}
\text{cov}_{nn}(e^{i\bm{q}\hat{\bm{X}}_k},e^{-i\bm{q}\hat{\bm{X}}_j})
\equiv \langle n|e^{i\bm{q}\hat{\bm{X}}_k}\, e^{-i\bm{q}\hat{\bm{X}_j}} |n\rangle - \langle n|e^{i\bm{q}\hat{\bm{X}}_k}|n\rangle \langle n|e^{-i\bm{q}\hat{\bm{X}}_j}|n\rangle
,\end{equation}
 and summation over initial states with factors $\omega_n$ 
is taken into account as follows:
\begin{equation}
\label{A41RchiA-CrossSection-Coh-vs-InCoh-Averaged-Covariance}
\langle\text{cov}(e^{i\bm{q}\hat{\bm{X}}_k},e^{-i\bm{q}\hat{\bm{X}}_j})\rangle_p 
\equiv \sum_n \omega_n \text{cov}_{nn}(e^{i\bm{q}\hat{\bm{X}}_k},e^{-i\bm{q}\hat{\bm{X}}_j}).
\end{equation}
In deriving formulas (\ref{A41RchiA-CrossSection-Coh-vs-InCoh-incoherent_term1})
and (\ref{A41RchiA-CrossSection-Coh-vs-InCoh-incoherent_term2}), 
the fact was used 
the "incomplete"\/ sum over $m$, i.e., 
$\displaystyle \sum_{m\ne n} f_{mn}^k f_{mn}^{j*}\, $, 
can be "completed"\/ and
represented as $\displaystyle\sum_{m} f_{mn}^k f_{mn}^{j*} - \sum_{m=n} f_{mn}^k f_{mn}^{j*}$.
  Using the completeness condition $\sum_m|m\rangle \langle m|=\hat{I}$, one arrives at expression (\ref{A41RchiA-CrossSection-Coh-vs-InCoh-nn-Covariance})
\begin{eqnarray*}
\sum_{m\ne n} f_{mn}^k f_{mn}^{j*} &=&  \sum_{m}\langle n|e^{i\bm{q}\bm{X}_k}|m\rangle\langle m|e^{-i\bm{q}\bm{X}_j}|n\rangle  -\sum_{m=n}\langle n|e^{i\bm{q}\bm{X}_k}|m\rangle\langle m|e^{-i\bm{q}\bm{X}_j}|n\rangle
=\\&=& \langle n|e^{i\bm{q}\hat{\bm{X}}_k} e^{-i\bm{q}\hat{\bm{X}}_j}|n\rangle-\langle n|e^{i \bm{q}\hat{\bm{X}}_k}|n\rangle \langle n|e^{-i\bm{q}\hat{\bm{X}}_j}|n\rangle.
\end{eqnarray*}
Expression (\ref{A41RchiA-CrossSection-Coh-vs-InCoh-Averaged-Covariance}) vanishes both for small transferred momenta $\bm{q}\to 0$ and for large $\bm{q}\to\infty$, i.e., 
$\lim_{\bm{q}\to 0}\langle\text{cov}(e^{-i\bm{q}\hat{\bm{X}}_j},e^{i\bm{ q}\hat{\bm{X}}_k})\rangle_p = 0$ and
$\lim_{\bm{q}\to \infty}\langle\text{cov}(e^{-i\bm{q}\hat{\bm{X}}_j},e^{i\bm {q}\hat{\bm{X}}_k})\rangle_p=0.$
A similar consideration is carried out for {\em different}\/ neutrons (the 3rd line in 
formulas (\ref{A41RchiA-CrossSection-Coh-vs-InCoh-Tssprime-mnen-with-lambdas_f})) and
is generalized to the joint case of protons and neutrons (the last line in formula
(\ref{A41RchiA-CrossSection-Coh-vs-InCoh-Tssprime-mnen-with-lambdas_f})).
Finally, the  "incoherent"\/ term (\ref{A41RchiA-CrossSection-Coh-vs-InCoh-Term-mn-definition-with-s-sprime}) 
appears as expression (\ref{eq:41chiA-CrossSection-Coh-vs-InCoh-CrossSection-Term-mn-with-s-sprime-final}) in the main text.

\bibliographystyle{JHEP} 
\bibliography{220519-ScatteringTheory}
\end{document}